\PassOptionsToPackage{dvipsnames,table}{xcolor}
\documentclass[a4paper,11pt]{article}
\usepackage{jheppub} 
\makeatletter
\def\@fpheader{}
\makeatother
\usepackage{float}
\usepackage{units}
\usepackage{amsmath}
\usepackage{graphicx}
\usepackage{dcolumn}
\usepackage{pbox}
\usepackage{amssymb}
\usepackage{slashed}
\usepackage{gensymb}
\usepackage{textcomp}
\usepackage{amssymb}
\usepackage{mathrsfs}
\usepackage[dvipsnames]{xcolor}

\usepackage{multirow}
\usepackage{lineno}
\usepackage{autobreak}
\usepackage{lipsum}
\usepackage{cancel}
\usepackage{xparse}
\usepackage{booktabs}
\usepackage{bm}
\usepackage{comment}
\usepackage{subcaption}
\usepackage{array}    
\usepackage{makecell}
\usepackage{bbding}
\usepackage{pifont}
\usepackage{wasysym}
\usepackage{tabu}

\NewDocumentCommand\semiloop{O{black}mmmO{}O{above}}
{
\draw[#1] let \p1 = ($(#3)-(#2)$) in (#3) arc (#4:({#4+180}):({0.5*veclen(\x1,\y1)})node[midway, #6] {#5};)
}

\definecolor{MyLightBlue}{rgb}{0.22,0.51,0.9}
\definecolor{BrickRed}{rgb}{0.8, 0.25, 0.33}
\RequirePackage{hyperref}
\hypersetup{colorlinks, citecolor=blue,linkcolor=BrickRed, urlcolor=MyLightBlue}

\newcommand{\be}{\begin{equation}}
\newcommand{\ee}{\end{equation}}

\newcommand{\minus}{\text{-}}

\newcommand{\de}{\delta}

\newcommand{\te}{\theta}

\newcommand{\La}{\Lambda}

\title{Fermion Mass Hierarchy and a High Quality Axion From Gauged \boldmath{$U(1)$} Flavor Symmetry}

\author[a]{\bf K.S. Babu,}
\emailAdd{babu@okstate.edu}
\affiliation[a]{Department of Physics, Oklahoma State University, Stillwater, OK 74078, USA}

\author[a]{\bf Sai Charan Chandrasekar,}
\emailAdd{sai.sekar@okstate.edu}

\author[b]{\bf Zurab Tavartkiladze}
\emailAdd{zurab.tavartkiladze@iliauni.edu.ge}
\affiliation[b]{Center for Elementary Particle Physics, ITP, Ilia State University, 0179 Tbilisi, Georgia}

\abstract{ We present a class of models based on a gauged $U(1)_F$ flavor symmetry that explains the hierarchical structure of fermion masses and mixings via the Froggatt-Nielsen (FN) mechanism, while also solving the strong CP problem by the Peccei-Quinn (PQ) mechanism. A global $U(1)_{\rm PQ}$ symmetry with a nonzero QCD anomaly emerges accidentally in this setup as a byproduct of the gauged $U(1)_F$ symmetry. The resulting axion is shown to be of high quality, with the axion potential safeguarded against quantum gravity corrections by the gauge symmetry.  Three models, which are generalizations of the Dine-Fischler-Srednicki-Zhitnitsky (DFSZ) axion model, are presented realizing this idea. The right-handed neutrino mass scale is identified as the Froggatt-Nielsen scale in these models.
We present explicit UV completions of the FN sectors of these models and show that they preserve the high quality of the axion. 
In these models, the axion acts as a flavon field, leading to testable predictions in flavor-changing decays of neutral mesons.  The axion also serves as the dark matter of the universe  with the right amount of relic abundance without causing cosmological domain wall problems.  Baryon asymmetry of the universe is realized via leptogenesis which is calculable in these models and  found to be of the right order of magnitude.
}

\begin{document}
\maketitle
\flushbottom

\section{Introduction}

Theoretical models that go beyond the highly successful Standard Model (SM) are often invoked to address certain peculiarities that are present but not addressed by the SM.  These include (i) the hierarchical structure of quark and lepton masses and mixing angles, including the neutrino sector (the flavor puzzle), (ii) the extreme smallness of the  CP violation parameter $\overline{\theta}$ in the strong interaction sector (the strong CP problem), (iii) the lack of a suitable particle dark matter (DM) candidate, and (iv) the absence of a mechanism to dynamically generate the observed baryon asymmetry of the universe.  The purpose of this paper is to provide a unified framework that addresses all four of these issues. The unifying theme of our framework is a gauged $U(1)_F$ flavor symmetry that distinguishes particles with identical SM quantum numbers.  This in turn leads to an understanding of the hierarchical structure of the fermion spectrum via the Froggatt-Nielsen (FN) mechanism ~\cite{Froggatt:1978nt}. Fermion masses arise as powers of a small parameter $\epsilon = \langle X\rangle/\Lambda_{\rm FN}$, where $\langle X \rangle$ is the vacuum expectation value (VEV) of a flavon field $X$ and $\Lambda_{\rm FN}$ is a flavor cutoff scale, with the powers determined by the respective $U(1)_F$ charges. Such a setup would allow for the fundamental Yukawa couplings to be all of order one, and yet explain the observed hierarchical fermion mass structure. For reviews, explicit models, and references of the FN mechanism, see, e.g., Ref.~\cite{Dimopoulos:1983rz,
Leurer:1992wg,Babu:2009fd,
King:2013eh,
Bauer:2016rxs,Calibbi:2016hwq,Ema:2016ops}. 

Our framework uses two Higgs doublets to generate the hierarchy in fermion masses and mixings. The choice of $U(1)_F$ charges, which should satisfy various anomaly cancellation conditions, is simpler with two Higgs doublets, rather than one.  We find that, quite generally, this setup has an accidental global $U(1)$ symmetry which has a nonzero QCD anomaly.  This global $U(1)$ is identified as the Peccei-Quinn (PQ) symmetry~\cite{Peccei:1977hh}, $U(1)_{\rm PQ}$, which leads to a light pseudoscalar field, the axion~\cite{Weinberg:1977ma,Wilczek:1977pj}, upon spontaneous symmetry breaking. Minimizing the QCD-induced potential for the axion field leads to $\overline{\theta} = 0$, thus solving the strong CP problem.  Our setup is a flavored generalization of the Dine-Fischler-Srednicki-Zhitnitsky (DFSZ) axion model~\cite{Dine:1981rt,Zhitnitsky:1980tq} with two Higgs doublets and two singlet scalars, but with some important differences as discussed below. In this paper, we shall not attempt to embed a $U(1)_F$ symmetry in the second popular axion model, the Kim-Shifman-Vainshtein-Zhakharov (KSVZ) model~\cite{Kim:1979if,Shifman:1979if}, although such attempts may also lead to interesting results.

Global symmetries, such as $U(1)_{\rm PQ}$, are subject to violations by quantum gravitational effects, which typically appear as higher dimensional operators in the effective Lagrangian with inverse powers of the Planck scale $M_{\rm Pl}$. In the  presence of such operators the axion potential will be tilted from its original form, which would minimize $\overline{\theta}$ at a value away from zero, thus spoiling the strong CP solution.  This roadblock has been termed the axion quality problem~\cite{Georgi:1981pu,Kamionkowski:1992mf,Barr:1992qq,Ghigna:1992iv}. Models that protect the axion potential from excessive corrections so that the strong CP solution is preserved are said to have  ``high quality axion".  In our framework, the axion is of high quality by virtue of the $U(1)_F$ gauge symmetry.  Quantum gravitational effects should preserve all gauge symmetries, including $U(1)_F$, which ensures that the dangerous operators that tilt the axion potential are under control. 

There have been attempts at solve the axion quality problem by realizing an accidental axion from a flavor universal $U(1)$ gauge symmetry~\cite{Barr:1992qq,Babu:1992cu,Duerr:2017amf,Bonnefoy:2018ibr,Bhattiprolu:2021rrj,Qiu:2023los,
Babu:2024udi,Babu:2024qzb} or a non-Abelian gauge symmetry~\cite{DiLuzio:2020qio,Ardu:2020qmo,Darme:2022uzl,DiLuzio:2025jhv}. Other attempts to solve this problem include composite axion~\cite{Randall:1992ut,Lillard:2018fdt,Gaillard:2018xgk,Vecchi:2021shj,Lee:2018yak,Cox:2023dou, Cox:2021lii,Nakai:2021nyf, Contino:2021ayn,Podo:2022gyj}, stabilized axion via discrete gauge symmetries~\cite{Babu:2002ic,Hook:2014cda}, mirror universe models~\cite{Berezhiani:2000gh, Hook:2019qoh},  multiple replication of the SM sector~\cite{Hook:2018jle, Banerjee:2022wzk},  extra dimensional~\cite{Choi:2003wr, Reece:2024wrn, Craig:2024dnl, Choi:2025lkg} and string theoretic constructions~\cite{Svrcek:2006yi,Burgess:2023ifd}. For other approaches to address the axion quality problem involving fifth force, heavy fields and small instanton contributions see Ref.~\cite{Zhang:2022ykd,Bonnefoy:2022vop,Kitano:2021fdl}. 

In our proposed framework axion arises accidentally from a gauged flavor symmetry. There have been attempts to identify the axion as arising from the flavor sector originally in Refs.~\cite{Davidson:1981zd,Wilczek:1982rv}. Axion with flavor-dependent couplings to matter have been proposed with continuous symmetries~\cite{Geng:1988nc,Berezhiani:1989fp,Calibbi:2016hwq,Ema:2016ops,Davidson:1983fy,Suematsu:2018hbu,Bjorkeroth:2018ipq}. While there have been attempts to address the axion quality through such flavor symmetries~\cite{Babu:1992cu,Cheung:2010hk,Bonnefoy:2019lsn,Darme:2022uzl,DiLuzio:2025jhv}, to our knowledge the models presented here are the first attempt that relies on minimal non-supersymmetric gauged $U(1)_F$ flavor symmetry that also explains the fermion mass hierarchy including the neutrinos.

While most Froggatt-Nielsen models that explain the hierarchical fermion spectrum assume a global $U(1)_F$ flavor symmetry~\cite{King:2013eh,
Bauer:2016rxs,Calibbi:2016hwq,Ema:2016ops}, here we insist on gauging the $U(1)_F$, which appears to be more natural and is necessary in our setup for solving the axion quality problem.  There have been a few attempts to realize the FN mechanism with gauged flavor $U(1)$ symmetry~\cite{Chen:2008tc,Tavartkiladze:2011ex,Smolkovic:2019jow,Tavartkiladze:2022pzf,Rathsman:2023cic}.  Inclusion of two Higgs doublets in the spectrum, as we do in our framework, leads to a simpler $U(1)_F$ charge assignment compared to other attempts.  The mass scale of the right-handed neutrinos employed for generating small neutrino masses via the seesaw mechanism is 
related to the Froggatt-Nielsen scale and the flavor symmetry breaking scales in our framework, which constrains the theory space for consistent models significantly.
There have been various attempts to use the anomalous $U(1)$ symmetry of string origin where the gauge anomalies are canceled by the Green-Schwarz mechanism~\cite{Green:1984sg} as an FN flavor symmetry~\cite{Ibanez:1994ig,Dudas:1995yu,Binetruy:1994ru,Nir:1995bu,Berezhiani:1996nu,
Choi:1996se,Grossman:1998jj,
Altarelli:1998sr,
Bando:2001bj,Shafi:2000su,Babu:2003zz,Babu:2004th}. Although they are interesting, the right-handed neutrino mass scale needed to fit the neutrino oscillation data and the axion quality cannot be simultaneously satisfied easily in such a setup, since the flavor cutoff scale $\Lambda_{\rm FN}$ is near the Planck scale.

The axion in our framework also serves as the dark matter in the universe~\cite{Preskill:1982cy,Abbott:1982af,Dine:1982ah} in a post-inflationary PQ breaking scenario that we adopt here. Furthermore, unlike the DFSZ axion model, the models presented here are devoid of a Domain Wall (DW) problem since the number of domain walls is $N_{\rm DW} = 1$ in this setup, which results in the rapid decay of the domain wall~\cite{Sikivie:1982qv}.  Finally, in our Froggatt-Nielsen setup, the baryon asymmetry of the universe is generated via leptogenesis, which is calculable, and shown to be of the right order of magnitude.

\begin{figure}[!htbp]
\centering
\includegraphics[width=0.9\textwidth]
{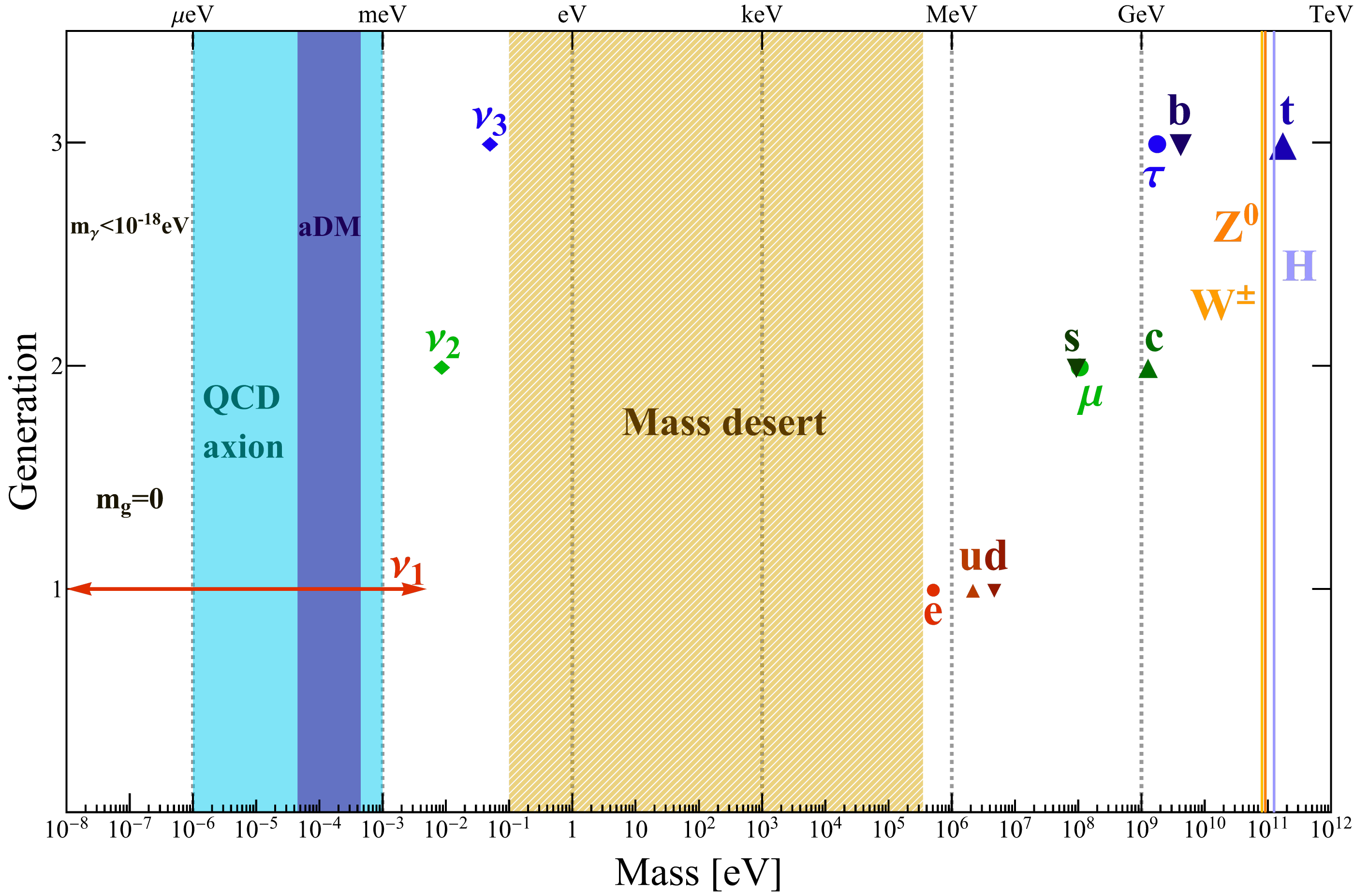}
\caption{A cartoon depicting the masses of the fundamental particles with their dependence on generation number. The red, green and blue icons denote the masses of the  first, second and third generation fermions~\cite{ParticleDataGroup:2024cfk,Esteban:2024eli,Antusch:2025fpm}. For the neutrino sector we have assumed a normal ordering of the masses. The masses of the Higgs boson and of the gauge bosons $W^{\pm},Z^0$   are shown in light purple, yellow, and orange bands respectively. The massless photon and gluon are depicted on the left side of the figure.  Also shown is the range of masses for a QCD axion in blue, with the dark blue band corresponding to axion dark matter denoted as ``aDM".  Note the presence of a mass desert in the range $(0.1~{\rm eV} - 0.5~{\rm MeV})$.}
\label{Fig: hierarchy}
\end{figure}

A pictorial representation of the various particle masses of the model is shown in Fig.~\ref{Fig: hierarchy}.  Here the fermion masses ranging from the top quark to the lightest neutrino $\nu_1$ are depicted, which shows a mass  desert in the range (0.1 eV $-$ $0.5$ MeV). A normal ordering of neutrino masses has been assumed for concreteness. Also shown in Fig.~\ref{Fig: hierarchy} is the QCD axion lying in the mass range ($10^{-6}-10^{-4})$ eV, with the blue shaded region corresponding to the mass window where the axion makes up the entire dark matter of the universe. Our unified framework based on a gauged $U(1)_F$ symmetry is capable of explaining most of the features of this figure, without assuming any hierarchy in the fundamental parameters.

The rest of the paper is organized as follows. In Sec. \ref{sec:sec2} we present our framework and discuss the general setup.  Fermion mass generation via the Froggatt-Nielsen mechanism is adopted here, with the spectrum consisting of the SM fermions, two Higgs doublets $H_u$ and $H_d$,  as well as two Higgs singlets $X$ and $S$. Anomaly cancellation conditions are outlined and the emergence of an accidental axion is explained. We then proceed to derive the couplings of the axion to the gluon and the photon as well as the fermions in this general setup. Flavor violation phenomenology is briefly discussed and the conditions for a high quality axion are derived here. We also discuss the parameter space where the axion can be the entire dark matter of the universe, and summarize briefly the baryogenesis mechanism. In Sec.~\ref{sec:ModelI} our Model I is presented and analyzed where the general results  of Sec.~\ref{sec:sec2} are utilized.  Here we also provide a UV completion of the Froggatt-Nielsen model employing primarily FN scalars, and show the robustness of the axion quality, including the FN fields and loop diagrams. Baryogenesis is shown to proceed successfully via resonant leptogenesis. In Sec.~\ref{sec:ModelII} the same analysis is repeated for Model II, with a different set of $U(1)_F$ charges and adopting a different way of UV completion with fermions. Here baryogenesis occurs via standard thermal leptogenesis. In Sec.~\ref{sec:ModelIII} a third model, Model III, with $SU(5)$-compatible $U(1)_F$ charges is presented and analyzed in the same fashion. In Sec.~\ref{sec:experimentalprobes} we discuss the experimental probes of a particular case from Model I with interesting features. In Sec. \ref{sec:Domainwallproblem} we give a comprehensive solution to the domain wall problem in this class of models. In Sec.~\ref{sec:Conclusions} we conclude and provide our outlook. In three appendices we provide discussions on the following: Appendix \ref{sec:AppendixA} has the UV-completions for Models II and III; Appendix \ref{sec:pseudoadd} discusses additional cases with high quality axion; and Appendix \ref{sec:NDWgen} shows the domain wall number to be one in Model I with scalar UV completion.

\section{General Setup}
\label{sec:sec2}

In this section, we describe the general setup for the class of models with a gauged $U(1)_F$ flavor symmetry that explains the fermion mass hierarchy via the Froggatt-Nielsen mechanism~\cite{Froggatt:1978nt}. These models also have a high quality axion which arises from an accidental Peccei-Quinn symmetry present in the setup.  Unlike most models in the literature that utilize  {\it global} flavor symmetries, here we make use of a {\it local}  $U(1)_F$ symmetry.  The requirement of anomaly cancellation would imply that the choices of $U(1)_F$ charge assignments are more restricted compared to the case of a global $U(1)$ symmetry. Only local symmetries are safe from gravity-induced corrections, which we shall apply to safeguard the quality of the axion.  The results presented here for a general framework will find direct application in the three explicit models proposed in the subsequent sections.

\subsection*{Flavor charge assignment}

The models we present are generalizations of the DFSZ axion model~\cite{Zhitnitsky:1980tq,Dine:1981rt} which introduces two Higgs doublets $H_u(1,2,\frac{1}{2})$ and $H_d(1,2,-\frac{1}{2})$ into the SM extended with a gauged $U(1)_F$ symmetry.  Here the quantum numbers (indicated in the brackets) refer to $SU(3)_c \times SU(2)_L \times U(1)_Y$. The $U(1)_F$ symmetry is spontaneously broken when two Standard Model singlet scalars $X$ and $S$ acquire nonzero vacuum expectation values (VEVs).  Very roughly, one linear combination of the phases of $X$ and $S$ becomes the longitudinal component of the $Z'$ gauge boson associated with $U(1)_F$, while the orthogonal combination becomes the axion. The existence of two singlet scalars is crucial to avoid a Weinberg-Wilczek-type weak-scale axion~\cite{Weinberg:1977ma,Wilczek:1977pj} which is excluded by experimental searches.

The $U(1)_F$ charge assignment of the fermion fields is listed in Table~\ref{tab:tab1}. The fermion spectrum includes the three families of quark and leptons of the Standard Model, along with three (and in some cases more than three, if needed) singlet right-handed neutrinos (RHNs) $N_i$.  In Table~\ref{tab:tab2} we list the scalar fields employed in the models along with their  $U(1)_F$ charges. Here, the singlet scalar $X$ plays the role of the flavon field used in the Froggatt-Nielsen mechanism, while $S$ is an additional singlet scalar that does not take part in the fermion mass generation. Its role is to facilitate the formation of a high-quality axion that is not a Weinberg-Wilczek-type weak scale axion.
\begin{table}[!htbp] 
   \small
   \centering
   \begin{tabular}{|c|c|c|c|c|c|c|}
   \hline\hline
   \textbf{Fermion} & $Q_i(3,2,\frac{1}{6})$ & $u^c_i(\overline{3},1,-\frac{2}{3})$ & $d^c_i(\overline{3},1,\frac{1}{3})$ & $L_i(1,2,-\frac{1}{2})$ & $e^c_i(1,1,1)$ & $N_i(1,1,0)$\\ 
   \hline
  $U(1)_F$ charge &  $(q_1,q_2,q_3)$ & $(u_1,u_2,u_3)$ & $(d_1,d_2,d_3)$ & $(l_1,l_2,l_3)$ & $(e_1,e_2,e_3)$ & $(n_1,n_2,n_3)$ 
     \\
   \hline\hline 
   \end{tabular}
   \caption{$U(1)_F$ charges of the three families of quarks and leptons, including singlet neutrinos $N_i$. All fermions are taken to be left-handed. The quantum numbers under $SU(3)_c \times SU(2)_L \times U(1)_Y$ of the fermions are also indicated
   (in the brackets of the first row).}
   \label{tab:tab1}
\end{table}
\begin{table}[!htbp]
   \small
   \centering
   \begin{tabular}{|c|c|c|c|c|}
   \hline \hline
   \textbf{Scalar} & $H_u(1,2,\frac{1}{2})$ & $H_d(1,2,-\frac{1}{2})$ & $X(1,1,0)$ & $S(1,1,0)$\\ 
   \hline
    $U(1)_F$ charge &  $h_u$ & $h_d$ & $q_X$ & $q_S$
     \\
   \hline \hline 
   \end{tabular}
   \caption{Scalar fields employed in the class of models, along with their $U(1)_F$ charges.}
   \label{tab:tab2}
\end{table}
The effective Yukawa Lagrangian of this class of models is given by\footnote{We found it economical to use $\widetilde{H_d}$ rather than $H_u$ in the $(L_i N_j$) couplings from the anomaly cancellation requirements, while reproducing desirable flavor textures for the neutrino masses.}
\begin{eqnarray}
{\cal L}_{\rm Yukawa} &=& y_{ij}^u Q_i u^c_j H_u \left(\frac{X^{(*)}}{\Lambda_{\rm FN}}\right)^{n^u_{ij}} +  y_{ij}^d Q_i d^c_j H_d \left(\frac{X^{(*)}}{\Lambda_{\rm FN}}\right)^{n^d_{ij}} +  y_{ij}^\ell L_i e^c_j H_d \left(\frac{X^{(*)}}{\Lambda_{\rm FN}}\right)^{n^\ell_{ij}} \nonumber \\
&+&  y_{ij}^{\nu} L_i N_j \widetilde{H_d} \left(\frac{X^{(*)}}{\Lambda_{\rm FN}}\right)^{n^{\nu}_{ij}} +  \frac{y_{ij}^N}{2} \Lambda_{\rm FN} N_i N_j \left(\frac{X^{(*)}}{\Lambda_{\rm FN}}\right)^{n^N_{ij}} + h.c.
\label{eq:Yuk}
\end{eqnarray}
Here we have defined $\widetilde{H_d} = i \tau_2 H_d^*$. The $y_{ij}^f$  are dimensionless effective couplings which are all taken to be of order unity. The exponents $n_{ij}^f$ are positive integers and $\Lambda_{\rm FN}$ is the cut-off scale for flavor, where additional Froggatt-Nielsen fields are expected to be present.  In each term in Eq.~(\ref{eq:Yuk}) either the field $X$ or its complex conjugate $X^*$ can couple, which is indicated as $X^{(*)}$.  It is evident from Eq.~(\ref{eq:Yuk}) that the overall scale of the Majorana mass of the $N_i$ fields has been identified as the flavor cut-off scale $\Lambda_{\rm FN}$. With such an identification, $\Lambda_{\rm FN}$ will be determined from neutrino oscillation data. In principle, the Majorana mass scale for $N_i$ need not be the same as $\Lambda_{\rm FN}$, but identifying the two scales is the simplest possibility, which we adopt in this paper. The positive integers $n_{ij}^f$ that appear in the exponents in Eq.~(\ref{eq:Yuk}) should be chosen to reproduce the observed fermion mass and mixing hierarchy without assuming any significant  hierarchy in the effective coupling coefficients $y_{ij}^f$.  
Invariance of the Lagrangian given in Eq.~(\ref{eq:Yuk}) under the $U(1)_F$ gauge symmetry with the charges listed in Table~\ref{tab:tab1} requires the following conditions be satisfied among the flavor charges and the positive integers $n_{ij}^f$:
\begin{subequations}
    \begin{align}
        \begin{split}
        \begin{aligned}
        q_i + u_j + h_u \pm q_X\, n_{ij}^u = 0,\label{eq:2.2}
        \end{aligned}
        \end{split}\\
        \begin{split}
        \begin{aligned}
        q_i + d_j + h_d \pm q_X\, n_{ij}^d = 0,\label{eq:2.3} 
        \end{aligned}
        \end{split}\\
        \begin{split}
        \begin{aligned}
        l_i + e_j + h_d \pm q_X\, n_{ij}^\ell = 0,\label{eq:2.4}
        \end{aligned}
        \end{split}\\
        \begin{split}
        \begin{aligned}
        l_i + n_j - h_d \pm q_X\, n_{ij}^\nu = 0, 
        \end{aligned}
        \end{split}\\
        \begin{split}
        \begin{aligned}
        n_i + n_j \pm q_X\, n_{ij}^N = 0.
        \end{aligned}
        \end{split}
    \end{align}
\end{subequations}
Here, the $\pm$ signs in the last terms correspond to the usage of  $X$ or $X^*$ field at the relevant locations.

\subsection{Anomaly cancellation}\label{subsec:Anomcancel}

For a gauged $U(1)_F$ symmetry, there are additional consistency requirements on the $U(1)_F$ charges of fermions arising from anomaly cancellation. These conditions read,
\begin{subequations}\label{eq:anomalies}
\begin{align}
  A[SU(3)_c^2 \times U(1)_F]&= \sum_i(2q_i+u_i+d_i) = 0, \label{eq:2.7} \\
  A[SU(2)_L^2 \times U(1)_F]&= \sum_i(3 q_i + l_i) = 0, \\
  A[U(1)_Y^2 \times U(1)_F] &= \frac{1}{6}\sum_i(q_i + 4 u_i + 2 d_i+ 3 l_i + 6 e_i) = 0,\\
  A[({\rm gravity})^2 \times U(1)_F] &= \sum_i(6 q_i + 3 u_i + 3 d_i + 2 l_i + e_i + n_i) = 0,\\
  A[U(1)_Y \times U(1)_F^2] &= \sum_i(q_i^2-2 u_i^2+d_i^2-l_i^2+e_i^2) = 0,\\
  A[U(1)_F^3] &= \sum_i(6q_i^3+3u_i^3+3 d_i^3+2 l_i^3+e_i^3+n_i^3) = 0~. 
\end{align}
\end{subequations}
It is challenging to find a set of flavor charges that leads to desirable mass matrix textures while satisfying these conditions. We shall present three models where this is achieved non-trivially. The $U(1)_F$ charges of Models I, II and III, given in Tables~\ref{tab:ModelIcharges}, \ref{tab:ModelIIcharges} and \ref{tab:ModelIIIcharges}, respectively, are chosen such that they satisfy all these conditions (and also lead to desirable fermion mass matrix textures).

\subsection{Accidental PQ symmetry and the axion field}

The Yukawa Lagrangian of Eq.~(\ref{eq:Yuk}) preserves three $U(1)$ symmetries that are flavor-universal. To see this, let us assign each field a $U(1)$ charge in Eq.~(\ref{eq:Yuk}), except for the $X$ scalar (which appears in varying powers) and the $S$ scalar (which does not appear in Eq.~(\ref{eq:Yuk})).  Since there are eight fields -- $(Q,u^c,d^c,L,e^c,N)$ fermions and $(H_u,H_d)$ scalars, there are eight separate $U(1)$s that can be defined, each acting on one of these fields.  The five Yukawa/mass terms of Eq.~(\ref{eq:Yuk}) would reduce these eight $U(1)$s down to three.  Two of these symmetries are identified as hypercharge $Y$ and baryon number $B$.  The third $U(1)$ symmetry has a QCD anomaly and can be identified as the Peccei-Quinn (PQ) $U(1)_{\rm PQ}$ symmetry.

The charge assignment under the global $U(1)_{\rm PQ}$ symmetry cannot be specified uniquely, since any linear combination of $Y$ and $B$ can be added to a given choice without making the QCD anomaly zero.  (Note that $A[SU(3)_c^2 \times U(1)_Y]$ and $A[SU(3)_c^2 \times U(1)_B]$ are both zero.) One particular choice of the PQ charges is shown in Table~\ref{tab:tab3}. The resulting QCD anomaly is $A[SU(3)_c^2 \times U(1)_{\rm PQ}] = 3$. Other choices of the PQ charges, such as $(Q,u^c,d^c,L,e^c,N,H_u,H_d,X,S) = (1,1,1,-2,4,0,-2,-2,4/n)$, also would have a nonzero QCD anomaly, viz., $A[SU(3)_c^2 \times U(1)_{\rm PQ}] = 6$.
\begin{table}[!htbp] 
   \label{tab:PQcharges}
   \small
   \centering
   \begin{tabular}{|c|c|c|c|c|c|c|c|c|c|c|}
   \hline\hline
   \textbf{Field} & $Q_i$ & $u^c_i$ & $d^c_i$ & $L_i$ & $e^c_i$ & $N_i$ & $H_u$ & $H_d$ & $X$ & $S$\\ 
   \hline
  $U(1)_{\rm PQ}$ charge &  $0$ & $1$ & $1$ & $-1$ & $2$ & $0$ & $-1$ & $-1$ & 0 & $2/n$ 
     \\
   \hline\hline 
   \end{tabular}
   \caption{One choice of the global $U(1)_{\rm PQ}$ charges of the fermions and scalars.}
   \label{tab:tab3}
\end{table}
Note that the scalar $S$ did not appear in the Yukawa Lagrangian of Eq.~(\ref{eq:Yuk}). Its role is to  give mass to the pseudoscalar Higgs field contained in $H_u$ and $H_d$.  This is achieved by including an interaction term in the Higgs potential given as
\begin{equation}
V_{\rm{HS}} \supset -\lambda_{\rm{HS}} \frac{H_u H_d S^n \left(X^{(*)}\right)^k}{M_{\rm Pl}^{n+k-2}} + \text{h.c}.
\label{eq:pot}
\end{equation}
Here we allow for either $X$ or $X^*$ to couple, as indicated by $X^{(*)}$.  Furthermore, $(n,k^\pm)$ are non-negative integers, where the superscript $^\pm$ in $k$ refers to the two cases with $X$ and $X^*$ appearing in Eq.~(\ref{eq:pot}). Specifically, if a choice lists ($n,k^+$) integers, then $X$ is used in the operator shown in Eq.~(\ref{eq:pot}). Invariance of Eq.~(\ref{eq:pot}) under the gauged $U(1)_F$ symmetry implies the constraint
\begin{equation}
h_u+h_d+nq_S\pm kq_X = 0.
\label{eq:cons}
\end{equation}
The PQ charge of $S$ field is determined from this coupling, which is equal to $2/n$, as listed in Table~\ref{tab:tab3}. Note that when $n+k \leq 2$ the interaction of Eq.~(\ref{eq:pot}) is renormalizable, and when $n+k \geq 3$ it is not.  In the latter case we use gravity-induced interactions to generate this term, which is economical. There is a restriction on the value of $n+|k|$ from the demand that the mass of the second Higgs doublet of the model should be above 100 GeV or so. Note that the pseudoscalar Higgs boson from the two Higgs doublets can acquire a mass only through the interaction term of Eq.~(\ref{eq:pot}). We find this constraint to be
\begin{equation}\label{eq:minnk}
n\geq 1,~~~{\rm min} \{n,|k|\} \leq 4~.
\end{equation}
The derivation of this constraint will be given after we identify the axion field and determine the axion decay constant $f_a$ in terms of the various VEVs. The specific flavor model and the value of $\Lambda_{\rm FN}$ decide the maximum value of $n+|k|$. In specific UV-completion of the FN sector with the use of Higgs doublets it turns out that the operator of Eq.~(\ref{eq:pot}) will be induced by the FN fields along with lower powers of $M_{\rm Pl}$. In this case certain powers of $M_{\rm Pl}$ will be replaced by powers of $\Lambda_{\rm FN} |\epsilon|^p$ where $p$ is a non-negative integer.  In such cases we shall keep the larger of the two contributions arising from Eq.~(\ref{eq:pot}) and from the UV-specific operators in computing the pseudoscalar Higgs mass. From $SU(2)_L$ gauge invariance it follows that this mass is also the mass of the heavy Higgs doublet of a two-Higgs doublet extension of the SM. The values $(n,k^{\pm})$ are fixed by $U(1)_F$ gauge invariance for a given model, thereby fixing the gauge charges $(q_S,\,q_X)$ of the $(S,\,X)$ fields. In the following sections, we thus list all such models corresponding to specific $(n,k^{\pm})$ for which the arising accidental axion is of high quality.

In order to identify the composition of the axion field, let us parametrize the various neutral Higgs fields of the model in the exponential form, assuming that all fields acquire VEVs: 
\begin{eqnarray}
H_u^0 = \frac{v_u}{\sqrt{2}} e^{\frac{i \eta_u}{v_u}}, ~~~H_d^0 = \frac{v_d}{\sqrt{2}} e^{\frac{i \eta_d}{v_d}}, ~~~
X = \frac{v_X}{\sqrt{2}} e^{\frac{i \eta_X}{v_X}},~~~S = \frac{v_S}{\sqrt{2}} e^{\frac{i \eta_S}{v_S}}~. 
\label{eq:expo}
\end{eqnarray}
Here we have frozen the radial components, which are not relevant for the present discussion. All of the vacuum expectation values ($v_u,v_d,v_X,v_S)$ can be made real, since the Higgs potential parameters are all real, with exception of $\lambda_{\rm HS}$ of Eq.~(\ref{eq:pot}), which can however be made real by field redefinitions. With the parametrization of the fields given in Eq.~(\ref{eq:expo}), the massive pseudoscalar field can be identified as
\begin{equation}
A=N_A \left(\frac{\eta_u}{v_u} + \frac{\eta_d}{v_d} \pm k  \frac{\eta_X}{v_X} + n \frac{\eta_S}{v_S}\right)~,
\label{eq:A}
\end{equation}
where $N_A$ is a normalization constant. In addition, the Goldstone bosons that are absorbed by the $Z$ boson and the $U(1)_F$ gauge bosons can be identified as
\begin{subequations}
\begin{align}
G_Z &= N_Z (v_u \eta_u - v_d \eta_d),\\
G_F &= N_F \left(h_u v_u \eta_u + h_d v_d \eta_d + q_X v_X \eta_X  + q_S v_S \eta_S   \right)~.
\label{eq:G}
\end{align}
\end{subequations}
Here again, $(N_Z, N_F)$ are normalization constants.
The massless combination among the $\eta_i$ fields, identified as the axion, is the combination orthogonal to $(A, G_Z, G_F)$ of Eqs. (\ref{eq:A})-(\ref{eq:G}).  Its composition can be written down as
\begin{equation}
a=\sum_{\alpha=u,d,X,S} K_\alpha \eta_\alpha
    \label{eq:axion}
\end{equation}
where we have defined
\begin{subequations}
    \begin{align}
    K_u &= N_a v_u v_d^2 (n\, q_X v_X^2 \mp k\, q_S v_S^2),\\ \label{eq:comp1}
    K_d &= N_a v_d v_u^2 (n\, q_X v_X^2 \mp k\, q_S v_S^2),\\ \label{eq:comp2}
    K_X &= N_a v_X \left(q_S\,v_S^2(v_u^2+v_d^2)-n\,(h_u+h_d)v_u^2v_d^2\right),\\ 
    K_S&= N_a v_S\left(\pm k(h_u+h_d)v_u^2v_d^2 - q_X v_X^2(v_u^2+v_d^2)   \right),
\label{eq:comp4}    
    \end{align}
\end{subequations}
with $N_a$ given by
\begin{eqnarray}
N_a &=& \left[v_u^2 v_d^2(v_u^2+v_d^2)\left\{n\,q_X v_X^2 \mp k\,q_S v_S^2\right\}^2+ v_X^2\left\{q_S v_S^2(v_u^2+v_d^2)-n\,(h_u+h_d)v_u^2v_d^2\right\}^2 \right. \nonumber \\
&&\left. + v_S^2 \left\{\pm k (h_u+h_d) v_u^2v_d^2-q_X v_X^2(v_u^2+v_d^2)\right\}^2\right]^{-1/2}~.
\label{eq:Na}
\end{eqnarray}
The upper and lower signs multiplying $k$ in these expressions should be applied depending on whether $X$ or $X^*$ appears in Eq.~(\ref{eq:pot}). The charges of the various scalar fields of course obey the constraint given in Eq.~(\ref{eq:cons}). 
Our phenomenological analysis will assume a hierarchy of VEVs $|v_u|, |v_d| \ll |v_S|, |v_X|$.  The ratio of the two larger VEVs, defined as
\begin{equation}
r\equiv  v_S/v_X
\label{eq:r}
\end{equation}
plays an important role in our analysis.  
The dependence of the axion field on this ratio $r$ would determine how flavorful the axion is.   If $r\ll 1$, axion is mostly contained in $\eta_S$, which is decoupled from the flavon.  In the opposite limit where $r \gg 1$, axion is mostly in $\eta_X$, which is mostly the flavon field. (There is also the radial mode of $X$ which is part of the flavon, with a mass of order $\Lambda_{\rm FN}$.)

\subsection{Couplings of the axion}
Having identified the axion field as given in Eqs. (\ref{eq:axion})-(\ref{eq:comp4}), we now proceed to determine its couplings to the fermions arising from the Yukawa Lagrangian of Eq.~(\ref{eq:Yuk}).  From there we also work out the axion couplings to the gluons and the photon.
\subsection*{Axion couplings to the fermions}\label{subsec:axionfermioncouplings}
The couplings of the charged fermions and the axion are given by
\begin{eqnarray}
{\cal L}_{\rm Yukawa}^{(a)} &=& u_i u_j^c M^u_{ij}\, {\rm exp}\,\left\{ia\left(\frac{K_u}{v_u} - \frac{K_X}{q_X v_X}(q_i+u_j+h_u)\right) \right\} \nonumber \\
&+&
d_i d_j^c M^d_{ij}\, {\rm exp}\,\left\{ia\left(\frac{K_d}{v_d} - \frac{K_X}{q_X v_X}(q_i+d_j+h_d)\right) \right\} \nonumber\\
&+&
\ell_i \ell_j^c M^\ell_{ij}\, {\rm exp}\,\left\{ia\left(\frac{K_d}{v_d} - \frac{K_X}{q_X v_X}(l_i+d_j+h_d)\right) \right\} + \text{h.c}.
\label{eq:aint1}
\end{eqnarray}
Here $M^{u,d,\ell}$ are the quark and lepton mass matrices, and we have made use of Eqs. (\ref{eq:2.2})-(\ref{eq:2.4}) in writing Eq.~(\ref{eq:aint1}). The axion field can be removed from Eq.~(\ref{eq:aint1}) by field-dependent redefinitions of the fermion fields. Derivative couplings of the axion with the fermions would then appear from the fermion kinetic energy terms.
Furthermore, we diagonalize the mass matrices by the biunitary transformations given by
\begin{equation}
V_u^T M^u V_{u^c} = \hat{M}^u, ~~~V_d^T M^d V_{d^c} = \hat{M}^d,~~~V_\ell^T M^\ell V_{\ell^c} = \hat{M}^\ell,
\end{equation}
where 
\begin{equation}
\hat{M}^u = {\rm diag}.\,(m_u,m_c,m_t),~~~ \hat{M}^d = {\rm diag}.\,(m_d,m_s,m_b),~~~\hat{M}^\ell = {\rm diag}.\,(m_e,m_\mu,m_\tau)
\end{equation}
are the diagonal mass matrices.  In this basis the axion couplings to the charged fermions take the form
\begin{eqnarray}
\hspace{-0.3cm}{\cal L}_{\rm kinetic}^{(a)} &=& \partial_\mu a \left[\left( \overline{u}_{iL} \gamma^\mu u_{jL} (V_u^\dagger P_Q V_u)_{ij} +  \overline{d}_{iL} \gamma^\mu d_{jL} (V_d^\dagger P_Q V_d)_{ij}
 +  \overline{\ell}_{iL} \gamma^\mu \ell_{jL} (V_\ell^\dagger P_L V_\ell)_{ij}\right)
\left(-\frac{K_X}{q_Xv_X}  \right) \right. \nonumber \\
&&\left. + 
\left( \overline{u}_{iR} \gamma^\mu u_{jR} (V_u^T P_u V_{u^c}^*)_{ij} +  \overline{d}_{iR} \gamma^\mu d_{jR} (V_{d^c}^T P_d V_{d^c}^*)_{ij}
 +  \overline{\ell}_{iR} \gamma^\mu \ell_{jR} (V_{e^c}^T P_{e} V_{e^c}^*)_{ij}\right)
\left(\frac{K_X}{q_Xv_X}  \right) \right.\nonumber\\
&&\left. -\overline{u}_{iR} \gamma^\mu u_{iR} \left(\frac{K_u}{v_u} - \frac{K_X h_u}{q_Xv_X} \right)- \overline{d}_{iR} \gamma^\mu d_{iR} \left(\frac{K_d}{v_d} - \frac{K_X h_d}{q_Xv_X} \right) - \overline{\ell}_{iR} \gamma^\mu \ell_{iR} \left(\frac{K_d}{v_d} - \frac{K_X h_d}{q_Xv_X} \right)
\right]~.\nonumber \\
\label{eq:aint2}
\end{eqnarray}
Here we have defined diagonal matrices $P_{Q,u^c,d^c,\ell,e^c}$ containing the flavor charges of the fermions as 
\begin{eqnarray}
P_Q =& {\rm diag}.\,(q_1,q_2,q_3),~~~P_{u} = {\rm diag}.\,(u_1,u_2,u_3),~~~P_{d} = {\rm diag}.\,(d_1,d_2,d_3), \nonumber \\ &P_L= {\rm diag}.\,(l_1,l_2,l_3),~~~
P_{e} = {\rm diag}.\,(e_1,e_2,e_3)
\end{eqnarray}
with the charges defined in Table~\ref{tab:tab1}. We note that Eq.~(\ref{eq:aint2}) contains both flavor conserving and flavor violating axion couplings.

\subsection*{Axion couplings to gluons and the photon}\label{subsec:axiongluonphotoncouplings}

From the interactions of the quarks with the axion, as given in Eq.~(\ref{eq:aint2}), one can readily compute the axion-gluon-gluon vertex which leads to the Lagrangian
\begin{equation}\label{eq:axiongluoncoupling}
{\cal L}_{aG\widetilde{G}} = \frac{g_s^2}{32 \pi^2} \left(\frac{a}{f_a}+\overline{\theta}\right) G_{\mu\nu}^a \widetilde{G}^{\mu\nu,a}
\end{equation}
with $\widetilde{G}^{\mu\nu,a} \equiv \frac{1}{2}\epsilon^{\mu\nu\alpha\beta} G_{\alpha \beta}^a$. 
Taking advantage of the shift symmetry $a \rightarrow a + \overline{\theta} \,f_a$ one can redefine this Lagrangian term to remove the term $\overline{\theta}$. QCD dynamics induces a potential for the axion by virtue of Eq.~(\ref{eq:axiongluoncoupling}), which when minimized would set the shifted $a$ field to zero, thus solving the strong CP problem.
Note that only the flavor diagonal couplings of Eq.~(\ref{eq:aint2}) will contribute to the $a G \widetilde{G}$ interaction vertex.  We find the axion decay constant $f_a$ in our general setup to be
\begin{eqnarray}
\frac{1}{f_a} = \frac{K_X}{q_X v_X}\sum_i(2 q_i+u_i+d_i) - 3\left(\frac{K_u}{v_u} + \frac{K_d}{v_d} - \frac{K_X}{q_Xv_X}(h_u+h_d)\right)~.
\label{eq:fa0}
\end{eqnarray}
The first term in Eq.~(\ref{eq:fa0}) above vanishes owing to the anomaly cancellation condition of Eq.~(\ref{eq:2.7}), leading to the relation
\begin{eqnarray}
\frac{1}{f_a} = -3 N_a \left[(v_u^2+v_d^2)(n q_X v_X^2\mp k q_S v_S^2)-\frac{h_u+h_d}{q_X} \left\{q_Sv_S^2(v_u^2+v_d^2)-nv_u^2 v_d^2(h_u+h_d)    \right\}  \right] .\nonumber \\
\label{eq:fa}
\end{eqnarray}
The constant $N_a$ is given in Eq.~(\ref{eq:Na}). We note that 
$f_a$ in (\ref{eq:fa}) is not positively defined.  Its sign depends on the chosen flavor-charge assignment, the signs of the VEVs, and the definition of the axion field. Throughout this paper, we report the numerical value of $|f_a|$, but we omit the absolute-value symbol for brevity.

The axion-photon coupling is found in a similar fashion to be:
\begin{equation}\label{axionphotoncoupling}
    g_{a\gamma}= \frac{8}{3}- \frac{\alpha_{em}}{2\pi f_a}\left( \frac{2}{3} \left(\frac{4 m_d+m_u}{m_u+m_d} \right) \right)=\frac{8}{3}-1.92(4),
\end{equation}
which is the same as in the DFSZ-I model~\cite{GrillidiCortona:2015jxo}. Here, the flavor-dependent part vanishes as a result of the vanishing of the mixed  $U(1)_F\times U(1)_{\rm EM}^2$ anomaly due to the gauging of the $U(1)_F$ symmetry.
\subsection*{Axion couplings to fermions in terms of \boldmath{$f_a$}}
With the identification of $f_a$ in terms of the $U(1)_F$ charges of the various fields as well as the VEVs of the scalars, we can re-express the couplings of the axion to fermions, Eq.~(\ref{eq:aint2}), in a more user-friendly form.  We first write the couplings of the axion to the $u$-quark, $d$-quark and the electron as
\begin{equation}
{\cal L}_{u,d,e}^{(a)} = -i a \left[m_u\, C_{ua}\, \overline{u} \gamma_5 u\, + m_d \,C_{da}\,\overline{d} \gamma_5 d \, + m_e \,C_{ea}\,\overline{e} \gamma_5 e \, \right]. 
\end{equation}
The coefficients $C_{fa}$ are then found to be
\begin{subequations}\label{eq:Cfa} 
    \begin{align}
        C_{ua} &= \frac{K_X}{q_Xv_X} \left\{(V_u^\dagger P_Q V_u)_{11} + (V_{u^c}^T P_u V_{u^c}^*)_{11}   \right\} - \left(\frac{K_u}{v_u} - \frac{K_X h_u}{q_Xv_X}  \right),  \\
        C_{da} &= \frac{K_X}{q_Xv_X} \left\{(V_d^\dagger P_Q V_d)_{11} + (V_{d^c}^T P_d V_{d^c}^*)_{11}   \right\} - \left(\frac{K_d}{v_d} - \frac{K_X h_d}{q_Xv_X}  \right), \\
        C_{ea} &= \frac{K_X}{q_Xv_X} \left\{(V_\ell^\dagger P_L V_\ell)_{11} + (V_{e^c}^T P_e V_{e^c}^*)_{11}   \right\} - \left(\frac{K_d}{v_d} - \frac{K_X h_u}{q_Xv_X}  \right)~.   
    \end{align}
\end{subequations}
Now, the VEVs $v_X$ and $v_S$ are much larger than the electroweak VEVs $v_{u,d}$.  In the approximation $v_{u,d} \ll v_{X,S}$, the expression for $f_a$ given in Eq.~(\ref{eq:fa}) simplifies to
\begin{equation}\label{eq:favX}
\frac{1}{f_a} \simeq \frac{-3 n \sqrt{q_X^2 +q_S^2r^2} 
 }{rq_X v_X } ~.
\end{equation}
where the ratio $r$ is defined in Eq.~(\ref{eq:r}). 
Note that when $r \ll 1$, $f_a \simeq -v_S/(3n)$. In this limit the axion is almost entirely contained in the $S$ field. In the opposite limit when $r \gg 1$, $f_a \simeq v_X/[3(\pm k + (h_u+h_d)/q_X)]$. In this limit the axion field is almost entirely contained in $X$, implying that the axion acts as a flavon. A theoretical bound on $r$, given as $r\in [10^{-8}, 10^8]$, arises from the requirement that the VEVs $v_S,v_X$ in our theory cannot exceed $M_{\text{Pl}}$, when an experimentally preferred range of  $f_a = (10^8-10^{12})$ GeV is used.

In the approximation $v_{u,d} \ll v_{X,S}$, the $C_{fa}$ couplings of Eq.~(\ref{eq:Cfa}) can be expressed as:
\begin{subequations}
    \begin{align}
        C_{ua} &\simeq  -\frac{\left[  r^2\left((V_u^\dagger P_Q V_u)_{11} + (V_{u^c}^T P_u V_{u^c}^*)_{11} + h_u  \right)  -\cos^2\beta\left(n\mp k \frac{q_S}{q_X} r^2 \right)\right]}{3nf_a \left[q_X^2 + q_S^2 r^2  
 \right]},\label{eq:caq}\\
        C_{da} &\simeq -\frac{\left[  r^2\left((V_d^\dagger P_Q V_d)_{11} + (V_{d^c}^T P_d V_{d^c}^*)_{11} + h_d  \right)  -\sin^2\beta\left(n\mp k \frac{q_S}{q_X} r^2 \right)\right]}{ 3nf_a \left[q_X^2 + q_S^2 r^2  
 \right]}, \\
        C_{ea} &\simeq -\frac{\left[  r^2\left((V_\ell^\dagger P_L V_\ell)_{11} + (V_{e^c}^T P_e V_{e^c}^*)_{11} + h_d  \right)  -\sin^2\beta\left(n\mp k \frac{q_S}{q_X} r^2 \right)\right]}{ 3nf_a \left[q_X^2 + q_S^2 r^2  
 \right]}~.
 \label{eq:cea}
    \end{align}
\end{subequations}
Here we have defined
\begin{equation}\label{eq:tanbeta}
    \tan\beta \equiv \frac{v_u}{v_d}~.
\end{equation}
Note that in the approximation $r \ll 1$, these couplings reduce to the standard DFSZ-I couplings, given as
\begin{eqnarray}\label{eq:DFSZIlimitmattercouplings}
C_{ua} \simeq \frac{1}{3 f_a} \cos^2\beta,~~~C_{da} \simeq  \frac{1}{3f_a} \sin^2\beta, ~~~C_{ea} \simeq \frac{1}{3f_a} \sin^2\beta~.
\end{eqnarray}
It is straightforward to extend Eqs. (\ref{eq:caq})-(\ref{eq:cea}) to the heavier quarks, which will be used in discussing the axion-nucleon coupling later in Sec.~\ref{sec:axionmattercouplings}.

\subsection*{Flavor violating axion couplings}\label{subsec:axionSMFVcouplings}

The axion interactions with fermions given in Eq.~(\ref{eq:aint2}) also contains flavor changing couplings.  We illustrate these couplings with one example, viz., $\overline{d} s a$ coupling that will be of phenomenological importance in Sec.~\ref{subsec:FVcons}.  Such couplings would be relevant for axion flavor physics and are constrained by processes such as $K^{+} \rightarrow \pi^{+}  a$ decay. We find the $\overline{d} s a$ couplings to be given by the pseudo-scalar and scalar terms in the Lagrangian:
\begin{equation}\label{eq:fvLagrangian}
    \begin{aligned}[b]
    \mathcal{L}_{\overline{d} s a}&=  \frac{ia\,q_Sr^2}{6\, n\,f_a \left[q_X^2 + q_S^2 r^2  
 \right]}
        (m_d+m_s)\left[(V_d^\dagger P_Q V_d)_{ds}  + (V_{d^c}^T P_d V_{d^c}^*)_{ds} \right] \overline{d} \gamma_5 s \\
        &+\frac{ia\,q_Sr^2}{6\, n\,f_a \left[q_X^2 + q_S^2 r^2  
 \right]}
        (m_d-m_s)\left[(V_d^\dagger P_Q V_d)_{ds} - (V_{d^c}^T P_d V_{d^c}^*)_{ds}\right]\overline{d} s ~.\\
    \end{aligned}
\end{equation}

We can similarly extend this to any of the massive pair of fermions $\overline{f}_i f_j$: 
\begin{equation}\label{FVLagrangian}
\begin{aligned}[b]
\mathcal{L}_{\bar{f}_if_ja}&=ia\left[(m_{f_i}-m_{f_j})\bar{f}fC^V_{f_if_j}+(m_{f_i}+m_{f_j})\bar{f}\gamma_5fC^A_{f_if_j} \right],\\
\end{aligned}
\end{equation}
with $C^V_{f_i f_j}$ and $C^A_{f_i f_j}$ defined as
\begin{subequations}\label{FVVAcouplings}
\begin{align}
\begin{split}
\begin{aligned}[b]
C^V_{f_if_j}& = \left( \frac{\left[(V_f^\dagger P_f V_f)_{f_if_j} - (V_{f^c}^T P_{f^c} V_{e^c}^*)_{f_if_j}\right]q_sr^2}{6\, n\,f_a \left(q_X^2 + q_S^2 r^2  \right)}\right), \\
\end{aligned}
\end{split}\\
\begin{split}
\begin{aligned}[b]
C^A_{f_if_j}& = \left( \frac{\left[(V_f^\dagger P_f V_f)_{l_il_j} + (V_{f^c}^T P_{f^c} V_{f^c}^*)_{f_if_j}\right]q_sr^2}{6\, n\,f_a \left(q_X^2 + q_S^2 r^2  
 \right)}\right).\\
\end{aligned}
\end{split}
\end{align}
\end{subequations}
Here, $C^V_{f_if_j}$ and $C^A_{f_if_j}$ are the vector and axial vector couplings derived from Eq.~(\ref{eq:fvLagrangian}) after applying the equations of motion. The two-body decays of pseudoscalar mesons into axion final states are sensitive to the $C^V_{f_i f_j}$ couplings, since the matrix element for the decay $\mathcal{M}(K^+\rightarrow\pi^+a)$ has a vanishing axial term, and the one for the $C^V_{f_i f_j}$ coupling is given by $\mathcal{M}(K^+\rightarrow\pi^+a)=C^V_{sd}\partial_{\mu}\langle{\pi(p_{\pi})}\overline{s}\gamma^{\mu}d\rangle{K(p_K)}=C^V_{sd}m_K^2(1-m_{\pi}^2/m_K^2)/(m_s-m_d)$. The factor $m_K^2(1-m_{\pi}^2/m_K^2)/(m_s-m_d)= B_sm_K$ arises from kinematics and from the evaluation of the meson transition amplitude in chiral perturbation theory. The flavor violating coupling in our framework can be obtained from Eq.~(\ref{eq:fvLagrangian}) and a general expression for the decay width can thus be written as:
\begin{equation}
\Gamma(K^{+} \rightarrow \pi^{+} a) \simeq \frac{m_K^3}{192 \pi f_a^2} \left( 1-\frac{m_{\pi}^2}{m_K^2} \right)^3 \left|\frac{\left[(V_d^\dagger P_Q V_d)_{ds} - (V_{d^c}^T P_d V_{d^c}^*)_{ds}\right]q_Sr^2}{
n\left(q_X^2 + q_S^2 r^2  
 \right)}\right|^2.
\end{equation}
For a specific texture, we first compute the mixing parameter $\kappa_{ds}(\epsilon)$ defined as $\kappa_{ds}(\epsilon)=\left[(V_d^\dagger P_Q V_d)_{ds} - (V_{d^c}^T P_d V_{d^c}^*)_{ds}\right](\epsilon)$, and derive a lower bound on $f_a$ from the experimental limit $\text{Br}(K^+ \rightarrow \pi^+ a) \lesssim 2.9 \times 10^{-11}$~\cite{NA62:2025upx}. The branching ratio can be calculated by noting that the leptonic channel $K^+\rightarrow \mu^+\nu_{\mu}$ constitutes $63.56\%$ of the decay width~\cite{ParticleDataGroup:2024cfk}, and thus the total decay rate is given by $\Gamma_{\rm tot}=\Gamma(K^+\rightarrow \mu^+\nu_{\mu})/0.6356\simeq  \frac{m_K}{0.6356\times8 \pi} \left( G_F f_k m_{\mu} |V_{us}|\right)^2$. 
For a given  $\kappa_{ds}(\epsilon)$ there is a lower bound on $f_a$ based on the flavor charges of the model and the VEV ratio $r=v_S/v_X$  given by:
\begin{equation}\label{eq:FVLowerbound}   
   f_a \gtrsim 3.05 \times 10^{11}\,\text{GeV} \,\, \frac{|q_S| r^2\sqrt{|\kappa_{ds}(\epsilon)|}}{
n\left(q_X^2 + q_S^2 r^2  
 \right)} \,\,.
\end{equation}
We also note that in all the models we consider, the parameter $\kappa_{ds}$ takes values $\kappa_{ds} (\epsilon)\sim \epsilon\simeq 0.2$ from the flavor charge assignment that lead to the desired mass matrix textures. The $\mathcal{O}(1)$ coefficients of $\kappa_{ds}$ are taken to be equal to $1$ in all the numerical analyses that follow. There are two other weaker bounds pertaining to the decay of mesons with the third generation quarks, viz., $B^{+} \rightarrow (K,\pi)^{+}  a$:
\begin{eqnarray}
\hspace{-0.5cm}\Gamma(B^{+} \rightarrow (K,\pi)^{+} a) &\simeq& \frac{m_B^3\left( f_{+}^{(K,\pi)}(0)\right)^2}{192 \pi f_a^2}  \left( 1-\frac{m_{(K,\pi)}^2}{m_B^2} \right)^3 \nonumber \\
&\times& \left|\frac{\left[(V_d^\dagger P_Q V_d)_{(d,s)b} - (V_{d^c}^T P_d V_{d^c}^*)_{(d,s)b}\right]q_Sr^2}{
n\left(q_X^2 + q_S^2 r^2  
 \right)}\right|^2,
\end{eqnarray}
where, $f_{+}^{(K,\pi)}(0)=(0.331,0.258)$ are the hadronic form factors of $K$ and $\pi$ mesons respectively~\cite{Ball:2004ye}. The branching ratio limit $\text{Br}(B^{+} \rightarrow (K,\pi)^+\,a)<1.0\times 10^{\minus 6}$ obtained by the BELLE-II experiment was interpreted in Ref.~\cite{Abumusabh:2025zsr}, which yields a bound:
\begin{equation}\label{eq:FVBLowerbound}   
   f_a \gtrsim (4.1,5.0) \times 10^{7}\,\text{GeV} \,\, \frac{|q_S| r^2\sqrt{|\kappa_{(d,s)b}(\epsilon)|}}{
 n\left(q_X^2 + q_S^2 r^2  
 \right)} \,\,,
\end{equation}
where again
\begin{equation}
    \kappa_{(d,s)b}(\epsilon)=\left[(V_d^\dagger P_Q V_d)_{(d,s)b} - (V_{d^c}^T P_d V_{d^c}^*)_{(d,s)b}\right](\epsilon)~.
\end{equation}
Similar bounds on $f_a$ based on the $B\rightarrow\pi a$ decays at BaBar was derived in Ref.~\cite{MartinCamalich:2020dfe}. Since the models developed here have $\kappa_{ds} \sim \epsilon$ and $\kappa_{sb} \sim \epsilon^2$, the bound derived on $f_a$  from  Eq.~(\ref{eq:FVBLowerbound}) is less stringent than the one arising from Eq.~(\ref{eq:FVLowerbound}).

A similar lower bound can be obtained from leptonic flavor violating (LFV) decays such as  $\mu^{+}\rightarrow e^{+}a$, but this time with a twist. The final leptonic state has a non-zero spin and thus these decays will be dependent on both the vector ($C^V$) and axial vector ($C^A$) couplings. Therefore, it is prudent to express them in terms of an effective coupling  $C_{l_il_j}=\sqrt{|C^V_{l_il_j}|^2+|C^A_{l_il_j}|^2}$~\cite{DiLuzio:2020wdo}.  The decay of $\mu^{+}\rightarrow e^{+}a$ provides a strong bound on $f_a$ but it is only sensitive to purely vector or axial vector currents owing to the genesis of this bound from the kinematically forbidden decay region of $\mu^{+}\rightarrow e^{+}\bar{\nu}\nu$ at TRIUMF~\cite{Carr:1983kp} and TWIST~\cite{TWIST:2014ymv}. The three body decay $\mu^{+}\rightarrow e^+ a \gamma$ with $\text{Br}(\mu^{+}\rightarrow e^+ a \gamma)<1.1\times 10^{- 9}$ at Crystal Box~\cite{Bolton:1988af} however, is independent of the chirality of the $C_{a\mu e}$ coupling, thereby giving us a conservative lower bound of
\begin{equation}\label{eq:FVmuLowerbound}   
   f_a \gtrsim 1.7 \times 10^{8}\,\text{GeV}\,\,  \frac{|q_S| r^2\sqrt{|\varkappa_{\mu e}(\epsilon)|}}{
n\left(q_X^2 + q_S^2 r^2  
 \right)} \,\,,
\end{equation}
where we have defined
\begin{equation}
\varkappa_{\mu e}(\epsilon)=\left[(V_e^\dagger P_L V_e)_{\mu e} + (V_{e^c}^T P_e V_{e^c}^*)_{\mu e}\right]^2+\left[(V_e^\dagger P_L V_e)_{\mu e} - (V_{e^c}^T P_e V_{e^c}^*)_{\mu e}\right]^2~.
\end{equation}
This bound on $f_a$ derived from the leptonic process  is still better than the bound in Eq.~(\ref{eq:FVBLowerbound}) in our setup, since the factors $\varkappa_{\mu e}$ and $\kappa_{sb}$ are comparable. Further details on such LFV bounds can be found in Ref.~\cite{Calibbi:2020jvd}. We will make use of these bounds to constrain models and  study their phenomenology further in Sec.~\ref{subsec:FVcons} where the flavor bounds on the axion compete with other experimental and astrophysical bounds.

\subsection{Quality of the axion: Tree and loop level corrections to the axion potential}\label{subsec:axionquality}
The effective instantonic coupling of the axion field $a(x)$ with the gluons due to the QCD anomaly in Eq.~(\ref{eq:axiongluoncoupling}) induces a potential $V(a)$ for the axion. This potential has been computed in chiral perturbation theory to be~\cite{Weinberg:1977ma,DiVecchia:1980yfw}:
\begin{equation}\label{eq:chiralaxionpotential}
    V(a)\simeq -m_{\pi}^2f_{\pi}^2\sqrt{1-\frac{4m_um_d}{(m_u+m_d)^2}\sin^2{\left(\frac{a}{2f_a}+\frac{\overline{\theta}}{2}\right)}}.
\end{equation}
This potential has a shift symmetry $a\rightarrow a+f_a \overline{\theta}$ and has its minimum at $\left(\frac{a}{f_a}+\overline{\theta}\right)=0$. Higher dimensional operators perturbatively generated by quantum gravity would shift this minimum away from zero. For instance, consider the $d=5$ operator arising from quantum gravity in the DFSZ model:
\begin{equation}\label{eq:PQbreakpotential}
V_{\cancel{\text{PQ}}}\supset  \frac{\xi}{M_{\rm Pl}} |\Phi|^4(e^{i\delta}\Phi+\text{h.c}.)\implies V^{(a)}_{\cancel{\text{PQ}}}=\frac{\kappa}{M_{\rm Pl}} \frac{f_a^5}{2^{5/2}}\cos{\left(\frac{a}{f_a}+\delta\right)},
\end{equation}
Here $\Phi=(\rho+f_a)/\sqrt{2}$ is the SM singlet scalar of DFSZ model whose VEV breaks $U(1)_{\rm PQ}$ spontaneously and $\xi,\delta$ are real parameters naturally of order unity.\footnote{We note here that we have not included $2!\, 3!$ factorials in the denominator of Eq.~(\ref{eq:PQbreakpotential}) which can potentially be present, corresponding to the identical $(\Phi^*)^2,\Phi^3$ fields. Introducing such factorials would help with having a higher quality axion. These factors are left out to stay conservative with axion quality.}
Such Planck suppressed operators would spoil the axion solution to the strong CP problem in a theory with just a global anomalous $U(1)_{\rm PQ}$ symmetry. Minimizing Eq.~(\ref{eq:chiralaxionpotential}) along with Eq.~(\ref{eq:PQbreakpotential}) would correspond to a gravity-induced shift in $\overline{\theta}$ given by
\begin{equation}\label{eq:bareaxionquality}
    \overline{\theta}\simeq\frac{\xi \sin \delta}{m_{\pi}^2f_{\pi}^2}\frac{(m_u+m_d)^2}{m_um_d}\frac{f_a^5}{4 \sqrt{2}M_{\rm Pl}}
\end{equation}
Using $M_{\rm Pl}=1.22\times10^{19}\,\,\text{GeV},f_{\pi}=93\,\,\text{MeV},m_{\pi}=140 \,\,\text{MeV}$, and $m_u/m_d=0.56$, and requiring that $\overline{\theta}\leq10^{-10}$ (due to the non-observation of an Electric Dipole Moment (EDM) for the neutron~\cite{Crewther:1979pi,Abel:2020pzs}), we would need
\begin{equation}\label{eq:barequalityconstraint}
    |\xi \sin \delta|\lesssim 1.6\times(10^{-39},10^{-44},10^{-54}), \,\,\text{for}\,\, f_a=(10^{9},10^{10},10^{11})\,\,\text{GeV}.
\end{equation}
This fine-tuning is even more severe than the one needed for the original strong CP problem.  This illustrates the 
need for a mechanism to protect the axion potential from excessive quantum gravity corrections. The present framework realizes this by identifying $U(1)_{\rm PQ}$ as an accidental symmetry originating from a $U(1)_F$ gauge symmetry, which is immune to quantum gravity corrections.

In our general setup, the residual PQ symmetry is explicitly broken by quantum gravitational operators~\cite{Holman:1992us,Kamionkowski:1992mf,Barr:1992qq,
Ghigna:1992iv,Babu:1992cu} (in addition to the QCD anomaly) which are of the form
\begin{equation}\label{eq:PQbreak}
    V_{\cancel{\text{PQ}}} \supset \xi e^{i\delta} \frac{ S^{m} X^{l}}{M_{\text{Pl}}^{(l+m)-4}}+ \text{h.c}.
\end{equation}
where $m,l$ are postive integers.
From the Higgs pseudoscalar mass operator of Eq.~(\ref{eq:cons}), we can establish without loss of generality that for $(h_u+h_d)/q_X=\mathfrak{p}/\mathfrak{q},\mathfrak{p}\in \mathbb{Z},\mathfrak{q}\in \mathbb{N}$, we have
\begin{equation}\label{eq:lm}
l= q_X(\pm k \mathfrak{q}+\mathfrak{p}),~m=n\mathfrak{q}.    
\end{equation}
This term in Eq.~(\ref{eq:PQbreak}) generates a shift in the axion potential and thereby a shift in $\overline{\theta}$. If we express the VEVs $v_S$ and $v_X$ in terms of $r,f_a$ by using Eq.~(\ref{eq:favX}), the induced shift $\Delta\overline{\theta}$ can be written as:
\begin{equation}\label{eq:axionquality}
\Delta\overline{\theta}_{\rm{tree}}\simeq \frac{\xi \sin \delta}{m_{\pi}^2f_{\pi}^2}\frac{(m_u+m_d)^2}{m_um_d}\left| \frac{-3\,n\sqrt{q_X^2 + q_S^2r^2} }{q_X\sqrt{2}}\right|^{l+m} \frac{(f_a)^{l+m}}{r^{l}M_{\text{Pl}}^{l+m-4}} .
\end{equation}
Normally, for the ranges $r\sim (0.1-10),f_a\sim(4.5\times10^{8}-  10^{12})~\rm{GeV}$, and $(l+m)\geq 10$, we have a high quality axion that is stable against perturbative quantum gravitational corrections. For values of $(l+m) < 10$ one should carefully examine the corrections to see if the quality of the axion is maintained.

Operators of the type shown in Eq.~(\ref{eq:PQbreak}) can also arise via loop diagrams which may have lower powers of $M_{\rm Pl}$ in the denominator. These operators arise from other scalar or fermionic PQ breaking operators which do not induce Eq.~(\ref{eq:PQbreak}) directly, but can combine to form loops with a final effective operator that violates PQ symmetry given as 
\begin{equation}\label{eq:generalaxionquality}
V_{\cancel{\text{PQ}}} \supset \Xi~ e^{i\delta }\frac{S^{\mathfrak{a}}X^{\mathfrak{b}}}{M_{\text{Pl}}^{(\mathfrak{a}+\mathfrak{b})-4}}\left( \frac{X}{\Lambda_{\rm FN}}\right)^{\mathfrak{c}}.
\end{equation}
Here, gauge invariance for Eqs. (\ref{eq:PQbreak})-(\ref{eq:generalaxionquality}) requires the relations $\mathfrak{a}q_S+(\mathfrak{b}+\mathfrak{c})q_X=lq_S+mq_X=0$ and $l=\mathfrak{a},(\mathfrak{b}+\mathfrak{c})=m$.  We have defined an operator-specific suppression factor $\Xi$ which would depend on the loop structure that will be specified case by case. Thus,
we can write a general expression for $\Delta\overline{\theta}$ using Eqs. (\ref{eq:axionquality})-(\ref{eq:generalaxionquality}):
\begin{equation}
\Delta\overline{\theta}_{\rm{loop}}\simeq \frac{\Xi \sin \delta}{m_{\pi}^2f_{\pi}^2}\frac{(m_u+m_d)^2}{m_um_d} \left| \frac{-3\,n\sqrt{q_X^2 + q_S^2r^2} }{q_X\sqrt{2}}\right|^{\mathfrak{a}+\mathfrak{b}+\mathfrak{c}}\frac{(f_a)^{\mathfrak{a}+\mathfrak{b}}}{r^{\mathfrak{a}}M_{\text{Pl}}^{\mathfrak{a}+\mathfrak{b}-4}}\left(\frac{f_a}{\Lambda_{\text{FN}}}\right)^{\mathfrak{c}} .
\label{teta-ma}
\end{equation}
For all the models that we present, we can calculate the shift in $\overline{\theta}$ for specific values of $(r,f_a)$ if we identify the leading order operator of the form in Eq.~(\ref{eq:generalaxionquality}). 
 
Along the same lines, we can estimate the mass of the pseudoscalar Higgs boson arising from the term in Eq.~(\ref{eq:pot}) in terms of $r,f_a,\tan \beta$:
\begin{equation}\label{Amass}
    M_{\text{A}_\text{H}}^2\simeq \lambda_{\text{HS}}\frac{1+\tan^2\beta}{\tan \beta}\left| \frac{-3\,n\sqrt{q_X^2 + q_S^2r^2} }{q_X\sqrt{2}}\right|^{n+k}\frac{|f_a|^{n+k}}{r^{k}M_{\text{Pl}}^{n+k-2}}.
\end{equation}
As noted before, due to $SU(2)_L$ invariance this will be also the mass of the heavy Higgs doublet in these models. 
This estimate will be useful in later sections to constrain the possible models of high quality axion. This will also be relevant for the discussion of baryon asymmetry of the universe via thermal leptogenesis.

\subsection{Axion dark matter constraints}\label{subsec:DMcons}

Here we consider the scenario of post-inflationary PQ breaking, which is more
viable cosmologically irrespective of specific inflationary models, as opposed to pre-inflationary PQ breaking owing to constraints from iso-curvature perturbations~\cite{Ema:2016ops}. The latest numerical simulations~\cite{Saikawa:2024bta,Benabou:2024msj} on the production of axion Dark Matter after inflation has revealed that the major contribution comes from the decay of global axionic strings and String-Wall networks in the case of $N_{\rm DW}=1$, whee $N_{\rm DW}$ denotes the number of domain walls connected by axion string networks.\footnote{In our class of models involving gauged and accidentally emerged global Abelian symmetries, the DM abundance appears to be dominated by the decay of ``global cosmic strings" with the ``gauged-cosmic strings" contribution to DM abundance being sub-dominant~\cite{Niu:2023khv}. It is thus reasonable to use the DM abundance obtained from the decay of cosmic strings produced by a single anomalous PQ symmetry. More details can be found in Sec.~\ref{sec:Domainwallproblem}. 
}
There seems to be some discourse in the community currently about the exact mechanism and range of axion DM production, but recent numerical simulations point to a very narrow range for the QCD axion to be the total DM in the universe. As we shall see in Sec.~\ref{sec:Domainwallproblem}, we have $N_{\rm DW}=1$ for all the models developed in this work. So, we combine the ranges of Ref.~\cite{Saikawa:2024bta,Benabou:2024msj} to conservatively bound the axion DM as
\begin{equation}
\begin{aligned}[b]
&\hspace{0.5cm}m_a\in[45,450]\,\,\mu\text{eV},\\
&f_a\in1.27\times[ 10^{10},10^{11}]\,\,\text{GeV}.
\end{aligned}   \label{eq:axionDMcons}
\end{equation}
We have used a fairly stringent bound on $m_a,f_a$, when compared to the earlier estimates from Ref.~\cite{Gorghetto:2020qws}, which predicts a considerably lower value of $f_a$ and a slightly heavier QCD axion. But it turns out that the lower values on $f_a$ are more constrained by flavor violating decays in our flavored axion setup and we cannot go to arbitrarily low $f_a$, even though this would benefit the quality of the axion greatly. A lower value in the range of $f_a\in[10^9,10^{10}]\,\,\text{GeV}$ is clearly viable for axion quality, and we choose the latest numerical bounds to demonstrate that our models can effectively have a high quality axion and DM candidate simultaneously. We should also note that any value of $f_a$ above the upper bound in Eq.~(\ref{eq:axionDMcons}) would produce too much axion DM that overcloses the universe~\cite{Preskill:1982cy, Abbott:1982af}. These bounds are used to check for a viable parameter space in the models of Secs.~\ref{sec:ModelI}, \ref{sec:ModelII} and \ref{sec:ModelIII} where we list in Tables~\ref{tab:ModelIqualitycases}, \ref{tab:ModelIIqualitycases} and \ref{tab:ModelIIIqualitycases} and Figs. ~\ref{fig:ModelIQualityPlot}, \ref{fig:ModelIIQualityPlot} and \ref{fig:ModelIIIQualityPlot}, the models which simultaneously provide a high quality axion as the full DM in the universe.

\subsection{Baryogenesis through leptogenesis}\label{sec:Baryogenesis}
Given the neutrino mass predictions in the models of Secs.~\ref{sec:ModelI}, \ref{sec:ModelII} and \ref{sec:ModelIII}, it is readily possible to assert whether these models are capable of generating the observed baryon asymmetry through thermal leptogenesis~\cite{Fukugita:1986hr,Pilaftsis:1997jf,Giudice:2003jh,Davidson:2008bu,Pilaftsis:2003gt,Buchmuller:2004nz} involving the heavy right-handed neutrinos present in the models.  Their presence is necessary for neutrino mass generation as well as for anomaly cancellation in our framework. We quickly provide a rundown of the common machinery needed to realize thermal leptogenesis. For the Dirac Yukawa coupling matrices $Y_{\nu}$ illustrated in Secs.~\ref{sec:ModelI}, \ref{sec:ModelII} and \ref{sec:ModelIII}, the light neutrino mass matrix is
defined as $M_{\nu}=v_{u,d}^2Y_{\nu}M_R^{-1}Y_{\nu}^T$
depending on which Higgs doublet is involved in the Dirac neutrino mass generation. Here $\hat{Y_\nu}= Y_\nu U_N$, where the unitary matrix $U_N$ diagonalizes the Majorana mass matrix of $N$-field via $U_N^T M_N U_N = {\rm diag}(\hat{M}_1, \hat{M}_2, \hat{M}_3)$ with $\hat{M}_i$ denoting the mass eigenvalues. Furthermore, the decay widths for $N_i \rightarrow L_j  + H_d^* + {\rm h.c.}$  are given by $\Gamma_i = \frac{\hat{M}_i}{4\pi}(\hat{Y}_\nu^\dagger \hat{Y}_\nu)_{ii}$. 

Since we would require ``resonant" and ``standard" thermal leptogenesis scenarios to realize the correct baryon asymmetry of the universe in different models, we introduce the CP asymmetries in each scenario.
The asymmetry arising from $\hat{N}_1$ decay ($\epsilon_1$) and its analog in $\hat{N}_2$ decay ($\epsilon_2$) in the resonant leptogenesis scenario are given by
\cite{Pilaftsis:1997jf,Pilaftsis:2003gt} 
\begin{equation} \label{eq:resonantCPassym}
\epsilon^{(1)}_{\text{CP}} = \frac{\text{Im}\left[ \{(\hat{Y}_\nu^\dagger \hat{Y}_\nu)_{21}\}^2\right]}{(\hat{Y}_\nu^\dagger \hat{Y}_\nu)_{11} (\hat{Y}_\nu^\dagger \hat{Y}_\nu)_{22}}\frac{\sin{2\phi_k}}{2}, ~~~~\epsilon^{(2)}_{\text{CP}} = \epsilon^{(1)}_{\text{CP}}(1\leftrightarrow2)~.
\end{equation}
where we have defined $\tan{\phi_k} = \frac{\Gamma_2}{2 (M_2-M_1)}$, such that the second kinematic factor in Eq.~(\ref{eq:resonantCPassym}) can be identified as 
\begin{equation}\label{eq:kinematicfactorres}
    \frac{\sin{(2\phi_k)}}{2}=\frac{(\hat{M}_2^2-\hat{M}_1^2)\, \hat{M}_1 \Gamma_2}{ (\hat{M}_2^2-\hat{M}_1^2)^2+ \hat{M}_1^2 \Gamma_2^2}.
\end{equation}
On the other hand, the total CP asymmetry in the standard thermal leptogenesis scenario is given by CP asymmetry generated in the decay of the lightest RHN $N_1$ given by~\cite{Covi:1996wh}
\begin{equation}\label{eq:VanillaCPasym}\epsilon_{\text{CP}}^{(1)}=\frac{1}{8 \pi}\sum_j\frac{ \text{Im}\left[ \{(\hat{Y}_\nu^\dagger \hat{Y}_\nu)_{1j}\}^2\right]}{(\hat{Y}_\nu^\dagger \hat{Y}_\nu)_{11}}g\left(x_j\right), 
\end{equation}
where
$x_j\equiv \hat{M}_j^2/\hat{M}_1^2$ and $g(x)=\sqrt{x}\left[\frac{1}{1-x}+1-(1+x)\log\left( \frac{1+x}{x}\right)\right]$ is the loop factor from the one loop contributions to the lepton CP asymmetry in $N_1 \rightarrow L + H_d^*$ and $N_1 \rightarrow \overline{L} + H_d$ decays.   The only surviving asymmetry is the one arising from the decay of the lightest RHN $N_1$ and hence we can effectively use $\epsilon^{(1)}_{\mathrm{CP}}$ for computing the baryon asymmetry, except in situations where $N_1$ is nearly degenerate with $N_2$, the second lightest state.
The efficiency factors associated with the production of lepton asymmetry from the decay of the lightest RHN ($N_1$) has a semi-analytic form given in Ref.~\cite{Giudice:2003jh}:\footnote{This approximation is obtained by solving the relevant Boltzmann equations numerically in the context of SM plus RHN. This should approximately be valid in our setup with two Higgs doublets as well, since the couplings of $H_d$ with the $b$-quark mimic those of $H$ in the SM with the top quark. The neutrino Dirac Yukawa couplings in our models require the participation of $\widetilde{H_d}$ alone with $H_u$ playing the role of a spectator, and so we can use the same expressions as standard thermal leptogenesis by the replacement of $v$ by $v_d$.}
\begin{equation}\label{eq:efffactor}
    \kappa_{f}=\left(\frac{3.3 \times 10^{-3}~\text{eV}}{\tilde{m}_1}+\left(\frac{\tilde{m}_1}{0.55 \times10^{-3}~\text{eV}} \right)^{1.16}\right)^{-1}.
\end{equation}
where $\tilde{m}_{1}= v_{d}^2 \frac{(\hat{Y}_{\nu}^{\dagger}\hat{Y}_{\nu})_{11}}{\hat{M}_1}$. The final baryon asymmetry relative to the comoving entropy density $(Y_{\Delta B}=(n_B-n_{\bar{B}})/s)$ is:
\begin{equation}\label{eq:Baryonasym}
Y_{\Delta B}\simeq \frac{135\, \zeta(3)}{4 \pi^4 g_{*}}a_{\rm{sph}}\epsilon_{\rm CP}^{(1)} \kappa_{f},     
\end{equation}
which has to be $Y_{\Delta B}\equiv (8.75\pm 0.23)\times 10^{-11}$~\cite{Planck:2018vyg}. This depends on the equilibrium number density of $N_i$ at $T\gg M_i$, when the number of relativistic degrees of freedom of the SM is $g_*\simeq106.75$; the efficiency factor $\kappa_{f}$ of lepton asymmetry production and the sphaleron factor $a_{\rm{sph}}=Y_{\Delta B}/Y_{\Delta(B-L)}=12/37$~\cite{Harvey:1990qw}. 

We now proceed to illustrate three models that have unique features and parameter spaces that shed light on several of the peculiarities of the SM that were described in the introduction. 
\section{Model I: A simple model with three right-handed neutrinos}\label{sec:ModelI}
We first present a minimal model that utilizes three right-handed neutrinos ($N_i$) to generate small neutrino masses, and also to cancel all the gauge anomalies listed in Eq.~(\ref{eq:anomalies}). The gauged $U(1)_F$ charges of the model are listed in Table~\ref{tab:ModelIcharges}. 
\begin{table}[!htbp]
\footnotesize 
$$
\begin{array}{|c||c|c|c|c|c|c|c|}

\hline
\vspace{-0.3cm}
 &  &  &  &  &  &  & \\

\vspace{-0.4cm}

{\rm Field}& \{Q_1, Q_2, Q_3\}&  \{u^c_1, u^c_2, u^c_3\} &
 \{d^c_1, d^c_2, d^c_3\}& \{L_1, L_2, L_3\} &  \{e^c_1, e^c_2, e^c_3\} & \{N_1, N_2, N_3\}&\{H_u, H_d, X, S \} \\

&  &  &  &  &  &  &\\

\hline

\vspace{-0.3cm}
 &  &  &  &  &  &  &\\

\vspace{-0.3cm}
\hspace{-0.5mm}Q_F\hspace{-0.5mm}&\hspace{-0.7mm} \{1,0,\minus 2\} \hspace{-0.7mm}& \hspace{-0.7mm} \{3,0,\minus 2\} \hspace{-0.7mm}  &\!\hspace{-0.7mm} \{\minus\frac{13}{3},\frac{8}{3}, \frac{8}{3}\} \hspace{-0.7mm}  &\{\minus\frac{2}{3},\frac{16}{3}, \minus\frac{5}{3} \} & \{\minus \frac{8}{3},\minus \frac{8}{3}, \frac{7}{3}\}&\{ 4, \minus 6,\minus 1\}& \{4, \minus\frac{2}{3},\, 1, \minus \frac{\pm 3 k + 10}{3n}\}\\

&  &  &  &  &  &  &\\
\hline
\end{array}$$
\caption{Anomaly-free flavor charge assignment for fermions and scalars $(H_u, H_d, X, S$) in Model I. All gauge anomalies listed in Eq.~(\ref{eq:anomalies}) can be verified to be vanishing.}
\label{tab:ModelIcharges}
\end{table}

With this charge assignment, the interactions Lagrangian relevant for the fermion masses and mixings can be written down:
\begin{equation}\label{eq:ModelI_Mass}
\begin{aligned}[b]
& \mathcal{L}_{\text{Yuk}}\supset Q^T H_u\begin{pmatrix}
        {\bar{\varepsilon}}^{8}  &  {\bar{\varepsilon}}^{5}  & {\bar{\varepsilon}}^{3}  \\
       {\bar{\varepsilon}}^{7} & {\bar{\varepsilon}}^{4}  & {\bar{\varepsilon}}^{2}  \\
       {\bar{\varepsilon}}^{5} & {\bar{\varepsilon}}^{2}  &1\\
        \end{pmatrix}u^c+
Q^T H_d \left( \begin{array}{ccc}
        \varepsilon^4 &  {\bar{\varepsilon}}^{~\!\!3}&  {\bar{\varepsilon}}^{~\!\!3} \\
       \varepsilon^5 &  {\bar{\varepsilon}}^{~\!\!2} &  {\bar{\varepsilon}}^{~\!\!2} \\
       \varepsilon^7 &1 &1
      \end{array}
\right) d^c+
L^T H_d\left( \begin{array}{ccc}
        \varepsilon^4 & \varepsilon^4 & {\bar{\varepsilon}} \\
       {\bar{\varepsilon}}^{~\!\!2} &{\bar{\varepsilon}}^{~\!\!2}& {\bar{\varepsilon}}^{~\!\!7}\\
       \varepsilon^5 &\varepsilon^5 &1
      \end{array}
\right) e^c \\
& \hspace{2.5cm}+L^T \tilde H_d\left( \begin{array}{ccc}
         {\bar{\varepsilon}}^{~\!\!4} & \varepsilon^6 & \varepsilon \\
       {\bar{\varepsilon}}^{~\!\!10} &1& {\bar{\varepsilon}}^{~\!\!5}\\
        {\bar{\varepsilon}}^{~\!\!3} &\varepsilon^7 & \varepsilon^2
      \end{array}
\right) N+
\Lambda_{\text{FN}}N^{T}\left( \begin{array}{ccc}
        {\bar{\varepsilon}}^{~\!\!8} & \varepsilon^2 &  {\bar{\varepsilon}}^{~\!\!3} \\
       \varepsilon^2 &  \varepsilon^{12} & \varepsilon^7 \\
       {\bar{\varepsilon}}^{~\!\!3} &\varepsilon^7 & \varepsilon^2
      \end{array}
\right) N+\text{h.c} ,
\end{aligned}
\end{equation}
where we have introduced the dynamical fields 
\begin{equation}
\varepsilon \equiv \frac{X}{\Lambda_{\rm{FN}}},~~~ \bar{\varepsilon} \equiv\frac{X^*}{\Lambda_{\rm{FN}}}.
\label{eq:ModelIepsilondef}
\end{equation}
In Eq.~(\ref{eq:ModelI_Mass}), $\Lambda_{\rm FN}$ is the flavor cut-off scale.  Here dimensionless couplings which are all taken to be of $\mathcal{O}(1)$ are not explicitly shown.  With the vacuum expectation value defined as
\begin{equation}
\langle \varepsilon \rangle =\langle {\bar{\varepsilon}} \rangle \equiv \epsilon,
\label{eq:ModelIXVEV}
\end{equation}
for the charged fermion Yukawa couplings and the CKM matrix elements we get from Eq.~(\ref{eq:ModelI_Mass}) the following order of magnitude estimates:
\begin{equation}    
Y_t\sim 1 ,~~ \frac{Y_u}{Y_c}\sim  \frac{Y_c}{Y_t} \sim \epsilon^4 ,~~~~\frac{Y_e}{Y_{\mu }} \sim \epsilon^2  ,~~ \frac{Y_{\mu }}{Y_{\tau }} \sim \epsilon^2~.
\label{eq:ModelIYukHier}
\end{equation}
\begin{equation}
 |V_{us}|\sim \epsilon ,~~~ |V_{cb}|\sim \epsilon^2 ,~~~ |V_{ub}|\sim \epsilon^3 .
\label{eq:ModelICKM}
\end{equation}
With a choice of $\epsilon\simeq 0.2$ all observables are satisfactorily reproduced. 
Thus, this setup provides an excellent fit to the charged fermion masses and CKM mixing angles with all Yukawa couplings being order one. We do allow for the ${\cal O}(1)$ couplings to vary modestly away from one, as is needed, for e.g., to fit the ratios $Y_e/Y_\mu$ in comparison with $Y_\mu/Y_\tau$.

\subsection{Model I: Neutrino phenomenology}\label{subsubsec:ModelIneutrinopheno}
Now we turn to the neutrino sector.
The neutrino Dirac and Majorana  mass matrices emerging from Eq.~(\ref{eq:ModelI_Mass}), to  a good approximation, can be parameterized as:
\begin{equation}
m_D\!\simeq \!\left( \! \begin{array}{ccc}
                 y_{11}^{\nu }\epsilon^4 & 0 &y_{13}^{\nu }\epsilon \\
                 0 & y_{22}^{\nu } & 0 \\
                 y_{31}^{\nu }\epsilon^3 & 0 & y_{33}^{\nu }\epsilon^2
               \end{array}
 \!\right) \!\frac{v_d}{\sqrt{2}}  ,~~~~
 M_R\!\simeq \!\left( \! \begin{array}{ccc}
                 0 &y_{12}^{N}\epsilon^2 & y_{13}^{N}\epsilon^3 \\
                y_{12}^{N}\epsilon^2 & 0 & 0 \\
                 y_{13}^{N}\epsilon^3 & 0 & y_{33}^{N}\epsilon^2
               \end{array}
 \!\right) \Lambda_{\rm FN}~.
\label{eq:ModelIneutrinomass}
\end{equation}
These matrices will lead, via the see-saw formula,  $M_{\nu }\simeq -m_DM_R^{-1}m_D^T$, for the light neutrino mass matrix of the form:
\begin{equation}
M_{\nu }\!=\!\left( \! \begin{array}{ccc}
                 \beta^2 & \gamma' & \beta \\
                 \gamma' & \gamma & \alpha \\
                 \beta & \alpha & 1
               \end{array}
 \!\right) \! m_0 .
\label{eq:ModelIlightneutrinomass}
\end{equation}
Here we have defined
$$
 m_0=-\frac{(y_{33}^{\nu }\epsilon )^2v_d^2}{2y_{33}^N\Lambda_{\rm FN}},~~~
\alpha=\frac{y_{22}^{\nu }(y_{31}^{\nu }y_{33}^N-y_{33}^{\nu }y_{13}^N)}
{(y_{33}^{\nu })^2y_{12}^N\epsilon },~~~
\beta=\frac{y_{13}^{\nu }}{y_{33}^{\nu }\epsilon },~~~
$$
\begin{equation}
\gamma=\left( \frac{y_{22}^{\nu }y_{13}^N}{y_{33}^{\nu }y_{12}^N\epsilon }\right)^{\!2},~~~
\gamma'=\frac{y_{22}^{\nu }(y_{11}^{\nu }y_{33}^N\epsilon^2-y_{13}^{\nu }y_{13}^N)}
{(y_{33}^{\nu }\epsilon )^2y_{12}^N }.
\label{eq:ModelImnuparams}
\end{equation}
The texture of Eq.~(\ref{eq:ModelIlightneutrinomass}) is rather predictive. The $(2,2)$ co-factor of $M_\nu$ is zero, i.e.,
\begin{equation}
M_{\nu }^{(1,1)}M_{\nu }^{(3,3)}-\left( M_{\nu }^{(1,3)}\right)^2+{\cal O}(m_0^2\epsilon^4)=0.
\label{eq:ModelInupred}
\end{equation}
The correction ${\cal O}( m_0^2\epsilon^4)$ in the L.H.S of Eq.~(\ref{eq:ModelInupred}), which is negligible,
 comes from the 1-2 rotation of ${\cal O} (\epsilon^2)$ of the charged lepton sector.
The relation in Eq.~ (\ref{eq:ModelInupred}), with the input from neutrino oscillation data, allows to have two predictive
relations. For instance, in  terms of two Majorana phases $(\alpha_1, \alpha_2)$
one can calculate the CP violating phase $\delta $ and the lightest neutrino mass. Such a scenario has been investigated in Ref.~\cite{Liao:2013rca} for both  normal ordering (NO) and inverted ordering (IO) of neutrino masses.
Here we present one set of input parameters for each case, which shows consistency with neutrino oscillation data.

\subsubsection*{Normal Ordering of Neutrino masses:}
\label{subsubsec:ModelINO}
For this case, we choose as input $y_{ij}^{\nu }, y_{ij}^{N}, v_d$ and $\Lambda_{\rm FN}$ to be:

$$
y_{ij}^{\nu }\!\simeq \!\left( \! \begin{array}{ccc}
                 1.9527e^{-i0.6391} & 0 &0.1782 \\
                 0 & 0.7071 & 0 \\
                 1.9534e^{-i0.0749}  & 0 & 3.2298
               \end{array}
 \!\right),~~
 y_{ij}^{N}\!\simeq \!\left( \! \begin{array}{ccc}
                 0 &1/3 & 0.2754 \\
               1/3  & 0 & 0 \\
                 0.2754 & 0 & 1
               \end{array}
 \!\right),~~
$$
\begin{equation}
v_d=5\sqrt{2}{\rm\,\, GeV},~~~~\Lambda_{\rm FN} =4.193\times 10^{11}~{\rm GeV},
\label{y-nuN}
\end{equation}
which yield for the neutrino masses, mixing angles and CP-violating phases the following output:
\begin{subequations}
\begin{align}
    \{m_1, m_2, m_3 \}&=\{0.003, ~0.0091,~ 0.0508\}{\rm \,\,eV},
\label{nu-masses-y}\\
    \{\sin^2\te_{12},  \sin^2\theta_{23}, \sin^2\theta_{13}\}&=\{0.3035,~ 0.455,~ 0.0224\},
\label{nu-mixings-y}\\
\{\de_{\rm CP},\alpha_1,\alpha_2\}& =\{1.259\pi ,~-1.352,~ 2.417\}.
\label{dero12}
\end{align}
\end{subequations}
From Eq.~(\ref{nu-masses-y}) we get
\begin{equation}
\Delta m_{\rm sol}^2=m_2^2-m_1^2=7.394\times 10^{-5}{\rm eV}^2,~~
\Delta m_{\rm atm}^2=m_3^2-m_2^2= 2.492\times 10^{-3}{\rm eV}^2~.
\label{nu-mass2dif-y}
\end{equation}
The output values given in Eq.~(\ref{nu-mixings-y}) and Eq.~(\ref{nu-mass2dif-y}) correspond to the best fit values of the neutrino data for the NO case~\cite{Esteban:2024eli}.
Here and below, for leptonic mixing matrix elements (including the Dirac and Majorana phases), we use the same parametrization as in Ref.~\cite{Esteban:2024eli}.
As we see, the couplings in Eq.~(\ref{y-nuN}) are within $[1/5,3]$, which we find to be satisfactory. From the value of $\Lambda_{\rm FN}$ we  get the estimate
$v_X\sim \sqrt{2}\Lambda_{\rm FN} \epsilon \simeq 1.2\times 10^{11}$~GeV.
This value goes well with a high quality axion which also serves as the entire DM of the universe.

Using Eq.~(\ref{y-nuN}) as input, for the heavy RHN masses we obtain:
\begin{equation}
 \{\hat M_1, \hat M_2, \hat M_3 \}=\{  0.5552,~ 0.5609,~ 1.683\}\times 10^{10}~{\rm GeV} .
\label{M123}
\end{equation}
The nearly degenerate $\hat M_1$ and $\hat M_2$ masses give good potential for realizing resonant leptogenesis.
Indeed, with Eq.~(\ref{M123}),  and inputs of Eq.~(\ref{y-nuN}), for the baryon asymmetry we obtain $Y_{\Delta B}\simeq 8.7\times 10^{-11}$
 as discussed in detail in  Sec.~\ref{subsec:ModelIleptogenesis}.

\subsubsection*{Inverted Ordering of Neutrino masses:}\label{subsubsec:ModelIIO}

For this case we take as input:
$$
y_{ij}^{\nu }\!\simeq \!\left( \! \begin{array}{ccc}
                 9.2152e^{i0.1076} & 0 &0.7527 \\
                 0 & 0.7071 & 0 \\
                 4.8151e^{-i0.2229}  & 0 & 0.7529
               \end{array}
 \!\right)_{\!\!ij},~~
 y_{ij}^{N}\!\simeq \!\left( \! \begin{array}{ccc}
                 0 &1 & 0.5692 \\
               1  & 0 & 0 \\
                 0.5692 & 0 & 0.8108
               \end{array}
 \!\right)_{\!\!ij},~~
$$
\begin{equation}
v_d=5\sqrt{2}{\rm~~ GeV},~~~~\Lambda_{\rm FN} =2.525\times 10^{11}~{\rm ~GeV},
\label{y-nuN-IO}
\end{equation}
which give as output:
\begin{subequations}
    \begin{align}
        \{m_1, m_2, m_3 \}&=\{0.0717, ~0.07225,~ 0.0523\}{\rm eV},
\label{nu-massesIO}\\
\{\sin^2\te_{12},  \sin^2\te_{23}, \sin^2\te_{13}\}&=\{0.303,~ 0.43,~ 0.0224\},
\label{nu-mixingsIO}\\
\{\de_{\rm CP},\alpha_1,\alpha_2\} &=\{1.9\pi,~ -3.228,~ -3.0958\}.
\label{dero12-IO}
    \end{align}
\end{subequations}
From Eq.~(\ref{nu-massesIO}) we get
\begin{equation}
\Delta m_{\rm sol}^2=m_2^2-m_1^2=7.394\times 10^{-5}~{\rm eV}^2,~~
\Delta m_{\rm atm}^2=m_2^2-m_3^2= 2.492\times 10^{-3}{\rm eV}^2~.
\label{nu-mass2difIO}
\end{equation}
Output values given in Eq.~(\ref{nu-mixingsIO}) and Eq.~(\ref{nu-mass2difIO}), except the value of the $\sin^2\te_{23}$ (which is within the $2\sigma $),
correspond to the best fit values of the neutrino data for the IO case~\cite{Esteban:2024eli}.

For the heavy RHN masses we obtain for the case of IO:
\begin{equation}
 \{\hat M_1, \hat M_2, \hat M_3 \}=\{  0.7922,~ 1.0135,~ 1.0402\}\times 10^{10}~{\rm GeV} .
\label{M123-IO}
\end{equation}
This case also gives the desirable baryon asymmetry, $Y_{\Delta B}\simeq 8.7\times 10^{-11}$,
as discussed in Sec.~\ref{subsec:ModelIleptogenesis}. Note that the $(1,1)$  entry of $y_{ij}^{\nu }$  is relatively large compared to unity in Eq.~(\ref{y-nuN-IO}) for the case of IO. A smaller value (along with other couplings) could have been achieved 
by properly increasing the value of $\Lambda_{\rm FN}$. Doing so would preserve the neutrino fit. However, leptogenesis requires such a large value for $y_{11}^{\nu }$. It is important to remember that the corresponding operator is generated from the FN sector via the diagram in 
Fig.~\ref{fig:ModelInuUV},
which involves five different Yukawa couplings. If three of these couplings are of order $2$, then $y_{11}^{\nu }\!\sim 2^3\!=8$.\footnote{Alternatively, if some of the bare masses of the FN fields are slightly below $\Lambda_{\rm FN}$, the effective coupling  $y^\nu_{11}$ would be larger than one.} Such values for the couplings are therefore fully consistent within the theoretical framework and we illustrate this consistency further in Sec.~\ref{subsec:ModelIaxionquality} where we explicitly show the dependence of the effective Yukawas on the fundamental Yukawa couplings.

\subsection{Model I: Resonant leptogenesis}\label{subsec:ModelIleptogenesis}
We now turn to the leptogenesis mechanism to create baryon asymmetry within Model I.
In the UV completion of this model, we used the effective couplings shown in Eq.~(\ref{eq:ModelIneutrinomass}), with the mass parameters having the hierarchy 
$(M_1,M_2,M_3)\sim \Lambda_{\rm FN}(\epsilon^2, \epsilon^3, \epsilon^2)$.  This structure of the Majorana mass matrix for $N_i$ naturally leads to two quasi-degenerate states composed of $(N_1, N_2)$ with nearly equal masses,  $\hat{M}_1 \simeq \hat{M}_2 \sim y_{12}^N \epsilon^2\Lambda_{\rm FN}$ and a third state $N_3$ with a mass nearly equal to $\hat{M}_3\sim y_{33}^N \epsilon^2\Lambda_{\rm FN}$ that is comparable to $\hat{M}_1$.  Denoting the three mass eigenstates of $N_i$ to be $\hat{N}_i$ (and the mass eigenvalues to be $\hat{M}_i$), the decays $\hat{N}_i \rightarrow L + H_d^*$ and $\hat{N}_i \rightarrow \overline{L} + H_d$ for $i = 1,2$ will generate a leptonic CP asymmetry that is resonantly enhanced. 
 
In the present case we note that all Majorana mass parameters $M_{1,2,3}$ appearing in $M_N$ of Eq.~(\ref{eq:ModelIneutrinomass}) can be made real by field rotations. The mass eigenvalues are easily worked out by standard perturbation theory, with the hierarchy $(M_R)_{12} \sim (M_R)_{33} \sim \epsilon^2 \Lambda_{\rm FN},\,(M_R)_{13} \sim \epsilon^3 \Lambda_{\rm FN}$ (see Eq.~(\ref{eq:ModelIneutrinomass})) to second order in $\frac{y_{13}^N}{y_{12}^N}\epsilon,\frac{y_{13}^N}{y_{33}^N}\epsilon$  to be
\begin{eqnarray}
\hat{M_1} &\simeq&  \Lambda_{\rm FN}\left( y_{12}^N + \frac{(y_{13}^N)^2}{2(y_{12}^N-y_{33}^N)}\epsilon^2 \right) \epsilon^2\nonumber\\
\hat{M_2} &\simeq& \Lambda_{\rm FN}\left( y_{12}^N + \frac{(y_{13}^N)^2}{2(y_{12}^N+y_{33}^N)}\epsilon^2 \right) \epsilon^2 \nonumber\\
\hat{M}_3 &\simeq&  \Lambda_{\rm FN}\left( y_{33}^N + \frac{(y_{13}^N)^2y_{33}^N}{(y_{33}^N)^2-(y_{12}^N)^2}\epsilon^2 \right) \epsilon^2~.
\label{eq:ModelIsplitting}
\end{eqnarray}
The unitary matrix $U_N$ that diagonalizes $M_N$ takes the form\footnote{The mass splitting $\hat{M}_1-\hat{M}_2$ depends quadratically on $\epsilon$, which is kept in Eq.~(\ref{eq:ModelIsplitting}); while for the mixing angles linear term in $\epsilon$ would suffice.} to linear order in
$y_{13}^N\epsilon$ 
\begin{eqnarray}
U_N \simeq \left(\begin{matrix}\frac{1}{\sqrt{2}} & -\frac{i}{\sqrt{2}} & \frac{y_{13}^Ny_{33}^N\epsilon}{(y_{33}^N)^2-(y_{12}^N)^2}
 \cr \frac{1}{\sqrt{2}} & \frac{i}{\sqrt{2}} & \frac{y_{12}^Ny_{13}^N\epsilon}{(y_{33}^N)^2-(y_{12}^N)^2}
 \cr \frac{y_{13}^N\epsilon }{\sqrt{2}(y_{12}^N-y_{33}^N)} & \frac{iy_{13}^N\epsilon }{\sqrt{2}(y_{12}^N+y_{33}^N)}  & 1 \end{matrix}\right)~.
\end{eqnarray}
This arises from three sequential rotations: First, a $45^o $ rotation is made in the 1-2 plane, followed by simultaneous small rotations in the 1-3 and 2-3 planes.  A third phase rotation by a diagonal phase matrix diag$(1, i,1)$ makes all the masses real and positive. (Note that the diagonal (1,1) and (2,2) entries of the resulting matrix do not correspond to the mass eigenvalues, which have been obtained to second order in
$y_{13}^N\epsilon $.)

We find the relations
\begin{eqnarray}
\hat{M}_1^2-\hat{M}_2^2 = 2 \epsilon^6 \frac{y_{12}^N(y_{13}^N)^2y_{33}^N}{(y_{12}^N)^2-(y_{33}^N)^2}\Lambda_{\rm FN}^2,~~~\Gamma_2 = \epsilon^2 \frac{|y_{22}^{\nu}|^2 y_{12}^N}{8 \pi}\Lambda_{\rm FN},
\end{eqnarray}
and consequently, $(\hat{M}_1^2 - \hat{M}_2^2)^2 \sim \epsilon^{12} \Lambda_{\rm FN}^4\sim \hat{M}_1^2 \Gamma_2^2 \sim \epsilon^8 \Lambda_{\rm FN}^4/(64 \pi^2) \sim \epsilon^{12} \Lambda_{\rm FN}^4$. The two terms in the denominator of the second factor of Eq.~(\ref{eq:resonantCPassym}) are then comparable.  The second factor approximates to its maximal value, which is $1/2$.  Using this approximation we can estimate the lepton asymmetry to be
\begin{equation}\label{eq:ModelICpassym}
\varepsilon_{\rm CP}\simeq
\frac{4 y_{12}^Ny_{13}^N}{(y_{12}^N)^2-(y_{33}^N)^2} \epsilon^6 \,\Im((y_{33}^{\nu })^*y_{31}^{\nu }+(y_{13}^{\nu})^*y_{11}^{\nu })~.
\end{equation}
 The suppression $\epsilon^6 \sim 6 \times 10^{-5}$, along with $\mathcal{O}(1)$ changes that can occur in varying the parameters
$y_{ij}^{\nu }$ and $y_{ij}^N$ would lead to the CP asymmetry being around few $\times 10^{-6}$.  This is in the right range to explain the observed baryon asymmetry of the universe.

We emphasize that the resonant leptogenesis can be realized in both - NO and IO cases of  neutrino oscillation  scenarios. The suppression factor for the CP asymmetry parameter is $\sim \epsilon^6$
for both cases. However, to get desirable baryon asymmetry within the IO case, we found that certain effective Dirac  Yukawa coupling has to be of order 8, as shown in the neutrino oscillation fits of Eq.~(\ref{y-nuN-IO}). We have argued that such modest deviations from unity in these effective Yukawa couplings are quite natural from the UV-completion perspective. Thus, we conclude that realistic neutrino masses and mixings, as well as baryogenesis via leptogenesis can be realized in NO and IO cases of Model I.

\subsection{Model I: Flavor UV completion and quality of the axion}\label{subsec:ModelIaxionquality}
We first present an explicit UV completion of the FN sector of Model I.
A sufficient set of Yukawa couplings to generate the charged fermion masses and CKM mixing angles combined with the neutrino parameters from Sec.~\ref{subsubsec:ModelIneutrinopheno} is found to be:

\begin{equation}\label{eq:ModelIUV}
\begin{aligned}[b]
& \mathcal{L}_{\text{Yuk}}\supset Q^T H_u\left( \begin{array}{ccc}
        0 & y^u_{12} {\bar{\varepsilon}}^{5}  &0  \\
       0& y^u_{22}{\bar{\varepsilon}}^{4}  & y^u_{23}{\bar{\varepsilon}}^{2}  \\
       y^u_{31}{\bar{\varepsilon}}^{5} & y^u_{32}{\bar{\varepsilon}}^{2}  &y^u_{33}
      \end{array}
\right)  u^c+
Q^T H_d\left( \begin{array}{ccc}
       y^d_{11}  \varepsilon^4 & 0 &  0 \\
      0 &  y^d_{22} {\bar{\varepsilon}}^{2} & y^d_{23} {\bar{\varepsilon}}^{2}\\
       0 & y^d_{32}  & y^d_{33} 
      \end{array}
\right)  d^c\\
&\hspace{4.0cm}+ L^T H_d\left( \begin{array}{ccc}
        y^l_{11} \varepsilon^4 & y^l_{12}\varepsilon^4 & 0 \\
       y^l_{21}{\varepsilon}^{5} &y^l_{22}{\bar{\varepsilon}}^{2}& 0\\
       0 &0 & y^l_{33}
      \end{array}
\right)  e^c +\text{h.c}.
\end{aligned}
\end{equation}
We adopt this effective Lagrangian for charged fermion mass generation along with Eq.~(\ref{eq:ModelIneutrinomass}) for the neutrino Dirac and Majorana mass matrices. 
The effective Yukawa couplings in Eq.~(\ref{eq:ModelIUV}) for the charged fermions can be generated by integrating out additional $H_{u}^{(q)}(1,2,\frac{1}{2}), H_d^{(q)}(1,2,-\frac{1}{2})$ Higgs doublet states with masses of the order $\Lambda_{\rm FN}$, whereas Dirac and Majorana effective Yukawa couplings given in Eq.~(\ref{eq:ModelIneutrinomass}) can be generated by integrating out heavy Higgs doublets and Majorana fermions with masses at the $\Lambda_{\rm FN}$ scale.

\begin{figure}[!htbp]
\centering
  \begin{minipage}{0.4\textwidth}
        \centering
        \includegraphics[width=\linewidth]{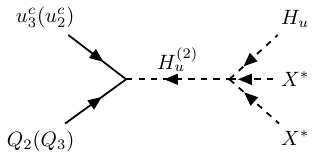}
    \end{minipage}
    \hfill
    \begin{minipage}{0.5\textwidth}
        \centering
        \includegraphics[width=\linewidth]{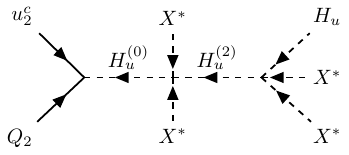}
\end{minipage}
\includegraphics[width=0.65\linewidth]{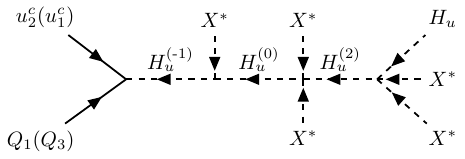}
\caption{Diagrams generating  up-type quark Yukawa operators of Model I given in Eq.~(\ref{eq:ModelIUV}).}
\label{fig:modelIuUV}
\end{figure}

\begin{figure}[!htbp]
\centering
\begin{minipage}{0.44\textwidth}
        \centering
        \includegraphics[width=\linewidth]{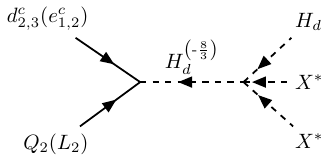}
    \end{minipage}
    \hfill
    \begin{minipage}{0.5\textwidth}
        \centering
        \includegraphics[width=\linewidth]{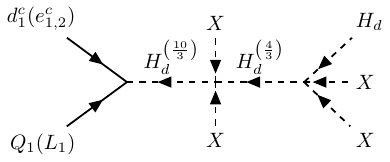}
\end{minipage}
\caption{Diagrams generating down type quark and charged lepton Yukawa operators for Model I given in Eq.~ (\ref{eq:ModelIUV}).}
\label{fig:ModelIdUV}
\end{figure}

\begin{figure}[!htbp]
\centering
\begin{minipage}{0.4\textwidth}
        \centering
        \includegraphics[width=\linewidth]{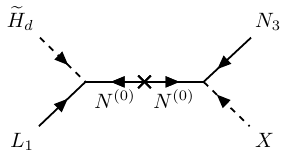}
    \end{minipage}
    \hfill
    \begin{minipage}{0.4\textwidth}
        \centering
        \includegraphics[width=\linewidth]{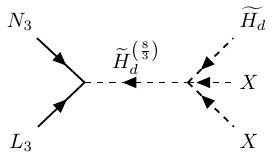}
\end{minipage}
\hfill
\begin{minipage}{0.6\textwidth}
        \centering
        \includegraphics[width=\linewidth]{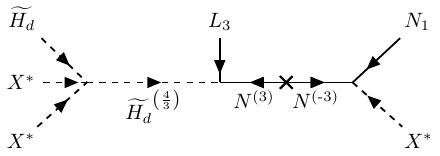}
    \end{minipage}
    \hfill
    \begin{minipage}{0.6\textwidth}
        \centering
        \includegraphics[width=\linewidth]{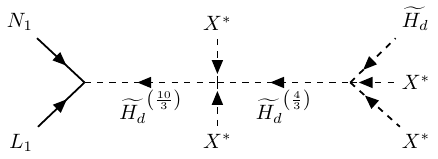}
\end{minipage}
\hfill
    \begin{minipage}{0.4\textwidth}
        \centering
        \includegraphics[width=\linewidth]{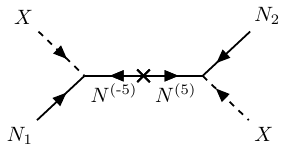}
\end{minipage}
\hfill
    \begin{minipage}{0.4\textwidth}
        \centering
        \includegraphics[width=\linewidth]{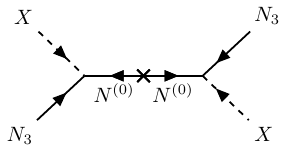}
\end{minipage}
\includegraphics[width=0.6\linewidth]{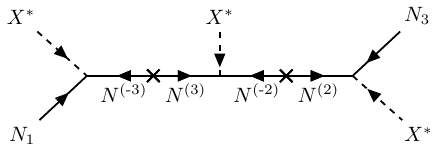}
\caption{Diagrams generating Dirac and Majorana Neutrino Yukawa operators for Model I given in Eq.~ (\ref{eq:ModelIneutrinomass}).}
\label{fig:ModelInuUV}
\end{figure}

We introduce three new $H_u^{(q)}$ Higgs doublet fields to generate the up-type quark masses shown in Fig.~\ref{fig:modelIuUV} with $U(1)_F$ charges denoted as:\footnote{For early attempts of UV-completing FN models with scalars, see Refs.~\cite{Bijnens:1986tt,Calibbi:2012yj,Calibbi:2012at}.}
\begin{equation}\label{eq:compHu}
    \{H_u^{(-1)},H_u^{(0)},H_u^{(2)}\}\supset \{ {\bar{\varepsilon}}^{5},  {\bar{\varepsilon}}^{4},  {\bar{\varepsilon}}^{2} \}H_u.
\end{equation}
We also introduce three $H_d^{(q)}$ Higgs doublets to generate the down-type quark and charged lepton masses shown in Fig.~\ref{fig:ModelIdUV} with the $U(1)_F$ charges given by: 
\begin{equation}\label{eq:compHd}
    \{H_d^{\left( \minus\frac{8}{3}\right)},H_d^{\left(\frac{4}{3}\right)},H_d^{\left( \frac{10}{3}\right)}\}\supset \{ {\bar{\varepsilon}}^{2},  {{\varepsilon}}^{2},  {{\varepsilon}}^{4} \}H_d.
\end{equation}
These Higgs fields are assumed to have positive squared masses, so that they only receive induced vacuum expectation values via their couplings with $H_u$ and $H_d$ Higgs doublets which have negative squared masses. These induced VEVs are suppressed by powers of $\epsilon=\langle X\rangle/\Lambda_{\rm FN}$, as indicated in Eqs.~(\ref{eq:compHu})-(\ref{eq:compHd}) in terms of powers of the spurion field $\varepsilon = X/\Lambda_{\rm FN}$.
For the Neutrino Yukawa sector we introduce three pairs of $U(1)_F$-charged and one neutral Majorana fermions denoted as
\begin{equation}\label{eq:ModelIRHNFNstates}
   N^{(\pm 5)}, N^{(\pm 3)},N^{(\pm 2)},N^{(0)}
\end{equation}
in order to generate both the Dirac and Majorana neutrino Yukawa operators shown in Fig.~\ref{fig:ModelInuUV}.
It is straightforward to integrate out the heavy doublets $H_u^{(q)}$ and $H_d^{(q)}$ which have masses of order $\Lambda_{\rm FN}$.  We denote the fundamental Yukawa couplings of the quarks with these heavy doublets as $(y^0_u)_{ij}$ and $(y^0_d)_{ij}$, and use the notation
\begin{equation}\label{eq:ModelIfundmassud}
\left\{M_{H_u^{(2)}}^2, \,M_{H_u^{(0)}}^2,\, M_{H_u^{(-1)}}^2, \,M_{H_d^{(-\frac{8}{3})}}^2,\,\,M_{H_d^{(\frac{10}{3})}}^2,\,\,M_{H_d^{(\frac{4}{3})}}^2,\, \mu^2 \right\} = \left\{c_1^u,\, c_2^u,\, c_3^u,\, c_1^d,\, c_2^d,\,c_3^d,\, \mu_0^2   \right\}\Lambda_{\rm FN}^2~.
\end{equation}
Here the cubic scalar coupling $\mu$ and the quartic scalar couplings relevant for integrating out the heavy fields  are defined as 
\begin{eqnarray}\label{eq:ModelIfundyukud}
 V &\supset& \lambda_1^u\, H_u^{(2) ^\dagger} H_u \,(X^*)^2+ \lambda_2^u\, H_u^{(0) ^\dagger} H_u^{(2)} \,(X^*)^2 + \lambda_1^d \,H_d^{\left(-\frac{8}{3}\right) ^\dagger} H_d \,(X^*)^2 + \lambda_2^d \,H_d^{\left( \frac{10}{3}\right) ^\dagger} H_d^{\left( \frac{4}{3}\right) } \,(X)^2 \nonumber \\
&+& \lambda_3^d\, H_d^{\left( \frac{4}{3}\right) ^\dagger} H_d \,(X)^2 + \mu\, H_u^{(-1) \dagger} H_u^{(0)} X^* + {\rm h.c.}
\end{eqnarray}
The effective Yukawa Lagrangian in the quark sector, obtained after applying equations of motion for the heavy Higgs doublets, reads (to leading order),
\begin{eqnarray}\label{eq:ModelILeffud}
{\cal L}_{\rm eff}^{u,d} &=& (y_u^{(0)})_{33}\, Q_3 u_3^c H_u - \left(\frac{\lambda_1^u} {c_1^u}\right) \left\{ (y_u^{(0)})_{32}\, \bar{\varepsilon}^2 Q_3 u_2^c H_u+ (y_u^{(0)})_{23}\,\bar{\varepsilon}^2 \,Q_2 u_3^c H_u\right\} \nonumber \\
&+&\left(\frac{\lambda_1^u \lambda_2^u (y_u^{(0)})_{22}}{c_1^u c_2^u}\right) \bar\varepsilon^4 Q_2 u_2^c H_u -\left(\frac{\mu_0^2 \lambda_1^u \lambda_2^u}{c_1^u c_2^u c_3^u} \right) \left\{(y_u^{(0)})_{12} \,\bar\varepsilon^5Q_1 u_2^c H_u + (y_u^{(0)})_{31} \,\bar\varepsilon^5 Q_3 u_1^c H_u \right\} \nonumber \\
&+& (y_d^{(0)})_{33} \,Q_3 d_3^c H_d + (y_d^{(0)})_{32}\, Q_3 d_2^c H_d -\left(\frac{d_2 \lambda_3^u(y_d^{(0)})_{22}}{c_1^d}\right)\,Q_2 d_2^c H_d \nonumber \\
&-& \left(\frac{2 d_3 \lambda_3^u(y_d^{(0)})_{23}}{c_1^d}\right)\,Q_2 d_3^c H_d + {\rm h.c.}
\label{eq:Leffuv}
\end{eqnarray}
The charged lepton and Dirac neutrino effective interactions can also be derived similar to the down-type quark interactions. 
These interactions can now be directly matched on to the effective Lagrangian of Eq.~(\ref{eq:ModelIUV}). We have, for example, 
\begin{equation}
y^u_{12} \equiv -\left(\frac{\mu_0^2 \lambda_1^u \lambda_2^u}{c_1^u c_2^u c_3^u}\right)(y_u^{(0)})_{12},
\end{equation}
and so forth. A few points are worth noting here. First, there are no factorials appearing in the denominators, even though various terms have powers of the same field\footnote{This fact is of importance when we calculate the shift in $\overline{\theta}$ due to PQ violating operators arising from loops of heavy FN states as they will have no extra suppression factors from such factorials.} encoded in $\bar\varepsilon^n$. The symmetry factors associated with identical particles is compensated by the number of Feynman diagrams.  Secondly, the effective Yukawa couplings can quite naturally be either a bit larger than one, or a bit smaller than one, even when all fundamental couplings are of the same order.  For example, for $y^u_{12}$ to be of order 8, it is sufficient to choose $c_1^u = c_2^u = c_3^u = 0.5$ (or $\mu_0^2 = \lambda_1^u = \lambda_2^u = 2$), which does not appear to be unnatural. Similarly, an effective coupling of order $0.1$ can also be realized naturally, by taking the reciprocal of the choice quoted.

In order to express the effective Yukawas in Eq.~(\ref{eq:ModelIUV}) for the Majorana neutrino sector in terms of the fundamental $\mathcal{O}(1)$ Yukawas of the FN sector, we write the masses of the heavy FN Majorana states as

\begin{equation}\label{eq:ModelIfundmassN}
\left\{M_{N^{(0)}},\,M_{N^{(2)}},\,M_{N^{(3)}},\,M_{N^{(5)}} \right\} = \left\{c_0^N,\, c_2^N,\, c_3^N,\, c_5^N\right\}\Lambda_{\rm FN}~.
\end{equation}
Now, the FN Yukawa couplings between the RHNs and the heavy Majorana FN fields are as follows:
\begin{equation}
\begin{aligned}[b]
        \mathcal{L}_{\rm FN}^{N}&\supset y_1^NN_1N^{(\minus5)}X+y_2^NN_2N^{(5)}X+y_3^NN_3N^{(0)}X+y_4^NN_1N^{(\minus3)}X^*\\
        &+y_5^NN^{(3)}N^{(\minus2)}X^*+y_6^NN_3N^{(2)}X^*+\text{h.c}.
\end{aligned}
\end{equation}
Integrating out the heavy fermions, we obtain the effective Majorana Yukawa Lagrangian to (leading order) as
\begin{eqnarray}\label{eq:ModelILeffnu}
    \mathcal{L}_{\text{eff}}^{\rm \nu,N}\supset \frac{y^{N}_{1} y^{N}_{2}}{c_5^N}N_1N_2\varepsilon^2- \frac{y_4^N y_5^N y_6^N} {c_2^Nc_3^N}N_1N_3\bar{\varepsilon}^3+\frac{\left( y_3^N\right)^2}{c_0^N}N_3N_3\varepsilon^2+\text{h.c.}
\end{eqnarray}
The effective Yukawas for the RHN sector in Eq.~(\ref{eq:ModelIUV}) can be read off from here as:
\begin{equation}\label{eq:FNRHNfunYuk}
    y_{12}^{N}\simeq \frac{y^{N}_{1} y^{N}_{2}}{c_5^N},~~~ y_{13}^{N}\simeq \frac{y_4^N y_5^Ny_6^N} {c_2^Nc_3^N},~~~y_{33}^N\simeq\frac{\left( y_3^N\right)^2}{c_0^N}.
\end{equation}
This has enough freedom in the choice of $\mathcal{O}(1)$ fundamental Yukawas that allows for the freedom in the effective Yukawa couplings (for instance, $y_{33}^N\sim0.2$ can be achieved by the choice of $y^{(3)}_{1N}\sim 0.9,~ y^{(2)}_{3N}\sim 0.5,~ y_{N}^{(32)}\sim 0.5,~ y_N^{(3)}\sim 1.1,~ y_N^{(2)}\sim 1.1$), effectively producing the satisfactory neutrino fits of Eqs. (\ref{y-nuN})-(\ref{y-nuN-IO}).

We note that the UV-complete theory presented here remains perturbative all the way to the Planck scale $M_{\rm Pl}$.  This can be seen by the running of the gauge couplings given by the one-loop renormalization group equation given by
\begin{equation}
16 \pi^2 \frac{dg_i}{d\,{\rm log}\,\mu} = b_i\, g_i^3
\end{equation}
with $b_3=-7, b_2=-2$ and $b_1 = +8$ with the new Higgs doublets added to the SM.  Thus we see that the $SU(3)_c$ and $SU(2)_L$ gauge couplings will remain asymptotically free, while the hypercharge gauge coupling will have a value of $\alpha_Y(M_{\rm Pl}) \simeq 0.019$ for a $\Lambda_{\rm FN}$ compatible with Eqs. (\ref{y-nuN})-(\ref{y-nuN-IO}).  As for the fundamental Yukawa couplings, they may be chosen at the scale $\Lambda_{\rm FN}$ to be slightly below unity, in which they would also remain perturbative up to the Planck scale.

\subsubsection*{Axion quality in Model I}
We now turn to address the quality of the axion in the fully UV-complete version of Model I. First we illustrate an instance of the high quality of the axion at tree-level for the case $(n,k)=(1,1^{+})$. In this case we have $q_S=-13/3$ in the normalization where $q_X=1$, see Eq.~(\ref{eq:cons}). The leading tree-level PQ violating operator induced by gravity is
\begin{equation}\label{eq:ModelIn1k1treequality}
V_{\cancel{\rm PQ}}\supset\frac{S^3X^{13}}{M_{\rm Pl}^{12}}+\text{h.c,}
\end{equation}
which has enough Planck suppression making its contribution to $|\Delta\overline{\theta}|$ very negligible. Other values of $(n,k)$ such as $(n,k) = (2,2^\pm)$ are also compatible with axion quality from the allowed tree-level operators.

\begin{figure}[!htbp]
\centering
\includegraphics[width=0.7\textwidth]{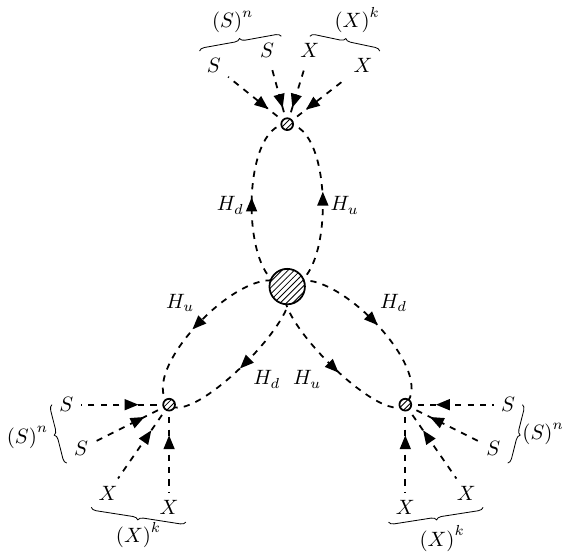}
\caption{Three-loop ``flower" diagram generating the operator of Eq.~(\ref{PQvil-Hud2}).}
\label{fig-3loop}
\end{figure}

However, once the Froggatt-Nielsen Higgs fields used for UV completion of Model I are included, there are more important loop-induced corrections arising from quantum gravity, which would disallow the choices of $(n,k) = (1,1^\pm)$ and $(2,2^\pm)$, as well as many other choices. We find the leading PQ-violating operator across all values of $(n,k)$ induced by quantum gravity to be
\begin{equation}
V_{\cancel{\text{PQ}}}\supset \frac{1}{M_{\rm Pl}^2}\left( H_uH_d^{\left(\frac{-8}{3}\right)}\right) \left( H_u^{(0)}H_d\right)^2+\text{h.c}.
\label{PQvil-HudFN}
\end{equation}
Using Eqs. (\ref{eq:compHu})-(\ref{eq:compHd}),  Eq.~(\ref{PQvil-HudFN}) gives the effective PQ-violating coupling
\begin{equation}
\frac{{\bar{\varepsilon}}^{10}}{M_{\rm Pl}^2}(H_uH_d)^3.
\label{PQvil-Hud1}
\end{equation}
Note that this operator has only two powers of $M_{\rm Pl}$ suppression. 
This operator, in combination with the coupling from Eq.~(\ref{eq:pot}) that induces the pseudoscalar Higgs mass, with $q_S=-1/[n\left( \pm k+10/3\right)]$, will induce an explicit PQ breaking operator via a three-loop diagram shown in Fig.~\ref{fig-3loop}. The hatched blob vertex in  Fig.~\ref{fig-3loop} corresponds to a Planck suppressed vertex generated by perturbative gravitational corrections.  The PQ-breaking scalar interactions involving $(X,S)$ fields arising from the loop diagram can be estimated to be
\begin{equation}
V_{\cancel{\text{PQ}}}^{\rm{ind}}\supset\left( \frac{\ln (\Lambda_{\rm FN}/M_{\text{A}_\text{H}})}{16\pi^2}\right)^3\frac{S^{3n}(X^{(*)})^{3|k|}}{M_{\text{Pl}}^{3n+3|k|-4}}\left( \frac{X}{\Lambda_{\rm FN}}\right)^{10}+\text{h.c}.
\label{PQvil-Hud2}
\end{equation}
The logarithmic enhancement factor $\Xi\sim \left(\ln (\Lambda_{\rm FN}/M_{\text{A}_\text{H}})/16\pi^2\right)^3$ in Eq.~(\ref{PQvil-Hud2}) comes from infrared regime of Fig \ref{fig-3loop}, where it happens that there are always two light Higgs states $H_u, H_d$ in each of the loops accompanied by the other heavy FN fields. So each loop contains $\int d^4k/(k^2+M_{H_u}^2)(k^2+M_{H_d}^2)$, which leads to the logarithmic enhancement. It should be noted that this estimate is valid when two external fields are identified with the light axion field with the rest of the external fields taking their vacuum values. In this case the external momenta (of the axion fields) can be set to zero and the estimate of Eq.~(\ref{PQvil-Hud2}) is valid. 
As it turns out, the operator of Eq.~(\ref{PQvil-Hud2}) is harmless for the axion quality only for $(n,k)=(3,1^{\pm})$. 
For values of $n+|k|<3$,
it can be seen from Eq.~(\ref{PQvil-Hud2}) that the Planck mass suppression will be too low to realize a high quality axion.   For example, if $n=1$, the Planck suppression factor is only $M_{\rm Pl}^{-2}$.
Using Eqs. (\ref{eq:generalaxionquality})-(\ref{teta-ma}) for the
operator of Eq.~(\ref{PQvil-Hud2}),
for $(n,|k|)=(3,1),~r\sim1,\, f_a\simeq 4.5\times 10^{10}$~GeV (so that the axion is the dark matter of the universe),
we obtain $|\Delta \overline{\theta}|\sim 4\times 10^{-18}$, which is fully consistent with neutron EDM limits. We will be interested in the cases when $v_X\sim v_S\sim (10^{10}- 10^{12})$~GeV which satisfy the lower bound $f_a\stackrel{>}{_\sim }4.5\times 10^8$~GeV arising {from the constraints on the cooling of SN 1987A~\cite{Raffelt:1990yz,Caputo:2024oqc,DiVecchia:2019ejf,DiLuzio:2021pxd,Carenza:2024ehj}. Larger values of $(n,k)$ are prohibited by Eq.~(\ref{eq:minnk}).

Therefore, the consistent choices for a high quality axion
are $(n, k)=(3,1^{+}), (3, 1^{-})$, and it will turn out that the latter case cannot have the axion as a DM candidate.  Therefore, for illustrative purposes, we will consider these options separately in detail. We will illustrate the leading and next-to-leading order loop contributions to $\Delta \overline{\theta}$ and demonstrate the quality of the axion in Model I. 

\subsubsection*{\boldmath{Case with $(n,k)=(3, 1^{+})$}}\label{subsubsec:ModelIcase1}
Corresponding to $(n,k)=(3,1^+)$ we have $q_S=-13/9$ from Eq.~(\ref{eq:cons}) and the charges in Table~\ref{tab:ModelIcharges} in 
the normalization with $q_X=1$.
Thus, the scalar operator that is readily PQ-violating in this case is 
\begin{equation}\label{eq:PQtreen3k1}
V_{\cancel{\text{PQ}}}\supset\frac{S^9 X^{13}}{M_{\rm Pl}^{18}}+\text{h.c},\end{equation} 
which is clearly safe for $\overline{\theta}$. 

We turn to the loop corrections arising from various scalar and fermionic loops that lead to dominant contributions to $\Delta \overline{\theta}$. The most relevant PQ-violating operator found to be the one arising from the UV completion sector and is given by
\begin{equation}
V_{\cancel{\text{PQ}}}\supset\frac{S^{*3}}{M_{\text{Pl}}^3}\left( H_u^{(0)}H_d\right) \left( H_u^{(-1)}H_d^{\left(\frac{-8}{3}\right)}\right)+\text{h.c}.
\label{PQviol-quart}
\end{equation}
Inserting the $H_u, H_d$ components of the FN Higgs fields given in Eqs.~(\ref{eq:compHu})-(\ref{eq:compHd}), Eq.~(\ref{PQviol-quart}) leads to the effective PQ-violating coupling
\begin{equation}
\frac{S^{*3}}{M_{\text{Pl}}^3}{\bar{\varepsilon}}^{~\!\!11}(H_uH_d)^2.
\label{PQvil-Hud4}
\end{equation}
While Eqs.~(\ref{eq:compHu})-(\ref{eq:compHd}) indicate the induced VEVs of the Higgs doublets in terms of $v_u$ and $v_d$, they also specify the weights with which $H_u$ and $H_d$ enter the FN states of the Higgs doublets. This allowed us to derive the effective operator given in Eq.~(\ref{PQvil-Hud4}). This operator, in combination with the pseudoscalar Higgs coupling of Eq.~(\ref{eq:pot}) (with $(n,k)=(3,1^{\pm})$)
at two-loop level generates the operator
\begin{equation}
V_{\cancel{\text{PQ}}}^{\text{ind}}\supset \left( \frac{\ln (\Lambda_{\rm FN}/M_{\text{A}_\text{H}})}{16\pi^2}\right)^2\frac{S^{9}X^2}{M_{\rm Pl}^7}\left( \frac{X}{\Lambda_{\rm FN}}\right)^{11}+ \text{h.c.}
\label{PQvln3k1-2loop}
\end{equation}
The corresponding diagram is shown in Fig.~\ref{fig:PQvil-Hud2}. 
\begin{figure}[!htbp]
\centering
\includegraphics[width=0.75\textwidth]{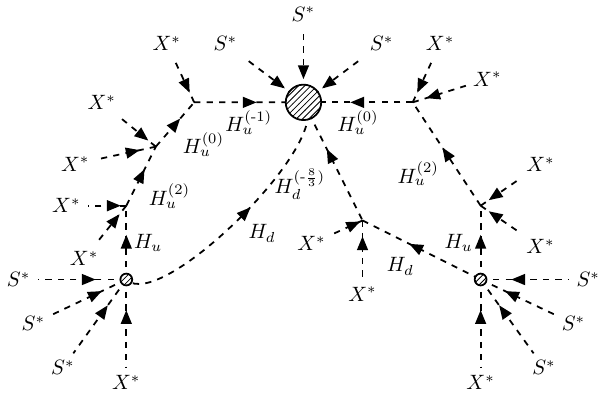}
\caption{Two-loop ``beetle" diagram generating $S^{9}X^{13}$ operator in Eq.~(\ref{PQvln3k1-2loop}) where the shaded blob is the PQ-violating interaction in Eq.~(\ref{PQviol-quart}).}
\label{fig:PQvil-Hud2}
\end{figure}
We can compare Eq.~(\ref{PQvln3k1-2loop}) with Eqs. (\ref{eq:generalaxionquality})-(\ref{teta-ma}) and identify $(\mathfrak{a},\mathfrak{b},\mathfrak{c})=(9,2,11)$. Here $\Xi\sim \left(\ln (\Lambda_{\rm FN}/M_{\text{A}_\text{H}})/16\pi^2\right)^2$ is the loop-specific factor with the logarithm arising from the presence of two light Higgs fields in each of the loops. The gravity-induced $\overline{\theta}$ can then be estimated to be $|\Delta \overline{\theta}|\sim 6.52 \times10^{-12}$ 
for $r\sim1,\,f_a\simeq 4.5\times 10^{10}$~GeV. This value is consistent with neutron EDM limits.

Another sub-leading PQ-violating operator arising from the RHN Majorana sector is
\begin{equation}
V_{\cancel{\text{PQ}}}\supset\frac{1}{M_{\rm Pl}^9}N_2N_2(S^*)^9X^*+ \text{h.c}.
\label{N2N2-n3k1}
\end{equation}
Using this operator together with the couplings of Eq.~(\ref{eq:ModelIUV}), the following operator involving only the $S$ and $X$ fields is generated through the one-loop diagram shown in Fig.~\ref{fig-n3k1-NN1loop}:
\begin{equation}
V_{\cancel{\text{PQ}}}^{\rm{ind}}\supset \left(\frac{1}{16\pi^2}\frac{\Lambda_{\rm FN}^2}{M_N^2}\right)\frac{S^9X^{4}}{M_{\rm Pl}^9}\left(\frac{X}{\Lambda_{\rm FN}}\right)^9+\text{h.c}.
\label{NN-1-loop}
\end{equation}
\begin{figure}[!htbp]
\centering
\includegraphics[width=0.5\textwidth]{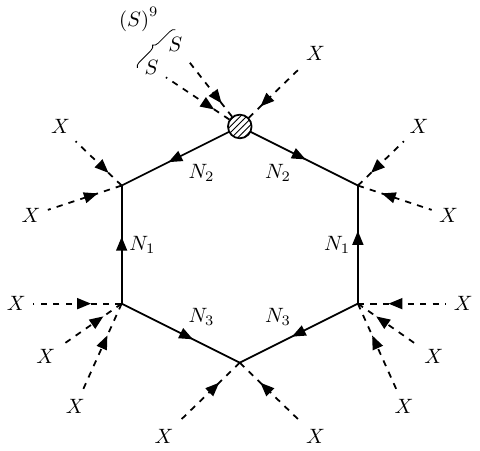}
\caption{One-loop ``hexagon" diagram generating the sub-leading $S^9X^{13}$ operator in Eq.~(\ref{NN-1-loop}) where the shaded blob is the PQ-violating interaction Eq.~(\ref{N2N2-n3k1}).}
\label{fig-n3k1-NN1loop}
\end{figure}
The $M_N$ in Eq.~(\ref{NN-1-loop}) stands for the (common) $N_{1,2,3}$ masses
$\sim M_N\sim \epsilon^2 \Lambda_{\rm FN}$ contributing a loop factor of $\Xi\sim (16\pi^2)^{-1}\Lambda_{\rm FN}^2/M_N^2$. The evaluation of the loop integral, with external momenta set to zero, yields the factor
$\int d^4k/(\cancel{k}+M_N)^6$, which leads to the $1/M_N^2$ term in Eq.~(\ref{NN-1-loop}).  Comparing Eq.~(\ref{NN-1-loop}) with Eqs.(\ref{eq:generalaxionquality})-(\ref{teta-ma}) we identify $(\mathfrak{a},\mathfrak{b},\mathfrak{c})=(9,4,9)$. 
For $r\sim1,\,f_a\simeq 4.5\times 10^{10}$~GeV, $M_N\sim \epsilon^2 10^{11}$~GeV,
this operator contributes to $\bar{\theta}$ 
by an amount 
$|\Delta \bar{\theta}|\sim  10^{-18}$, which is well within the experimental limit from  neutron EDM.

Our systematic investigation of loop diagrams has found no other diagram that shifts $\overline{\theta}$ more than the diagrams that we have discussed. We thus conclude that the case of $(n,k) = (3,1^+)$ is a fully consistent scenario with a high quality axion.

\subsubsection*{\boldmath{Case with $(n,k)=(3, 1^-)$}}\label{subsubsec:ModelIcase2}
In this case, with the
normalization $q_X=1$, we have $q_S=-7/9$ from Eq.~(\ref{eq:cons}) and the charges in Table~\ref{tab:ModelIcharges}. The scalar operator that is readily PQ-violating is 
\begin{equation}\label{eq:PQtreen3k1_}
V_{\cancel{\text{PQ}}}\supset\frac{S^9\left( X\right)^{7}}{M_{\rm Pl}^{12}}+\text{h.c}.\end{equation} 
which again, is clearly safe for $\overline{\theta}$. For this case, the most relevant PQ-violating operator is:
\begin{equation}
V_{\cancel{\text{PQ}}}\supset\frac{S^{*3}}{M_{\text{Pl}}^3}(H_u^{(0)}H_d)(H_u^{(-1)}H_d)+\text{h.c}.
\label{PQvioln3k-1-quart}
\end{equation}
This operator, in combination with the PQ conserving coupling
\begin{equation}
\frac{S^{3}}{M_{\text{Pl}}}\left( H_u^{(-1)}H_d^{\left(\frac{10}{3}\right)}\right)+\text{h.c},
\label{Hud-n3k-1}
\end{equation}
at two-loop level generates the operator: 
\begin{equation}
V_{\cancel{\text{PQ}}}^{\rm{ind}}\supset\left(\frac{1}{(16\pi^2)^2}\frac{|X|^2}{\Lambda_{\rm FN}^2}\right)\frac{S^{9}}{M_{\text{Pl}}^5}\left( \frac{X}{\La_{\rm FN}}\right)^{7}+\text{h.c}.
\label{PQvln3k-1-2loop}
\end{equation}
The corresponding diagram is shown in Fig.~\ref{fig:PQvln3k-1-2loop}.
\begin{figure}[!htbp]
\centering
\includegraphics[width=0.75\textwidth]{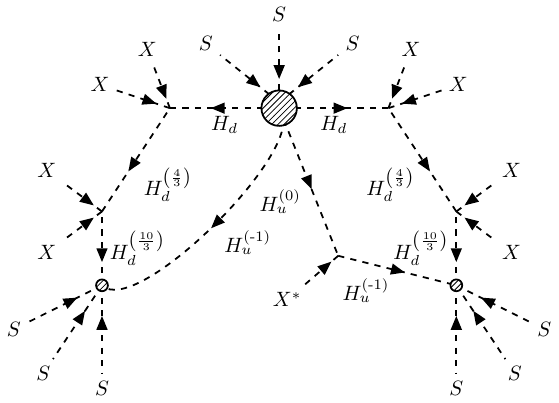}
\caption{Two-loop ``beetle" diagram generating $(S)^{9}(X)^{7}$ operator in Eq.~(\ref{PQvln3k-1-2loop}) where the central large shaded blob is the PQ-violating interaction in Eq.~(\ref{PQvioln3k-1-quart}) and the smaller shaded blobs are the PQ-conserving interaction in Eq.~(\ref{Hud-n3k-1}).}
\label{fig:PQvln3k-1-2loop}
\end{figure}
The loop factor in this case is simply $\Xi\sim\epsilon^2\left(16\pi^2\right)^{-2}$, since there is only one light Higgs doublet in each loop and thus no infrared enhancement. Using Eqs. (\ref{eq:generalaxionquality})-(\ref{teta-ma}) and $(\mathfrak{a},\mathfrak{b},\mathfrak{c})=(9,0,7)$ (note that the $|X|^2$ does not contribute to the phase (axion)  and the factor $\epsilon^2$ has been included in $\Xi $), the operator of Eq.~(\ref{PQvln3k-1-2loop}), 
for $r\sim0.1,f_a\simeq 2.4\times 10^{9}$~GeV - (incompatible with axion as DM), gives  $|\Delta \overline{\theta}|< 10^{-10}$ which is safe for the axion quality. Thus we see that while axion quality can be maintained in this case, there is not enough axion DM.  

Another sub-leading PQ-violating operator is
\begin{equation}
V_{\cancel{\text{PQ}}}\supset \frac{S^{*6}X^*}{M_{\text{Pl}}^5}\left( H_u^{(-1)}H_d^{\left(\frac{-8}{3}\right)}\right)+\text{h.c}.
\label{PQviol-sqr}
\end{equation}
This operator, in combination with the coupling in Eq.~(\ref{Hud-n3k-1}) at one-loop level generates the purely $(S,X)$ operator
\begin{equation}
V_{\cancel{\text{PQ}}}^{\rm{ind}}\supset\left(\frac{1}{16\pi^2}\right)\frac{S^{9}X}{M_{\text{Pl}}^6}\left( \frac{X}{\Lambda_{\text{FN}}}\right)^{6}+\text{h.c}.
\label{PQvln3k-1-1loop}
\end{equation}
The corresponding diagram is shown in Fig.~\ref{fig-PQvln3k-1-1loop}.
\begin{figure}[!htbp]
\centering
\includegraphics[width=0.5\textwidth]{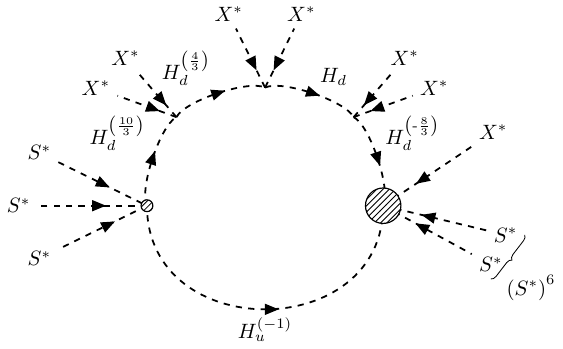}
\caption{One-loop ``Seraphim" diagram generating a sub-leading $S^{9}X^{7}$ operator in Eq.~(\ref{PQvln3k-1-1loop}) where the large shaded blob is the PQ-violating interaction in Eq.~(\ref{PQviol-sqr}) and the smaller shaded blob is the PQ conserving interaction in Eq.~(\ref{Hud-n3k-1}).}
\label{fig-PQvln3k-1-1loop}
\end{figure}
Here again the one-loop factor is trivial, $\Xi\sim(16\pi^2)^{-1}$, without any logarithmic enhancement, as there is only a single light Higgs doublet in the loop. The operator of Eq.~(\ref{PQvln3k-1-1loop}), 
 for the selection of $r$ and $f_a$ 
given after Eq.~(\ref{PQvln3k-1-2loop}), is completely safe from the viewpoint of the high quality axion. Nevertheless, larger $f_a$ values are prohibited, since the operator of Eq.~(\ref{PQvln3k-1-2loop}) imposes a strict upper bound that makes this scenario incompatible with axion being the entire DM of the universe.

It is interesting to note that both the cases $(n,k)=(3,1^{\pm})$, produce leading corrections to the axion potential generated from two-loop diagrams whereas, the one-loop diagrams are sub-leading. Hence, it is not readily apparent that the axion potential receives quantum gravitational corrections in a descending order of the perturbative expansion and we need to evaluate all possible leading order corrections irrespective of the order of the loop at which they are generated. Writing all the flavor invariants and taking into account the flavor invariant PQ violating couplings involving the $S$ field with the least suppression of $M_{\text{Pl}}$ lets us enumerate all possible diagrams. From these, we narrow down to the ones that can form loops without the use of additional $M_{\text{Pl}}$ suppressed vertices to find the leading PQ-violating loop contribution to $\Delta\overline{\theta}$. This is how we arrive at the cases tabulated in Tables~\ref{tab:ModelIqualitycases}, \ref{tab:ModelIIqualitycases} and \ref{tab:ModelIIIqualitycases}.

\subsection{Model I: Summary of results}\label{subsec:ModelIsummary}

The couplings $y_{ij}^{\nu}$, $y_{ij}^{N}$, and the scale $\Lambda_{\rm FN}$ must be chosen carefully to accommodate neutrino data and achieve the desired baryon asymmetry. All of these are accomplished within the frameworks of both NO and IO of neutrino masses. The specific selections are provided in Eq.~(\ref{y-nuN}) and Eq.~(\ref{y-nuN-IO}) respectively, which yield $Y_{\Delta B}\simeq 8.7\times 10^{-11}.$
\begin{table}[!htbp]
    \hspace{0.1cm}
    \renewcommand{\arraystretch}{1.5} 
    \begin{minipage}{0.5\textwidth}
    \hspace{0.5cm}\begin{tabular}{|c|c|c|c|c|c|}
        \hline
        $(n,k)$&  $q_S$& $r$&$M_{\text{A}_\text{H}}[\text{GeV}]$ &\text{High Quality} &\text{Axion DM} \\
        \hline
         $(3,1^+)$&  $-13/9$&$[10^{\minus4},10]$&$[5\times10^4,1.1\times10^6]$ &\textcolor{Blue}{\text{\ding{52}}} &\textcolor{Blue}{\text{\ding{52}}}\\
        \cline{1-6}
         $(3,1^-)$& $-7/9$&$[10^{\minus 4},5]$&$[1.8\times10^3,3.9\times10^{6}]$ & \textcolor{Blue}{\text{\ding{52}}}&\textcolor{Red}{\text{\ding{56}}} \\
        \hline
    \end{tabular}
    \end{minipage}
    \hfill
\caption{Allowed values of $(n,k)$, $r$, $q_S$ and the range of the pseudoscalar Higgs mass $M_{\text{A}_\text{H}}$ that solve the flavor puzzle, fit neutrino data and solve strong CP with high quality axion to all orders in Model I within $\tan\beta\in [50,140]$.}\label{tab:ModelIqualitycases}
\end{table}
The list of viable values of $(n,k),r\text{ and} M_{\text{A}_\text{H}}$ that are possible in Model I are given in Table~\ref{tab:ModelIqualitycases}. This list is not exhaustive and additional possible cases are listed in Appendix \ref{subsec:pseudoaddI} where we consider additional contributions to the pseudoscalar Higgs mass from for specific cases similar to $(n,k)=(3,1^-)$. The additional contribution to $M_{\text{A}_\text{H}}$ in Eq.~(\ref{eq:Amassadd}) is reflected in the range quoted in Table~\ref{tab:ModelIqualitycases} for the case $(3,1^{-})$. In the current and upcoming models, the ranges in the Tables~\ref{tab:ModelIqualitycases}, \ref{tab:ModelIIqualitycases} and \ref{tab:ModelIIIqualitycases} are obtained with the following additional considerations. The couplings of the pseudoscalar Higgs mass can be chosen in a reasonable range of $\lambda_{\text{HS}}\in\{1/5,2\}$ and $\tan\beta$ in the allowed range for each specific model and sub case. Also, whenever axion can be the entire DM, $f_a$ is in the axion DM range discussed in Sec.~\ref{subsec:DMcons} and whenever it is unfeasible, the supremum of the lower bound of $f_a\geq 4.5 \times 10^{9}\,\text{GeV}$ from SN 1987A~\cite{DiLuzio:2020wdo,Raffelt:1990yz,DiVecchia:2019ejf,Caputo:2024oqc} and the model-dependent flavor violating lower bound from Sec.~\ref{subsec:axionSMFVcouplings} are considered as requirements for a successful model in our framework.
\begin{figure}[!htbp]
\centering
\includegraphics[width=1.0\textwidth]{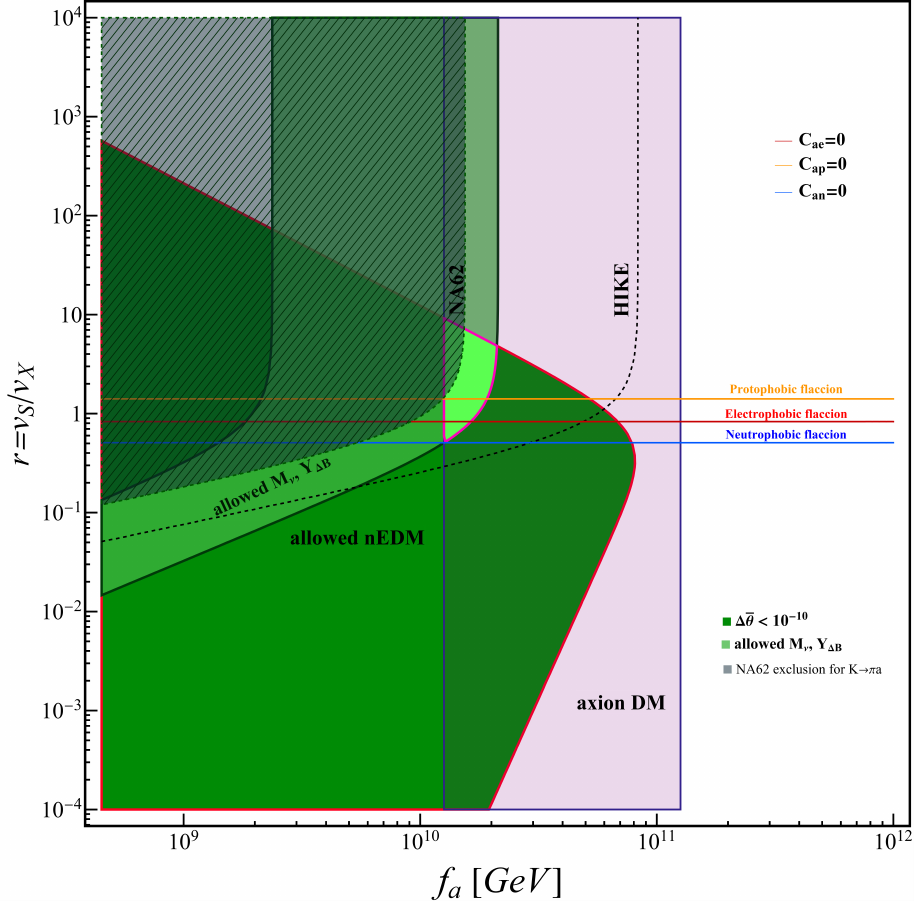}
\caption{Viable range of $(r,f_a)$ and high $\tan \beta\in [50,140]$ depicting the region for a high quality axion for the values of $(n,k)=(3,1^{+})$ in green. The lighter green band denotes the parameter space for allowed values of $\Lambda_{\text{FN}}$, that provide a good fit to neutrino oscillation data and satisfactory baryon asymmetry in  Eq.~(\ref{Eq:ModelILambdaFN}). The purple region denotes the allowed axion DM range from Sec.~\ref{subsec:DMcons}. The dark-green grayed and hatched out region denotes the restricted regions from current $K\rightarrow \pi a$ bounds from Sec.~\ref{subsec:FVcons} at NA62~\cite{NA62:2025upx}. The black dashed line denotes the future sensitivity of HIKE~\cite{HIKE:2023ext}. The red, yellow and blue lines mark the values of $r$ which exhibit electrophobia, protophobia and neutrophobia respectively illustrated in Sec.~\ref{subsec:FVcons}. The pink bordered brightest green region finally provides the allowed parameter space, where all possible features listed above are simultaneously feasible.}
\label{fig:ModelIQualityPlot}
\end{figure}
In Fig.~\ref{fig:ModelIQualityPlot},  we have allowed for the range 
\begin{equation}\label{Eq:ModelILambdaFN}
\Lambda_{\text{FN}}\in \left[\frac{1}{3},3\right]\times 4.2\times10^{11}\,\,\text{GeV}\end{equation} for Model I based on the neutrino and baryon asymmetry fit given in Sec. \ref{subsubsec:ModelINO}. This reasonable range of values for the FN scale is possible due to the inherent freedom allowed by the FN mechanism in choosing the effective Yukawas as a combination of the fundamental Yukawas at the flavor scale, as illustrated earlier in Eqs. (\ref{eq:ModelIfundyukud})-(\ref{eq:FNRHNfunYuk}). So, it is safe to say the above range lies within the constraints of our framework. The brightest green preferred region in Fig.~\ref{fig:ModelIQualityPlot} is the intersection of high quality axion (green), neutrino and baryon asymmetry fit (dark green) and axion DM (purple). In the next section, we will discuss a different model with other interesting features and a different UV completion.

\section{Model II: A case with a high scale flavor symmetry}\label{sec:ModelII}
Choosing a different type of UV completion to the earlier model results in enhancing the quality of the axion in some cases as we will see further. We use Vector-Like Fermions (VLFs) to fully UV complete this model at $\Lambda_{\text{FN}}$. This time, we have four RHNs instead of the three usual RHNs to cancel the gravitational and cubic anomalies in Sec.~\ref{subsec:Anomcancel}. We should also note that not all solutions give an accidental axion and that additional constraints were imposed to obtain rational fractional charges for $q_S$ to arrive at the anomaly free charge assignment in Table~\ref{tab:ModelIIcharges}. 
\begin{table}[!htbp]
{\footnotesize
$$
\begin{array}{|c||c|c|c|c|c|c|c|}

\hline
\vspace{-0.3cm}
 &  &  &  &  &  &  & \\

\vspace{-0.4cm}

{\rm Field}& \{Q_1, Q_2, Q_3\}&  \{u^c_1, u^c_2, u^c_3\} &
 \{d^c_1, d^c_2, d^c_3\}& \{L_1, L_2, L_3\} &  \{e^c_1, e^c_2, e^c_3\} & \{N_1, N_2, N_3,N_4\}&\{H_u, H_d, X, S \} \\

&  &  &  &  &  &  &\\

\hline

\vspace{-0.3cm}
 &  &  &  &  &  &  &\\

\vspace{-0.3cm}
\hspace{-0.5mm}Q_F\hspace{-0.5mm}&\hspace{-0.7mm} \{\minus \frac{2}{3},\frac{1}{3}, \frac{7}{3}\} \hspace{-0.7mm}& \hspace{-0.7mm} \{\minus \frac{22}{3},\minus\frac{13}{3}, \minus\frac{7}{3}\} \hspace{-0.7mm}  &\!\hspace{-0.7mm} \{2,4,4\} \hspace{-0.7mm}  &\{\minus\frac{4}{3},\minus\frac{7}{3}, \minus\frac{7}{3} \} & \{\frac{8}{3},\frac{20}{3}, \frac{26}{3}\}&\{  \minus 2,\minus 6,\minus 6,8\}& \{0, \minus\frac{19}{3},\, 1, \minus \frac{\pm3 k \minus 19}{3n}\}\\

&  &  &  &  &  &  &\\
\hline
\end{array}$$
\caption{Anomaly free flavor charge assignment for fermions and scalars $(H_u, H_d, X, S$) in Model II.}
\label{tab:ModelIIcharges}
}
\end{table}
The final textures of the fermion sector,
emerging from the charge assignment given in 
Table~\ref{tab:ModelIIcharges}, are:
\begin{equation}\label{eq:ModelII_Mass}
\begin{aligned}[b]
\mathcal{L}_{\text{Yuk}}\supset Q^TH_u&\begin{pmatrix}
\varepsilon^8 & \varepsilon^5 & \varepsilon^3 \\
\varepsilon^7 & \varepsilon^4 & \varepsilon^2 \\
\varepsilon^5 & \varepsilon^2 & 1
\end{pmatrix}u^c\:
+Q^TH_d
\begin{pmatrix}
\varepsilon^{5} & \varepsilon^{3} & \varepsilon^{3} \\
\varepsilon^{4} & \varepsilon^{2} & \varepsilon^{2} \\
\varepsilon^{2} & 1 & 1
\end{pmatrix}d^c+L^TH_d\begin{pmatrix}
\varepsilon^{5} & \varepsilon & \overline{\varepsilon} \\
\varepsilon^{6} & \varepsilon^{2} & 1 \\
\varepsilon^{6} & \varepsilon^{2} & 1
\end{pmatrix}e^c  \\
&\hspace{-0.7cm} + L^T\widetilde{H_d}\begin{pmatrix}
\overline{\varepsilon}^{3} & \varepsilon & \varepsilon & \overline{\varepsilon}^{13} \\
\overline{\varepsilon}^{2} & \varepsilon^{2} & \varepsilon^{2} & \overline{\varepsilon}^{12} \\
\overline{\varepsilon}^{2} & \varepsilon^{2} & \varepsilon^{2} &\overline{\varepsilon}^{12} 
\end{pmatrix}N+\Lambda_{\rm{FN}}~N\begin{pmatrix}
 \varepsilon^4 &\varepsilon^8 &\varepsilon^8 &\overline{\varepsilon}^{6} \\
\varepsilon^8 &\varepsilon^{12} & \varepsilon^{12}& \overline{\varepsilon}^{2} \\ 
\varepsilon^8 &\varepsilon^{12} &\varepsilon^{12}&  \overline{\varepsilon}^{2}\\
\overline{\varepsilon}^{6} & \overline{\varepsilon}^{2} & \overline{\varepsilon}^{2} & \overline{\varepsilon}^{16}
\end{pmatrix}N +\text{h.c}.
\end{aligned}
\end{equation}
The constraints for the anomaly cancellation search were to fix the $\epsilon$ suppression of diagonal entries of the up quark mass matrix; and the upper diagonal entries of the down quark mass matrix to get the right fit for the CKM mixing elements and the quark mass hierarchy.

In the charged lepton Yukawa matrix of  Eq.~(\ref{eq:ModelII_Mass}), in the $(1,2)$
entry we need to have a slightly suppressed effective Yukawa coupling 
$y_{12}^l<1/5$ (of Eq.~(\ref{eq:modelIIUV}))
in order to obtain a satisfactory $Y_e/Y_{\mu}$ ratio. 
With this arrangement from the charged fermion Yukawa matrices, we find:
\begin{equation}
\begin{aligned}[b]
    &Y_{t}\sim 1,\,\,\, \frac{Y_u}{Y_c}\sim \frac{Y_c}{Y_t}\sim \epsilon^4, \,\,\, \frac{Y_{e}}{Y_{\mu}}\sim\frac{Y_{d}}{Y_s}\sim \epsilon^3,\,\,\,\frac{Y_{\mu}}{Y_{\tau}}\sim\frac{Y_{s}}{Y_b}\sim \epsilon^2  \\
    &\hspace{2.0cm}|V_{us}|\sim \epsilon,\,\,\,\,\,\,\,|V_{cb}|\sim \epsilon^2,\,\,\,\,\,\,|V_{ub}|\sim \epsilon^3,\\
\end{aligned} 
\end{equation} 
which is the right fit.

\subsection{Model II: Neutrino phenomenology}\label{subsec:ModelIIneutrinopheno}

The light neutrino mass matrix $M_{\nu}$ can be induced by effectively utilizing only two RHNs $N_2,N_3$ to facilitate the see-saw mechanism where the renormalizable couplings to $N_1,N_4$ can be prevented by imposing a discrete global $\mathbb{Z}_{2}$ symmetry that takes $(N_1,N_4)\rightarrow-(N_1,N_4)$ simultaneously. This discrete symmetry is broken by quantum gravitational effects which plays a vital role in realizing the working neutrino mass texture.  The light neutrino mass matrix can be constructed if we choose the UV completion described in Sec.~\ref{subsec:ModelIIUVcompletion}, where the Dirac and Majorana mass matrices are as follows. Via the renormalizable FN sector we will have the couplings
\begin{equation}
\Lambda_{\rm FN}\bar \varepsilon^6N_1N_4 +\Lambda_{\rm FN}\varepsilon^{12}N_{2,3}N_{2,3}.
\label{eq:ModelIIRHNrenormalizable}
\end{equation}
On the other hand, the relevant totally Planck suppressed operators will be
\begin{equation}
\frac{\Lambda_{\rm FN}^2}{M_{\rm Pl}}\bar \varepsilon^2N_{2,3}N_4 .
\label{eq:ModelIIRHNPlancksuppressed}
\end{equation}
Besides this, the two neutral UV completion FN states $N^{(0)}_{1,2}$ will couple with $N_1$ via a partially Planck suppressed operator:
\begin{equation}
\frac{\Lambda_{\rm FN}^2}{M_{\rm Pl}}\varepsilon^2N_1 N^{(0)}_{1,2}.
\label{eq:ModelIIRHNmixed}
\end{equation}
With these mass entries, the whole $4\times 4$ RHN mass matrix will be
\begin{equation}
M_N=\Lambda_{\rm FN} \begin{pmatrix}
\left(\frac{(y^N_{12})^2}{y^N_{22}}\!+\!\frac{(y^N_{13})^2}{y^N_{33}}\right)\left(\frac{\Lambda_{\rm FN}}{M_{\rm Pl}}\right)^2\epsilon^4 & y^N_{12} \left(\frac{\Lambda_{\rm FN}}{M_{\rm Pl}}\right)\epsilon^8  & y^N_{13} \left(\frac{\Lambda_{\rm FN}}{M_{\rm Pl}}\right)\epsilon^8   & y^N_{14} \bar \epsilon^6 \\
 y^N_{12} \left( \!\frac{\Lambda_{\rm FN}}{M_{\rm Pl}}\!\right)\epsilon^8  & y^N_{22} \epsilon^{12} & 0 & y^N_{24} \left( \!\frac{\Lambda_{\rm FN}}{M_{\rm Pl}}\!\right)\bar \epsilon^2 \\
y^N_{13} \left( \!\frac{\Lambda_{\rm FN}}{M_{\rm Pl}}\!\right)\epsilon^8  & 0 & y^N_{33} \epsilon^{12} &  y^N_{34} \left( \!\frac{\Lambda_{\rm FN}}{M_{\rm Pl}}\!\right)\bar \epsilon^2 \\
 y^N_{14}\bar \epsilon^6 &  y^N_{24} \left( \!\frac{\Lambda_{\rm FN}}{M_{\rm Pl}}\!\right)\bar \epsilon^2& y^N_{34} \left( \!\frac{\Lambda_{\rm FN}}{M_{\rm Pl}}\!\right)\bar \epsilon^2 & 0
\end{pmatrix}
,
\label{eq:ModelIIMajoranamass}
\end{equation}
where we have chosen a basis where the middle $2\times 2$ block is diagonal. With this rotation of $N_2, N_3$ states, the structure
of the remaining terms and also the Dirac Yukawa matrix form will not be changed as $N_2,N_3$ possess the same flavor charges. Note that since we have two neutral FN states $N^{(0)}_{1,2}$, the determinant of the upper $3\times 3$ block is zero which leads to the decoupling of 2 RHNs, effectively allowing only the remaining 2 RHNs to participate in the seesaw.

Turning to the Dirac Yukawa couplings, we choose a basis in which the charged lepton mass matrix is diagonal. 2-3 and 1-3 rotations of the latter will not change the structure of the Dirac Yukawa. However, as noted above, we are selecting $y^l_{12}\lesssim 1/5$
and therefore the 1-2 rotation can be large. Denoting this 1-2 rotation angle by $\theta_e$,
in this basis, the neutrino Dirac Yukawa matrix can be parameterized as:
\begin{equation}
\hat Y_{\nu }\!=
\begin{pmatrix}
     \cos \te_e &e^{i\eta }\sin \theta_e & 0\\
     \minus e^{-i\eta }\sin \theta_e &\cos \theta_e & 0\\
     0 &0& 1\\
\end{pmatrix}      
\begin{pmatrix}
    \bar y_{11}^{\nu } \frac{\Lambda_{\rm FN}}{M_{\rm Pl}}\bar \epsilon^3   & \bar y_{12}^{\nu }\epsilon &  \bar y_{13}^{\nu } \epsilon &0\\
   \bar y_{21}^{\nu } \frac{\Lambda_{\rm FN}}{M_{\rm Pl}}\bar \epsilon^2     &  \bar y_{22}^{\nu } \epsilon^2 & \bar y_{23}^{\nu } \epsilon^2 &0 \\
   \bar y_{31}^{\nu } \frac{\Lambda_{\rm FN}}{M_{\rm Pl}}\bar \epsilon^2     &  \bar y_{32}^{\nu } \epsilon^2 & \bar y_{33}^{\nu } \epsilon^2  &0\\
\end{pmatrix}.
\label{eq:ModelIIDiracYuk}
\end{equation}
The Planck suppressed vertices above, given in Fig.~\ref{fig:modelIInuDUV} are induced by the breaking of the $\mathbb{Z}_2$ symmetry (by Planck suppressed operators) in the Dirac sector but it turns advantageous for the neutrino fit.
Here we present one selection which is reasonable:
\begin{equation}
\begin{aligned}[b]
& \epsilon=0.22,~~ v_d=4~{\rm GeV},~~\Lambda_{\rm FN} =3.8814 \times 10^{16}~{\rm GeV},\\
&\hspace{2.8cm}\theta_e =1.3352,~~~\eta =\pi ,\\ 
\end{aligned}\label{eq:ModelIIparams}
\end{equation}
and the $\mathcal{O}(1)$  dimensionless effective couplings 
$\bar y^{\nu }$  and $y^N$ are taken to be: 
\begin{equation}
\bar{y}_{ij}^{\nu }=\begin{pmatrix}
 2.07 & 0.2849e^{-0.078615i} & 0.2135e^{3.0494i} & 0 \\
 0.3530e^{-1.751i} & 0.4729e^{3.0948i} & 0.8147 & 0 \\
 0.33805e^{-2.2828i} & 1.856e^{-2.6804i} &0.3308 & 0
\end{pmatrix},
\label{Eq:ModelIIDiracYukfit}
\end{equation}
\begin{equation}
    y_{ij}^N=\begin{pmatrix}
       i & 1.6735e^{0.25\pi i} & 0.6247e^{0.25\pi i} & 2.5577i \\
1.6735e^{0.25\pi i} & 3.1118 & 0 &0.45876e^{-1.58665i} \\
 0.6247e^{0.25\pi i} & 0 & 3.9026& 0.9797\\
 2.5577i & 0.4588e^{-1.5867i} & 0.9797 & 0
 \end{pmatrix}.
 \label{Eq:ModelIIMajYukfit}
\end{equation}
Utilizing the see-saw mechanism
$M_{\nu}=-\frac{1}{2}v_d^2\hat Y_{\nu }M_N^{-1}\hat Y_{\nu }^T$, we get NO neutrinos with the following masses
\begin{equation}
\{m_1, m_2, m_3 \}=\{0, ~0.0086,~ 0.0507\}{\rm eV},
\label{eq:ModelII-neutrinomassfit}
\end{equation}
with the  bfv's of the oscillation parameters
\begin{equation}
\{\sin^2\te_{12},  \sin^2\te_{23}, \sin^2\te_{13}\}=\{0.3035, 0.455, 0.0224\},
\label{eq:ModelII-neutrinomixings}
\end{equation}
and the solar and atmospheric mass differences being
\begin{equation}
\Delta m_{\rm sol}^2=m_2^2-m_1^2=7.42\times 10^{-5}{\rm eV}^2~~,
\Delta m_{\rm atm}^2=m_3^2-m_2^2= 2.492\times 10^{-3}{\rm eV}^2.
\label{nu-mass2difNO-II}
\end{equation}
The masses of the heavy RHNs with the two lightest (coming from $N_{2,3}$ couplings) dominating the seesaw and the heavier two RHNs dominated by Planck suppressed operators are
\begin{equation}
\{M_1, M_2, M_3, M_4 \}=\{1.246\times10^{9},~2.072\times10^{9}, ~1.303\times 10^{13},~1.304\times 10^{13} \}~{\rm GeV}.
\label{eq:ModelII-Majoranamassfit}
\end{equation}
As we see from Eqs. (\ref{Eq:ModelIIDiracYukfit},~\ref{Eq:ModelIIMajYukfit}), the values of $|\bar y_{ij}^N|$ and $|\bar y_{ij}^{\nu }|$ are selected within the range $(0.2- 4)$ which gives very good fit to neutrino oscillation data and is consistent with our framework.

\subsection{Model II: Standard thermal leptogenesis}\label{subsec:ModelIIleptogenesis}

 Model II has a $4\times4$ texture for heavy RHNs, given in Eq.~(\ref{eq:ModelIIMajoranamass}), that is hierarchical and hence we can proceed with standard thermal leptogenesis with the final lepton asymmetry produced by the lighter of the lightest $N_2,N_3$ RHN
 states. It is readily apparent that the mass of the lightest RHN $M_1$ should satisfy the Davidson-Ibarra (DI) bound $M_1>10^{9} \, \text{GeV}$~\cite{Davidson:2002qv}. In this diagonal basis as before, we compute Eq.~(\ref{eq:VanillaCPasym}) where the Yukawa ratio of the CP asymmetry is generated solely from the second and third columns of the Dirac Yukawa from Eq.~(\ref{Eq:ModelIIDiracYukfit}) as these correspond to the lightest RHN mass states.  
 
We can have successful leptogenesis if the lightest RHN decays into the heavy Higgs $H\sim \widetilde{H_d}$ so that the suppression factor of $1/\tan^2\beta$ in Eq.~(\ref{eq:Baryonasym}) is absent. This enhancement factor will be present only if $M_{\text{A}_\text{H}}\sim M_H>M_1$, when it is kinematically forbidden for $N$ to decay to only the heavy higgs $H$. The DI bound, the lack of suppression from the kinematically forbidden decay  (we have $M_1>M_{\text{A}_\text{H}}$) and the neutrino fit in Sec.~\ref{subsec:ModelIIneutrinopheno} provides a conservative range for the flavor scale:
\begin{equation}\label{eq:modelIILambda_FN}
    \Lambda_{\text{FN}}\in [10^{16},10^{17}] \,\text{GeV}. 
\end{equation}
Here, given the above flavor scale, the $y_{ij}^{\nu}$ Dirac Yukawa coupling instance in Eq.~(\ref{Eq:ModelIIDiracYukfit}) and a reasonable value of $v_d\sim3-4$ GeV, we achieve the final baryon asymmetry from Eq.~(\ref{eq:Baryonasym}):
\begin{equation}\label{eq:ModelIIBasym}
    Y_{\Delta B}\simeq \frac{135\, \zeta(3)}{4 \pi^4 g_{*}}\frac{12}{37}\left(\epsilon_{\rm CP}^{(1)} \kappa_{f1}\right)\simeq8.71 \times 10^{-11},
\end{equation} 
to be a satisfactory value (up-to $\mathcal{O}(1)$ corrections).
\subsection{Model II: Quality of the axion}\label{subsubsec:ModelIIAxionquality}
As illustrated in Sec.~\ref{subsec:ModelIIUVcompletion}, the Yukawa Lagrangian for Model II can be UV completed by integrating out VLFs with masses at the flavor scale $\Lambda_{\rm FN}$. The effective charged fermion textures that provide the correct masses and mixings are 
\begin{equation} \label{eq:modelIIUV}
\begin{aligned}[b]
\mathcal{L}_{\text{yuk}}\supset Q^T H_u&\begin{pmatrix}
y^{u}_{11}\, \varepsilon^8 & y^{u}_{12}\,\varepsilon^5 & \textcolor{blue}{y^{u}_{13}\, \varepsilon^3} \\
y^{u}_{21}\, \varepsilon^7 & \textcolor{blue}{y^{u}_{22}\, \varepsilon^4} & \textcolor{blue}{y^{u}_{23}\,\varepsilon^2} \\
\textcolor{blue}{y^{u}_{31}\,\varepsilon^5} & \textcolor{blue}{y^{u}_{32}\,\varepsilon^2} & \textcolor{blue}{y^{u}_{33}}\,
\end{pmatrix}u^c+Q^T
H_d \begin{pmatrix}
y^{d}_{11}\,\varepsilon^5 & y^{d}_{12}\,\varepsilon^3 & \textcolor{blue}{y^{d}_{13}\,\varepsilon^3} \\
y^{d}_{21}\,\varepsilon^4 & \textcolor{blue}{y^{d}_{22}\,\varepsilon^2} & \textcolor{blue}{y^{d}_{23}\, \varepsilon^2} \\
\textcolor{blue}{y^{d}_{31}\,\varepsilon^{2}} & \textcolor{blue}{y^{d}_{32}}\, & \textcolor{blue}{y^{d}_{33}}\,
\end{pmatrix}d^c\\
&\hspace{1.cm}+L^T H_d\begin{pmatrix}
\textcolor{blue}{y^{l}_{11}\,\varepsilon^5} & y_{12}^l \varepsilon & y_{13}^l \overline{\varepsilon}  \\
y^{l}_{21}\,\varepsilon^6 &\textcolor{blue}{y^{l}_{22}\, \varepsilon^2} & \textcolor{blue}{y^{l}_{23}}\, \\
y^{l}_{31}\,\varepsilon^6 & \textcolor{blue}{y^{l}_{32}\,\varepsilon^2} & \textcolor{blue}{y^{l}_{33}}\,
\end{pmatrix}e^c+ \text{h.c}.\:\\
\end{aligned}
\end{equation}
where, the entries in blue are the decisive entries which determine the masses and mixings.\footnote{We additionally note that the rest of the non-highlighted couplings provide negligible contributions to the charged fermion masses and mixings and are anyways induced by the set of VLFs given in Sec.~\ref{subsec:ModelIIUVcompletion} needed to UV complete the necessary blue highlighted entries.} Additionally, the UV completion for the already illustrated Dirac and Majorana neutrino textures in Eqs. (\ref{eq:ModelIIDiracYuk})-(\ref{eq:ModelIIMajoranamass}) respectively is shown in Figs.~\ref{fig:modelIInuDUV} and \ref{fig:modelIInunuUV}.

Now, we analyze the quantum gravitational corrections to the axion potential arising from leading contributions from the SM and extended UV sector in Sec.~\ref{subsec:ModelIIUVcompletion}. We eliminate possible values of $(n,k)$ for which axion quality is lost. As a representative illustration for a VLF UV completion and in the normalization with $q_X=1$, we study several cases of $(n,k)$ in the following sections.
\subsubsection*{Case with \boldsymbol{$(n,k)=(3,1^{-})$}.}
 From Eq.~(\ref{eq:cons}), we have $q_S=22/9$ and the leading PQ-violating scalar term in this case is \begin{equation}V_{\cancel{\text{PQ}}}\supset\frac{S^9\left( X^*\right)^{22}}{M_{\rm Pl}^{27}}+\text{h.c},\end{equation} which is obviously harmless for the neutron EDM limits ( i.e., $\Delta\overline{\theta}\ll10^{-10}$). The leading contribution comes from two terms in the RHN sector and its UV completion (shown in Fig.~\ref{fig:modelIInunuUV}).  The resulting two-loop ``\emph{moth}" diagram in Fig.~\ref{fig:MII-loop-corrections} can be formed by utilizing the PQ conserving couplings $N_{2} N_{2,3}^{(\minus5)} X^{11}/\Lambda_{\rm FN}^{10}$ and $ N_{1}^{(3)}N_{1}^{(3)}(X^*)^6/\Lambda_{\rm FN}^5$ from Eq.~(\ref{eq:ModelIIMajoranamass}), and the PQ-violating couplings 
\begin{equation}
\label{eq:ModelII_RHN1axion_quality}
    \begin{aligned}[b]
        &\hspace{1.0cm}V_{\cancel{\text{PQ}}}\supset N_{2} N_{2} N_{2,3}^{(\minus5)} N_{2,3}^{(\minus 5)} \frac{S^9}{M_{\rm Pl}^{11}}+ N_{4} N_{4} N_{1}^{(3)} N_{1}^{(3)} \frac{(S^*)^9}{M_{\rm Pl}^{11}}+\text{h.c}
        \end{aligned}
\end{equation}
which induces the respective PQ violating operators
\begin{equation}
    \begin{aligned}\label{eq:ModelII_RHN1axion_quality1}
        & \hspace{1.0cm}V_{\cancel{\text{PQ}}}^{\rm ind}\supset\Xi_1 \frac{S^9(X^*)^{6}}{M_{\rm Pl}^{11} }\left(\frac{X^*}{\Lambda_{\rm FN}}\right)^{16}+\Xi_2 \frac{S^9(X^*)^{8}}{M_{\rm Pl}^{13}}\ \left(\frac{X^*}{\Lambda_{\rm FN}}\right)^{14}+\text{h.c},\\
        &\text{where,}~~\Xi_1\sim\left(\frac{1}{16\pi^2}\right)^2,~\Xi_2\sim\left(\frac{1}{16\pi^2}\right)^2 \left(1+ \epsilon^{24}\left(\frac{\Lambda_{\rm FN}}{M_4}\right)^2
    (y_{14}^Ny_{12}^{N*})^2y_{22}^N\right).
    \end{aligned}
\end{equation}
\begin{figure}[!htbp]
\centering
\includegraphics[width= 0.75\textwidth]{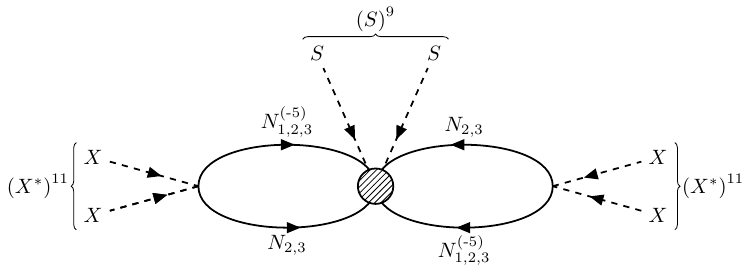}
\caption{Two-loop ``\emph{moth}" diagram generating the first effective PQ-violating operator in Eq.~(\ref{eq:ModelII_RHN1axion_quality1}) from the RHN sector where the shaded blob is the first PQ violating operator in Eq.~(\ref{eq:ModelII_RHN1axion_quality}).}
\label{fig:MII-loop-corrections} 
\end{figure}
 For the first operator with the loop specific factor $\Xi_1$, there are at most two light RHN states ($N_2$) in each loop in Fig.~\ref{fig:MII-loop-corrections} and the loop integral goes as $\int d^4k/(\cancel{k}+M_{1,2})^2$, which contributes no enhancement or suppression in the IR. As for the second operator and its loop factor $\Xi_2$ in Eq.~(\ref{eq:ModelII_RHN1axion_quality1}), the diagram is similar to Fig.~\ref{fig:MII-loop-corrections} with the exception of having the loop with $(N_4N_4)$ states be composed of a combination of a diamond 1-loop diagram produced by the operator $(\Lambda_{\rm FN}/M_{\rm Pl})^2N_4N_4(X^*)^{16}/\Lambda_{\rm FN}^{15}$ (see also Fig.~\ref{fig:modelIInunuUV} to form said operator) and a hexagon 1-loop diagram (like in Fig.~\ref{fig-n3k1-NN1loop}). The second loop of the second term in Eq.~(\ref{eq:ModelII_RHN1axion_quality1}) consists of all heavy RHN states connecting $N_1^{(3)}N_1^{(3)}$ contributing a trivial loop factor. We consider these two leading 1-loop operators as the hexagon 1-loop part has additional $\Lambda_{\rm FN}^2/M_4^2$ enhancement (second term in the expression for $\Xi_2$ in Eq.~(\ref{eq:ModelII_RHN1axion_quality1})) like in Eq.~(\ref{NN-1-loop}) and can be slightly leading in certain regions of the parameter space.\footnote{All higher polygon operators are well suppressed by appropriate powers of $\epsilon$ and Planck suppressed vertices. These terms are the higher terms of the Coleman-Weinberg (CW) potential contribution to the axion mass and hence are condensed into the logarithm of the CW potential justifying only the leading terms we have listed.}  We use again, Eqs.(\ref{eq:generalaxionquality})-(\ref{teta-ma}) with $(\mathfrak{a,b,c})=(9,6,16)$ for the first term and $(\mathfrak{a,b,c})=(9,8,14)$ for the second term in Eq.~(\ref{eq:ModelII_RHN1axion_quality1}) to arrive at a value which is perfectly safe for neutron EDM limits. Here, $f_a$ is also allowed in the axion DM range given in Eq.~(\ref{eq:axionDMcons}). 

The RHN sector has more freedom to form flavor invariants due to them being neutral under the SM gauge group and thus loop contributions involving the RHN sector are more dominant for axion quality. But we can also form flavor invariants using the charged SM right-handed fermions and their extended UV sectors which are more constrained by the SM gauge symmetry. We identify the operator with the lowest
power of $1/M_{\rm Pl}$, arising from loops involving FN fields to be from the extended up quark sector (see also Eq.~(\ref{eq:ModelIIuUVfields})). Thus, we present this as an illustration of sub-leading PQ violation from the charged fermion sector
\begin{equation}\label{eq:ModelII_UVaxion_quality}
     V_{\cancel{\text{PQ}}}\supset\frac{S^9}{M_{\rm Pl}^{14}} \left[ u_1^cF_u^{\left(\frac{1}{3}\right)}\right]^{2} \left[u_1^cF_u^{\left(\frac{\minus2}{3}  \right)}\right]+\text{h.c}. \implies V_{\cancel{\text{PQ}}}^{\rm{ind}}\supset\Xi\frac{S^{9}(X^*)^{9}}{M_{\rm Pl}^{14}}\left(\frac{X^*}{\Lambda_{\rm FN}}\right)^{13}+\text{h.c}.
\end{equation}
The corresponding diagram is the three-loop ``\emph{dragonfly}" diagram in Fig.~\ref{fig:MIIUV-loop-corrections} with a straightforward loop specific factor of $\Xi\sim(16\pi^2)^{-3}$, since all the internal states are heavy VLFs at the $\Lambda_{\rm FN}$ scale with at most one light state in each loop. Using  $(\mathfrak{a,b,c})=(9,9,13)$ in Eqs.~(\ref{eq:generalaxionquality})-(\ref{teta-ma}), we verify that the induced PQ-violation in Eq.~(\ref{eq:ModelII_UVaxion_quality}) is yet again ultra safe for the quality of the axion and is sub-leading to Eq.~(\ref{eq:ModelII_RHN1axion_quality1}) for this case.
\begin{figure}[!htbp]
\centering
\includegraphics[width= 1.0\textwidth]{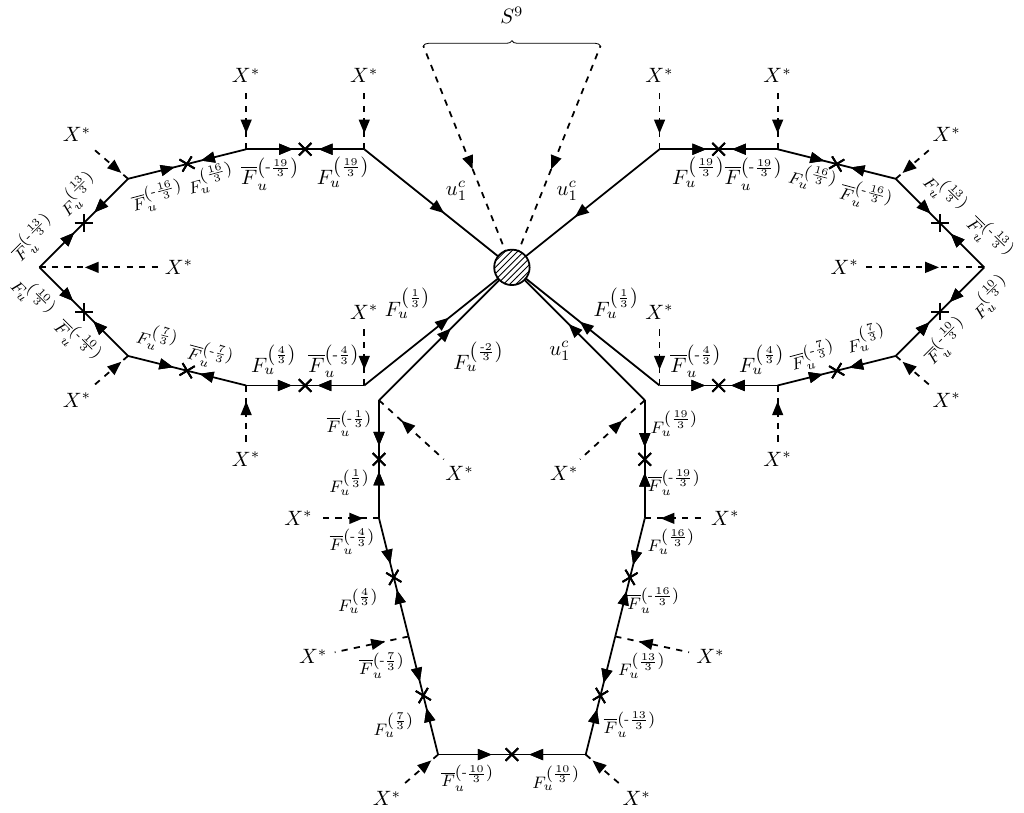}
\caption{Three-loop ``\emph{dragonfly}" diagram generating the PQ-violating operator in Eq.~ (\ref{eq:ModelII_UVaxion_quality}) through the VLF UV completion fields given in Sec.~\ref{subsec:ModelIIUVcompletion}.}
\label{fig:MIIUV-loop-corrections} 
\end{figure}
A way to see that further operators with smaller Planck suppression are not possible is to consider all possible Lorentz, SM and flavor invariants that can generally be written that violate PQ symmetry (involving the $S$ field). These should be of the form (with the appropriate Planck suppression) \begin{equation}\label{eq:ModelIIUVflavinvariants}
\left(\left[\overline{F}_{u,d,l}^{(q_1)}F_{u,d,l}^{(q_2)}\right]\left[\overline{F}_{u,d,l}^{(q_3)}F_{u,d,l}^{(q_4)}\right]\cdots\right) S^{3n}.
\end{equation} 
Observing the flavor charges of the SM right handed charged fermions in Table~\ref{tab:ModelIIcharges} and the UV fields in Eqs. (\ref{eq:ModelIIuUVfields})-(\ref{eq:ModelIIdUVfields})-(\ref{eq:ModelIIlUVfields}), we can conclude that such pairs of $\left[\overline{F}_{u,d,l}^{(q_1)}F_{u,d,l}^{(q_2)}\right]$ will always have $q_1+q_2\in\mathbb{Z}$. So, identifying the maximum and minimum possible value for the sum $q_1+q_2$ for each of the charged fermion sectors allows us to predict the minimum number of such pairs required to form a PQ violating operator of the form given in Eq.~(\ref{eq:ModelIIUVflavinvariants}).\footnote{Loops using invariants with $\overline{F}_uF_d$ are possible but occur with more than one PQ violating vertex. These are thus further highly Planck suppressed than the ones with invariants from the same sector like the one mentioned in Eq.~(\ref{eq:ModelII_UVaxion_quality}) and are negligible.} This informs us that there are no further leading operators from the charged fermion sector other than Eq.~(\ref{eq:ModelII_UVaxion_quality}) for the case of $(n,k)=(3,1^{-})$ since the minimum number of pairs in this case is three (refer to the numerator of $q_S=22/9$, with $q_1+q_2\in\{-8,-7,\cdots,7\}$ from Eq.~(\ref{eq:ModelIIuUVfields}) making $22+(-8-7-7)=0$) which results in a three loop diagram given in Fig.~\ref{fig:MIIUV-loop-corrections}. Similar statements about the operators from the up, down and charged lepton UV sectors in Eqs. (\ref{eq:ModelIIuUVfields})-(\ref{eq:ModelIIdUVfields})-(\ref{eq:ModelIIlUVfields}) respectively can be made for the rest of the below cases. It can be safely concluded that their contribution will be sub-leading to the contributions from the RHN UV sector for each case.

Noting from the previous illustrations that the tree level PQ-violating operators are obviously safe for the axion quality, we now provide the leading PQ-violating operators which usually arise at higher loop levels.
\subsubsection*{Case with \boldsymbol{$(n,k)=(2,2^-)$}.}
In this case, from Eq.~(\ref{eq:cons}), we have $q_S=25/6$. Using the same logic as the earlier case, the leading PQ-violating operators arise from the following PQ-violating couplings 
 \begin{equation}
 \begin{aligned}[b]
&V_{\cancel{\text{PQ}}}\supset N_{2} N_{2} N_1^{(\minus6)} N_1^{(\minus7)} \frac{S^6}{M_{\text{ Pl}}^{8}} + N_{4} N_{4} N_{1}^{(4)} N_{1}^{(5)} \frac{(S^*)^6}{M_{\rm Pl}^{8}}+\text{h.c}\\
&\hspace{1.25cm}\implies V^{\text{ind}}_{\cancel{\text{PQ}}}\supset \Xi\frac{S^6 (X^*)^{6}}{M_{\text{Pl}}^8}\left(\frac{X^*}{\Lambda_{\text{FN}}}\right)^{19}+ 
\text{h.c}.
\end{aligned}
\end{equation}
The corresponding two-loop moth diagram is similar to Fig.~\ref{fig:MII-loop-corrections} with a similar loop factor $~\Xi=\Xi_1+\Xi_2$ as explained earlier for such $(N_2N_2)$ and $(N_4N_4)$ loops above and under Eq.~(\ref{eq:ModelII_RHN1axion_quality}) respectively.~\footnote{We note that, $N_1^{(q)},N_{2,3}^{(q)}$ states mix with a Planck suppression factor (see Eqs.~(\ref{eq:ModelIIRHNPlancksuppressed})-(\ref{eq:ModelIIRHNmixed})). So, we use states of the same subscript to avoid further Planck suppression and arrive at the leading operator.} For this case, the values of 
$(\mathfrak{a},\mathfrak{b},\mathfrak{c})=(6,6,19)$ can be substituted into Eqs.(\ref{eq:generalaxionquality})-(\ref{teta-ma}) to confirm that $\Delta\overline{\theta}\ll10^{-10}$.
    
\subsubsection*{Case with \boldsymbol{$(n,k)=(3,1^+)$}.}
In this case from Eq.~(\ref{eq:cons}), we have $q_S=16/9$. Again the leading PQ-violating operator arises from 
\begin{equation}
\begin{aligned}[b]  
&V_{\cancel{\text{PQ}}}\supset N_4N_4 \frac{(S^*)^9}{M_{\text{ Pl}}^{8}}+\text{h.c}\implies V^{\text{ind}}_{\cancel{\text{PQ}}}\supset (\Xi_1+\Xi_2) \frac{S^9 (X^*)^5}{M_{\rm Pl}^{10}}
    \left( \frac{X^*}{\La_{\rm FN}}\right)^{11}+\text{h.c},\\
    &\hspace{01.5cm}\text{where,}~~~\Xi_1\sim \frac{1}{16\pi^2}
    ,~~~~\Xi_2\sim \frac{\epsilon^{24}}{16\pi^2} \left(\frac{\Lambda_{\rm FN}}{M_4}\right)^2
    (y_{14}^Ny_{12}^{N*})^2y_{22}^N
\end{aligned}
\end{equation}
through a one-loop diagram consisting of $(N_4N_4)$ states whose loop factor $\Xi=(\Xi_1+\Xi_2)$ was explained under Eq.~(\ref{eq:ModelII_RHN1axion_quality}). For this case, the values of 
$(\mathfrak{a},\mathfrak{b},\mathfrak{c})=(9,5,11)$ (the reason for this selection is similar to the one explained after 
Eq.~(\ref{PQvln3k-1-2loop})) can be substituted into Eqs.~(\ref{eq:generalaxionquality})-(\ref{teta-ma}) to see that $\Delta\overline{\theta}$ is well within the neutron EDM limits. Using the above facts, we list the viable cases that satisfy all requirements of Model II in Table~\ref{tab:ModelIIqualitycases}.

\subsection{Model II: Summary of results}
Once again, following the procedure illustrated in Sec.~\ref{subsec:ModelIaxionquality}, we choose appropriate $\mathcal{O}(1)$ fundamental Yukawa couplings to achieve the desired fermion mass hierarchies in Eqs.~(\ref{Eq:ModelIIDiracYukfit})-(\ref{Eq:ModelIIMajYukfit})-(\ref{eq:modelIIUV}).
Finally, based on the above fits and the range of values in Eqs.~(\ref{eq:ModelIIparams})-(\ref{eq:modelIILambda_FN}), we can have successful leptogenesis, high quality axion and axion as DM simultaneously for the cases with $(n,k)=\{(2,2^-),(3,1^{\pm})\}$, as they satisfy all the requirements posited above for the allowed parameter space provided in Table~\ref{tab:ModelIIqualitycases}.
\begin{table}[!htbp]
    \renewcommand{\arraystretch}{1.4} 
    \begin{minipage}{0.5\textwidth}
    \begin{tabular}{|c|c|c|c|c|c|}
        \hline
        $(n,k)$ & $q_S$ &$r$&$M_{\text{A}_\text{H}}\,[\text{GeV}]$ &\text{High Quality} &\text{Axion DM} \\
        \hline
         $(3,1^+)$ & $16/9$&$ [5\times10^{\minus6}, 6 \times 10^{\minus 3} ]$&$[6.7 \times 10^{5}, 5.2 \times 10^7]$ &\textcolor{Blue}{\text{\ding{52}}} &\textcolor{Blue}{\text{\ding{52}}} \\
        \cline{1-6}
         $(2,2^{-})$ & $25/6$&$[8 \times 10^{\minus6},10^{\minus 4}]$& $[7.9 \times 10^{7},6.1 \times 10^{8}]$&\textcolor{Blue}{\text{\ding{52}}} &\textcolor{Blue}{\text{\ding{52}}}\\
        \cline{1-6}
        $(3,1^{-})$ & $22/9$ &$[5\times10^{\minus6}, 6 \times 10^{\minus 3} ]$&$[6.7 \times 10^{5}, 5.2 \times 10^7]$ & \textcolor{Blue}{\text{\ding{52}}}&\textcolor{Blue}{\text{\ding{52}}} \\
        \hline
    \end{tabular}
    \end{minipage}
    \hfill
\caption{Allowed values of $(n,k)$, $r$, $q_S$ and the range of the pseudoscalar Higgs mass $M_{\text{A}_\text{H}}$ that solve the flavor puzzle, fit neutrino data, leptogenesis and solve strong CP with high quality axion to all orders in Model II within $\tan\beta\in [50,140]$.}\label{tab:ModelIIqualitycases}
\end{table}
Graphically, the same information is illustrated and all the relevant parameter spaces where the model exhibits different features is presented case by case for each value of $(n,k)$ in Fig.~\ref{fig:ModelIIQualityPlot} (see also the caption). The brightest colored regions in Fig.~\ref{fig:ModelIIQualityPlot} for $(2,2^-)$ (teal) and the overlapping regions for $(3,1^+)$ (light blue) and $(3,1^-)$ (dark blue) are the intersections of the preferred parameter space $(r,f_a)$ of the high quality axion, neutrino fit and baryon asymmetry and axion DM for each of the cases tabulated in Table~\ref{tab:ModelIIqualitycases}. Additionally, the higher value of $\Lambda_{\rm FN}$ in Model II further opens up new possible cases and are elaborated upon further in Appendix~\ref{subsec:pseudoaddII}.
\begin{figure}[!htbp]
\centering
\hspace{-1.0cm}\includegraphics[width=1.0\textwidth]{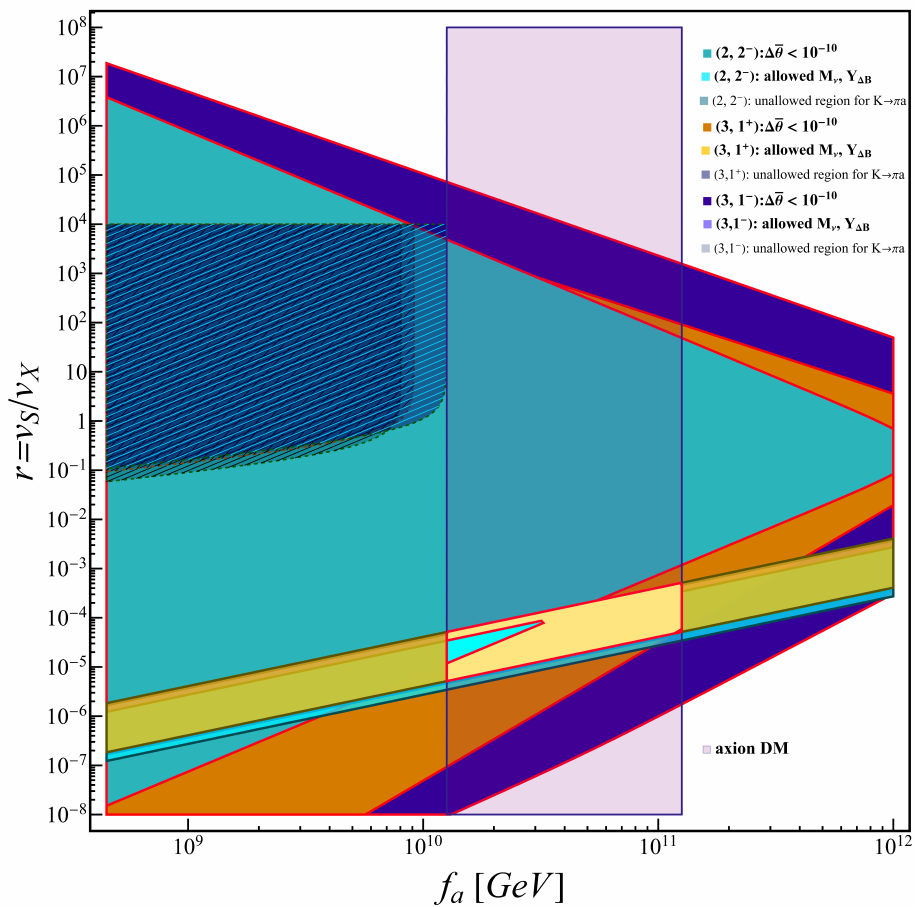}
\caption{Viable range of $(r,f_a)$ and high $\tan \beta\in [50,140]$ depicting the region for a high quality axion for the values of $(n,k)=(2,2^-),(3,1^{\pm})$ in teal, orange and purple respectively. \textcolor{blue}{The overlaid colors depict their parameter spaces to be subspaces of the underlaid colored spaces. The respective lighter straight bands denote the parameter space for allowed values of $\Lambda_{\text{FN}}$ that provide a good fit to neutrino oscillation data and satisfactory baryon asymmetry from Eqs.~(\ref{eq:ModelIIparams})-(\ref{eq:modelIILambda_FN}) for each model. The purple region shows the allowed axion DM range from Sec.~\ref{subsec:DMcons} for all models.} The hatched regions denote the restricted regions from $K\rightarrow \pi a$ bounds from Sec.~\ref{subsec:FVcons}. The red bordered brightest teal, blue and dark blue (overlaps with the brightest blue) regions finally provide the allowed parameter space, where all possible features listed above are simultaneously feasible for each case of $(n,k)$ listed in the legends.}
\label{fig:ModelIIQualityPlot}
\end{figure} 
\section{Model III: \boldmath{$SU(5)$} GUT compatible flavor model}\label{sec:ModelIII}
We now present our third UV-complete model with a flavor charge assignment that is $SU(5)$ Grand Unified Theory (GUT) compatible and anomaly-free which also leads to realistic fermion masses and mixings. The charges of fermions and scalar fields of the model are listed in Table~\ref{tab:ModelIIIcharges}.  It can be verified that all the anomaly cancellation conditions of Eq.~(\ref{eq:anomalies}) are satisfied by this choice of flavor charges. 

\begin{table}[!htbp]
{\footnotesize
$$
\begin{array}{|c||c|c|c|c|c|c|c|}
\hline
\vspace{-0.3cm}
 &  &  &  &  &  &  & \\

\vspace{-0.4cm}

{\rm Field} & \{Q_1, Q_2, Q_3\}&  \{u^c_1, u^c_2, u^c_3\}&
  \{d_1^c, d_2^c, d_3^c\} &  \{L_1, L_2, L_3\}&\{e^c_1, e^c_2, e^c_3\} & \{N_1, N_2, N_3\} & \{H_u, H_d, X,S \}\\

&  &  &  &  &  & & \\

\hline

\vspace{-0.3cm}
 &  &  &  &  &  &  & \\

\vspace{-0.3cm}
\hspace{-0.5mm}{\rm Charge}\hspace{-0.5mm}&\hspace{-0.7mm} \{\frac{5}{3}, \minus\frac{1}{3}, \minus \frac{7}{3}\} \hspace{-0.7mm}\!& \hspace{-0.7mm} \{\frac{5}{3}, \minus\frac{1}{3}, \minus\frac{7}{3}\} \hspace{-0.7mm} &\{\frac{5}{3}, \frac{2}{3}, \frac{2}{3}\}  &\{\frac{5}{3}, \frac{2}{3}, \frac{2}{3}\}& \{\frac{5}{3}, \minus\frac{1}{3}, \minus\frac{7}{3}\} &\{\minus 6, \minus 9,\, 10\} & \{\frac{14}{3}, \frac{14}{3},\, 1, \minus \frac{\pm 3 k + 28}{3n}\}\\
&  &  &  &  &  & & \\
\hline

\end{array}$$
\caption{$SU(5)$-compatible anomaly free $U(1)_F$ charge assignment for fermions and scalars $(H_u, H_d,X,S)$ in Model III.}
\label{tab:ModelIIIcharges} 
}
\end{table}
The resulting effective Yukawa Lagrangian for Model III takes the form
\begin{equation}
\begin{aligned}[b]    
& \mathcal{L}_{\text{Yuk}}\supset Q^TH_u \left( \begin{array}{ccc}
        {\bar{\varepsilon}}^{8}  &  {\bar{\varepsilon}}^{6}  & {\bar{\varepsilon}}^{4}  \\
       {\bar{\varepsilon}}^{6} & {\bar{\varepsilon}}^{4}  & {\bar{\varepsilon}}^{2}  \\
       {\bar{\varepsilon}}^{4} & {\bar{\varepsilon}}^{2}  &1
      \end{array}
\right)  u^c+
Q^T H_d\left( \begin{array}{ccc}
        \bar{\varepsilon}^8 &  {\bar{\varepsilon}}^{7}&  {\bar{\varepsilon}}^{7} \\
       \bar{\varepsilon}^6 &  {\bar{\varepsilon}}^{5} &  {\bar{\varepsilon}}^{5} \\
       \bar{\varepsilon}^4 & \bar{\varepsilon}^3 & \bar{\varepsilon}^3
      \end{array}
\right)   d^c+
L^T H_d\left( \begin{array}{ccc}
        \bar{\varepsilon}^8 & \bar{\varepsilon}^6 & {\bar{\varepsilon}^4} \\
       \bar{\varepsilon}^7 &{\bar{\varepsilon}}^{5}& {\bar{\varepsilon}}^{3}\\
       \bar{\varepsilon}^7 &\bar{\varepsilon}^5 &\bar{\varepsilon}^3
      \end{array}
\right) e^c\\
& \hspace{2.0cm}+ 
L^T\tilde H_d \left( \begin{array}{ccc}
        \varepsilon^{9}  & \varepsilon^{12}   & \bar{\varepsilon}^7 \\
      \varepsilon^{10}  & \varepsilon^{13} & \bar{\varepsilon}^6\\
      \varepsilon^{10}  & \varepsilon^{13} & \bar{\varepsilon}^6 \\
      \end{array}
\right) N + 
\Lambda_{\rm FN} \, N^{T} \left( \begin{array}{ccc}
        \varepsilon^{12} & \varepsilon^{15} & \bar{\varepsilon}^4\\ 
        \varepsilon^{15} & \varepsilon^{18} &  \bar{\varepsilon}  \\
  \bar{\varepsilon}^4 &  \bar{\varepsilon} & \bar{\varepsilon}^{20}
      \end{array}
\right) N + \text{h.c}.
\label{eq:ModelIIImass}
\end{aligned}
\end{equation}
 ``Lopsided" mass matrices of this type  for quarks and charged leptons were proposed to realize large leptonic mixing and small quark mixing in unified theories in Ref.~\cite{Babu:1995hr,Sato:1997hv,Elwood:1998kf}. Such textures were also analyzed in the context of anomalous $U(1)$ flavor symmetry in Ref.~\cite{Babu:2003zz,Babu:2004th}. Here, the texture is realized from a gauged $U(1)_F$ symmetry. As before, $\mathcal{O}(1)$ coefficients multiplying each entry in the mass matrices of Eq.~(\ref{eq:ModelIIImass}) are not shown but should be understood.

The mass matrices of Eq.~(\ref{eq:ModelIIImass}) can explain in a natural way the various mass hierarchies observed in the fermion spectrum. Specifically, they lead to the following features:
\begin{eqnarray}\label{eq:ModelIIImassratios}
&~&m_t:m_c:m_u \sim 1: \epsilon^4: \epsilon^8,~~~m_b:m_s:m_d \sim m_\tau: m_\mu:m_e \sim 1:\epsilon^2:\epsilon^5,~~~m_b/m_t\sim \epsilon^3,\nonumber \\
&~&\hspace*{1.5cm} V_{cb} \sim \epsilon^2, ~~~V_{us} \sim \epsilon^2,~~~V_{ub} \sim \epsilon^4,~~~\theta_{23}^\ell \sim 1,~~~\theta_{12}^\ell \sim \epsilon,~~~\theta_{13}^\ell \sim \epsilon~.
\label{su5-hier}
\end{eqnarray}
Note the suppressed values of $\tau$ lepton and $b$-quark
Yukawa couplings, suggesting the low values of the $\tan \beta$.
In Eq.~(\ref{su5-hier}),  $\theta_{ij}^\ell$ represent the charged lepton contributions to the leptonic mixing angles, which should be combined with the contributions from the light neutrino sector, which will be discussed below. All of these features are in reasonably good agreement with observations, when $\mathcal{O}(1)$ coefficients are included.

\subsection{Model III: Neutrino phenomenology}

We shall assume that the Yukawa coupling $N_2 N_3 X^*$ (the (2, 3) entry of the Majorana mass matrix in Eq.~(\ref{eq:ModelIIImass}) is absent due to a discrete symmetry such as a $\mathbb{Z}_2$ under which only $N_3$ is odd.  Such discrete symmetries are however, broken by quantum gravity, and thus we expect terms in the Lagrangian of the type
\begin{eqnarray}
{\cal L}_{\rm Yuk} \supset N_2 N_3 X^* \left(c_1 \frac{|X|^2}{M_{\rm Pl}^2} + c_2 \frac{|S|^2} {M_{\rm Pl}^2}\right) + \text{h.c}.
\label{eq:ModelIIIMajorananeutrinomass}
\end{eqnarray}
The (2, 3) entry of the Majorana mass matrix then becomes of order
\begin{equation}
(M_N)_{23} = \Lambda_{\rm FN} \epsilon \left(c_1 \frac{v_X^2}{M_{\rm Pl}^2} + c_2 \frac{v_S^2} {M_{\rm Pl}^2}\right) \simeq  y_{23}^N \,\epsilon^3 \left(\frac{\Lambda_{\rm FN}}{M_{\rm Pl}} \right)^2,
\end{equation}
where in the second half we assumed $v_X \gg v_S$, which is suggested by a high quality axion, as shown later in this section.

In the explicit UV completion of the Dirac neutrino sector we shall not employ the $N_1$ state. Furthermore, in the Majorana neutrino mass matrix the UV completion generates only the (1, 3) and (1, 1) elements, with the (1, 3) element arising partially from Planck-suppressed operators as it decouples from the $N_2,N_3$ states in the seesaw mechanism.  Thus, the Dirac\footnote{The benefit of choosing this texture with the first column zeroes is twofold, the UV completion of the Dirac sector is far simpler, requiring lesser fields and $y_{11}^{\nu}=0$ prohibits one of the light neutrinos from being too heavy (of $\mathcal{O}(\epsilon^6)$).} and the Majorana neutrino mass matrices take the form
\begin{eqnarray}\label{eq:ModelIIINmassmatrices}
M_{\nu^D} = 
\begin{pmatrix} 0~~ & y^{\nu}_{12}\,\, \epsilon^{12}~~ & 0 \cr 0~~ & y^{\nu}_{22}\,\,\epsilon^{13}~~ & y^{\nu}_{23}\,\,\overline{\epsilon}^{6} \cr 
0~~ & y^{\nu}_{32}\,\,\epsilon^{13}~~ & y^{\nu}_{33}\,\,\overline{\epsilon}^{6}   \end{pmatrix}  v_d~,~ M_N = \Lambda_{\rm FN} \left(\begin{matrix}   y^N_{11} \epsilon^{12}  & 0 & y_{13}^N \epsilon^4 \left(\frac{\Lambda_{\rm FN}}{M_{\rm Pl}}  \right)^2 \cr
0 & 0 & ~y^N_{23} \epsilon^3 \left(\frac{\Lambda_{\rm FN}}{M_{\rm Pl}}  \right)^2 \cr
 y_{13}^N \epsilon^4 \left(\frac{\Lambda_{\rm FN}}{M_{\rm Pl}}  \right)^2 &y^N_{23} \epsilon^3 \left(\frac{\Lambda_{\rm FN}}{M_{\rm Pl}}  \right)^2 & 0 \end{matrix} \right).\nonumber \\
\end{eqnarray}
The light neutrino mass matrix in this case will have the form (to leading order in $\epsilon$)
\begin{eqnarray}\label{eq:ModelIIInumass}
\hspace{-0.1cm}M_\nu^{\rm light} \sim  \left(\frac{\epsilon^{15}v_d^2}{y_{23}^N\Lambda_{\rm FN}} \right) \left(\frac{M_{\rm Pl}}{\Lambda_{\rm FN}}\right)^2 \left( \begin{matrix}  2 ~ y_{12}^{\nu} y_{13}^{\nu}~\epsilon & y_{12}^{\nu} y_{23}^{\nu} & y_{12}^{\nu} y_{33}^{\nu} \cr y_{12}^{\nu} y_{23}^{\nu} & 2~y_{22}^{\nu} y_{23}^{\nu}~\epsilon &  (y_{23}^{\nu} y_{32}^{\nu}+y_{22}^{\nu} y_{33}^{\nu})~\epsilon \cr y_{12}^{\nu} y_{33}^{\nu} & (y_{23}^{\nu} y_{32}^{\nu}+y_{22}^{\nu} y_{33}^{\nu}) ~\epsilon  &  2~y_{32}^{\nu} y_{33}^{\nu}~\epsilon \end{matrix} \right),
\end{eqnarray}
with $v_d \sim 10^2\,\text{GeV}$ (corresponding to $\tan\beta \sim 1$). This texture would result in the light neutrino masses consistent with $m_0\sim\left(\frac{\epsilon^{15}v_d^2}{y_{23}^N\Lambda_{\rm FN}} \right) \left(\frac{M_{\rm Pl}}{\Lambda_{\rm FN}}\right)^2\sim 0.05$~eV and is compatible with an IO mass spectrum. Analytical expressions for the solar $\Delta m_{\rm sol}^2$ and atmospheric $\Delta m_{\rm atm}^2$ mass differences and their ratios $R=\Delta m_{\rm sol}^2/\Delta m_{\rm atm}^2$ can be obtained by diagonalizing Eq.~(\ref{eq:ModelIIInumass}) perturbatively in $\epsilon$ as
 \begin{equation}\label{eq:ModelIIInumassdiffs}
 \begin{aligned}[b]
     \Delta m_{\rm sol}^2&=m_2^2-m_1^2=4~ m_0^2~\epsilon ~|y_{12}^{\nu}||(y_{12}^{\nu})^*y_{13}^{\nu}+(y_{22}^{\nu})^*y_{23}^{\nu}+(y_{12}^{\nu})^*y_{13}^{\nu}|\sqrt{|y_{23}^{\nu}|^2+|y_{33}^{\nu}|^2},\\
     \Delta m_{\rm atm}^2&=m_3^2-m_2^2=-m_0^2~\epsilon ~|y_{12}^{\nu}|^2 (|y_{23}^{\nu}|^2+|y_{33}^{\nu}|^2),\\
    R&=\frac{\Delta m_{\rm sol}^2}{\Delta m_{\rm atm}^2}=-\frac{4~ \epsilon ~|(y_{12}^{\nu})^*y_{13}^{\nu}+(y_{22}^{\nu})^*y_{23}^{\nu}+(y_{12}^{\nu})^*y_{13}^{\nu}|}{ |y_{12}^{\nu}| \sqrt{|y_{23}^{\nu}|^2+|y_{33}^{\nu}|^2}}=4 ~\epsilon ~|X|\sim\frac{1}{33.7}.
 \end{aligned}   
 \end{equation}
Here we have defined the parameter $X$ as 
\begin{equation}\label{eq:X}
    X=(y_{12}^{\nu})^*y_{13}^{\nu}+(y_{22}^{\nu})^*y_{23}^{\nu}+(y_{12}^{\nu})^*y_{13}^{\nu}=\frac{-R}{4\epsilon} |y_{12}^{\nu}| \sqrt{|y_{23}^{\nu}|^2+|y_{33}^{\nu}|^2},
\end{equation}} 
which appears as a common factor in the mass difference ratio and CP asymmetry for leptogenesis. Thus $X$ is constrained from both requirements, as we will encounter below. Using this, we can obtain a general expression for the flavor scale $\Lambda_{\rm FN}$ in this model to be
\begin{equation}\label{eq:ModelIIIlambdaFN}
    \Lambda_{\rm FN} = (5.8\times10^{13}~\text{GeV}) \left(\frac{\epsilon}{0.22}\right)^{\frac{15}{3}}\left(\frac{v_d}{100 ~\text{GeV}}\right)^{\frac{2}{3}}\left(\frac{M_{\rm Pl}}{1.22 \times 10^{19} ~\text{GeV}}\right)^{\frac{2}{3}}\frac{4~\epsilon~|X|}{|y_{23}^N|R},
\end{equation}
from which we estimate the RHN masses as $M_{1}\sim \epsilon^{12}\Lambda_{\rm FN},~ M_{2,3}\sim y_{23}^N~\epsilon~ \Lambda_{\rm FN}^3/M_{\rm Pl}^2$. For $\epsilon\sim 0.3$ and $y_{23}^N\sim 1$, we have $M_{2,3}\sim76$ TeV. From the discussions around Eqs. (\ref{eq:ModelIfundmassN})-(\ref{eq:ModelILeffnu}), the fundamental Yukawas can be chosen such that $y_{23}^N\sim 0.3$ for an enhancement of $M_{2,3}$ to even $760$ TeV. This demonstrates again, the freedom in the choice of $\mathcal{O}(1)$ fundamental couplings and their consistency within our theoretical framework.

\subsection{Model III: Resonant leptogenesis}
Leptogenesis in Model III proceeds again via a resonant scenario owing to two naturally quasi-degenerate states composed of $(N_2,N_3)$ whereas $N_1$ decouples from the seesaw mechanism. The decay widths of $N_i \rightarrow L + H_d^*$ and $N_i \rightarrow \overline{L} + H_d$ decays will be affected by the value of $\tan\beta$ in this model, which is of order one, see the charged fermion mass spectrum in Eq.~(\ref{eq:ModelIIImassratios}). Since $H_d=\cos{\beta}~ h-\sin{\beta}~ H$, where $h$ is the lighter Higgs, all couplings to the light Higgs, $(\hat{Y}_{\nu})_{ij}$, will be scaled by $\cos{\beta}~(\hat{Y}_{\nu})_{ij}$. The unitary matrix that diagonalizes the Majorana mass matrix $M_N$ in Eq.~(\ref{eq:ModelIIINmassmatrices}) is
\begin{eqnarray}
U_N \simeq \left(\begin{matrix}1 & 0 & 0
 \cr0 & \frac{1}{\sqrt{2}} & -\frac{i}{\sqrt{2}}
 \cr0 & \frac{1}{\sqrt{2}}  & -\frac{i}{\sqrt{2}} \end{matrix}\right)~.
\end{eqnarray}
The mass splitting of the quasi-degenerate states, $\hat{M}_3^2-\hat{M}_2^2$, which arises from the small mixing  generated by the seesaw of the 1-3 block involving the $N_1,N_3$ states, and the decay width of $N_2$  can be calculated to be
\begin{eqnarray}
\hat{M}_3^2-\hat{M}_2^2 = \frac{(y_{13}^N)^2}{\epsilon^4 ~y_{11}^N}\frac{\Lambda_{\rm FN}^5}{M_{\rm Pl}^4}  ,~~\Gamma_2 = \frac{\epsilon^{13}\cos^2{\beta}~ \Lambda_{\rm FN}^3}{8 \pi M_{\rm Pl}^2} y_{23}^N(|y_{23}^{\nu}|^2+|y_{33}^{\nu}|^2).
\end{eqnarray}
The kinematic factor in Eq.~(\ref{eq:kinematicfactorres}) tends to $1/2$ for specific values of the Yukawa couplings which will be taken into account for the final baryon asymmetry fit. Using the relation for $R$ in Eq.~(\ref{eq:ModelIIInumassdiffs}) and expressing $X=|X|e^{i\phi}$, the total CP asymmetry in this model from Eq.~(\ref{eq:resonantCPassym}) is thus given by
\begin{equation}\label{eq:ModelIIICpassym}
    \varepsilon_{\rm CP} = \sum_{i=1}^2 \varepsilon_{\text{CP}}^{(i)}   \simeq
\frac{8~\epsilon^7~\text{Im}((y_{12}^{\nu})^*y_{13}^{\nu}+(y_{22}^{\nu})^*y_{23}^{\nu}+(y_{12}^{\nu})^*y_{13}^{\nu})}{|y_{23}^{\nu}|^2+|y_{33}^{\nu}|^2}\sin{2\phi_k}= \frac{2 R \sin{\phi}\sin{2\phi_k}~\epsilon^6|y_{12}^{\nu}|}{\sqrt{|y_{23}^{\nu}|^2+|y_{33}^{\nu}|^2}} .
\end{equation}
The dilution factor from Eq.~(\ref{eq:efffactor}) can be estimated to be
\begin{equation}\label{eq:ModelIIIefffactor}
    \kappa_{f(2,3)}\simeq\left(\frac{0.55\times10^{-3}}{0.05}\frac{|y_{12}^{\nu}|\epsilon^6}{\sqrt{|y_{23}^{\nu}|^2+|y_{33}^{\nu}|^2}}\right)^{1.16}.
\end{equation}
Thus, a closed form expression for the final baryon asymmetry in terms of the Yukawa couplings, $\epsilon$, $\Delta m_{\rm atm}^2$, and $\Delta m_{\rm sol}^2$ can be obtained: 
\begin{equation}\label{eq:ModelIIIBaryonasymm}
    Y_{\Delta B}\simeq2\times10^{-7}\epsilon^{12.96}\sin{\phi}\sin{2\phi_k}\left(\frac{R}{1/33.7}\right)\left(\frac{0.05~\text{eV}}{|\Delta m_{\rm atm}^2|^{\frac{1}{2}}}\right)^{1.16}\left(\frac{|y_{12}^{\nu}|}{\sqrt{|y_{23}^{\nu}|^2+|y_{33}^{\nu}|^2}}\right)^{2.16}.
\end{equation}
A satisfactory fit for neutrino data and observed baryon asymmetry can be achieved simultaneously for example, for the following instance of parameters: $|y_{12}^{\nu}|=7,~\sqrt{|y_{23}^{\nu}|^2+|y_{33}^{\nu}|^2}=1/7,~|y_{11}^N|=|y_{23}^N|=6,~|y_{13}^N|=1/5,~\epsilon=0.3,~\beta=\pi/4,~v_d=100~\text{GeV}$ and $\tan{\phi_k}=1.01$.\footnote{This value of $\tan{\phi_k}=1.01$ is consistent with the rest of the parameters as it is implicitly dependent on them through Eq.~(\ref{eq:kinematicfactorres}) and Eq.(\ref{eq:ModelIIICpassym}).} These are consistent within our theoretical framework as illustrated in Sec.~\ref{subsec:ModelIaxionquality}.

\subsection{Model III: Quality of the axion}\label{subsec:ModelIIIaxionquality}
The UV completion of Model III can be achieved by introducing Higgs doublet scalars with masses of order $\Lambda_{\text{FN}}$ similar to Sec.~\ref{sec:ModelI}.  A complete set of fields and the resulting diagrams are shown in Sec.~\ref{subsec:ModelIIIUVcompletion}. The effective textures of the charged fermions with flavor charges in Table~\ref{tab:ModelIIIcharges} that provide the correct masses and mixings are: 
\begin{equation} \label{eq:modelIIIUV}
\begin{aligned}[b]
\mathcal{L}_{\text{yuk}}\supset Q^T H_u&\begin{pmatrix}
0 & 0 & y^{u}_{13}\, \overline{
\varepsilon}^4 \\
0 & y^{u}_{22}\, \overline{\varepsilon}^4 & y^{u}_{23}\,\overline{\varepsilon}^2 \\
y^{u}_{31}\,\overline{\varepsilon}^4 & y^{u}_{32}\,\overline{\varepsilon}^2 & y^{u}_{33}\,
\end{pmatrix}u^c+Q^T
H_d \begin{pmatrix}
0 & y^{d}_{12}\,\overline{\varepsilon}^7 & y^{d}_{13}\,\overline{\varepsilon}^7 \\
0 & y^{d}_{22}\,\overline{\varepsilon}^5 & y^{d}_{23}\,\overline{\varepsilon}^5 \\
y^{d}_{31}\,\overline{\varepsilon}^4 & y^{d}_{32}\overline{\varepsilon}^3\, & y^{d}_{33}\, \overline{\varepsilon}^3
\end{pmatrix}d^c\:\\
&\hspace{1.5cm}+L^T H_d\begin{pmatrix}
0 & 0 & y^{d}_{13}\,\overline{\varepsilon}^4 \\
y^{d}_{21}\,\overline{\varepsilon}^7 & y^{d}_{22}\,\overline{\varepsilon}^5 & y^{d}_{23}\,\overline{\varepsilon}^3 \\
y^{d}_{31}\,\overline{\varepsilon}^7 & y^{d}_{32}\overline{\varepsilon}^5\, & y^{d}_{33}\, \overline{\varepsilon}^3
\end{pmatrix}e^c\:+ \text{h.c}.\\
\end{aligned}
\end{equation}
We can now analyze the quality of the axion thoroughly along the same lines as Model I from Sec.~\ref{sec:ModelI}. We note from previous illustrations that the tree-level operators are clearly safe for axion quality. So we start first by identifying the leading contribution with reduced powers of $1/M_{\rm Pl}$ suppression. Taking into account Eqs. (\ref{eq:ModelIIIUVufields})-(\ref{eq:ModelIIIUVdfields})-(\ref{eq:ModelIIIUVuDfields}) gives the effective PQ-violating coupling
\begin{equation}
\frac{{\bar{\varepsilon}}^{28}}{M_{\rm Pl}^2}(H_uH_d)^3.
\label{eq:ModelIIIPQvilHud}
\end{equation}
This operator, in combination with the coupling from Eq.~(\ref{eq:pot}) with $q_S=-1/n\left( \pm k+28/3\right)$
through the three-loop flower diagram in Fig.~\ref{fig-3loop}, generates the PQ-violating operator which involves only the $(X,S)$ fields, given by
\begin{equation}
V_{\cancel{\text{PQ}}}^{\rm{ind}}\supset\left( \frac{\ln \left(\Lambda_{\rm FN}/M_{\text{A}_\text{H}}\right)}{16\pi^2}\right)^3\frac{S^{3n}(X^{(*)})^{3|k|}}{M_{\text{Pl}}^{3n+3|k|-4}}\left( \frac{X}{\Lambda_{\rm FN}}\right)^{28}+\text{h.c.}
\label{eq:ModelIIIPQvil-Hud}
\end{equation}
This diagram inevitably rules out models with the values of $(n,k)=(1,1^{\pm}),(2,2^{\pm})$ etc. However again, this model allows both values $(n,k)=(3,1^{\pm})$ as the acceptable cases of a high quality axion, as we shall demonstrate now.

\subsubsection*{Case with \boldsymbol{$(n,k)=(3,1^{-})$}}
As earlier, in the normalization $q_X=1$, using Eq.~(\ref{eq:cons}) we have $q_S=-25/9$. The leading gravity-induced  PQ-violating operator which contains only the $(X,S)$ fields is:
\begin{equation}\label{eq:ModelIII1treePQ}
    V_{\cancel{\text{PQ}}}\supset\frac{S^9\left( X\right)^{25}}{M_{\rm Pl}^{30}}+ \text{h.c},
\end{equation}
which is extremely safe for the quality of the axion. The only way to construct a PQ-violating term of the type given in Eq.~(\ref{eq:ModelIII1treePQ}) and a lower inverse Planck suppression is to make use of the $d=5$ PQ conserving and $d=7$ PQ-violating operators 
\begin{equation}\label{eq:ModelIIIPQV}
\begin{aligned}
    V&\supset \frac{S^3}{M_{\rm{Pl}}} H_uH_d^{\left(\frac{11}{3}\right)} +\frac{(S^*)^3}{M_{\rm Pl}^{3}}\left( H_u^{\left( \frac{8}{3} \right)} H_d^{\left( \minus\frac{16}{3} \right)} \right) \left( H_u^{\left( \frac{8}{3} \right)} H_d^{\left(\minus \frac{25}{3} \right)}\right) + \text{h.c}.\\
\end{aligned}
\end{equation}
and using the fact that $\widetilde{H_u}^{(q)}\equiv H_d^{(-q)}$, to get the leading operator
\begin{equation}\label{eq:ModelIII2loopn3k-1PQ}
    V_{\cancel{\text{PQ}}}^{\text{ind}}\supset \Xi~\frac{S^{9}}{M_{\rm Pl}^5}\left(\frac{X}{\Lambda_{\text{FN}}}\right)^{25} + \text{h.c},
\end{equation}
\begin{figure}[!htbp]
\centering
\includegraphics[width= 0.75\textwidth]{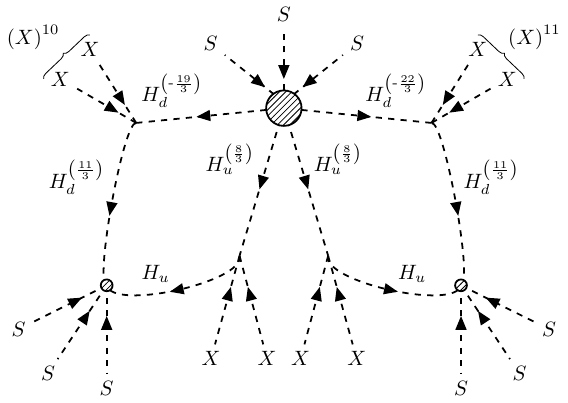}
\caption{Two-loop beetle diagram generating the PQ-violating operator in Eq.~(\ref{eq:ModelIII2loopn3k-1PQ}) from the scalar UV completion fields given in Sec.~\ref{subsec:ModelIIIUVcompletion} with $H_d^{\left(\minus\frac{19}{3} \right)}\equiv \widetilde{H_u}^{\left( \frac{19}{3}\right)},H_d^{\left(\minus\frac{22}{3} \right)}\equiv \widetilde{H_u}^{\left( \frac{22}{3}\right)}$. The larger shaded blob is the PQ violating interaction given as the second term of Eq.~(\ref{eq:ModelIII2loopn3k-1PQ}) and the smaller shaded blobs are the PQ conserving interactions given as the first term of Eq.~(\ref{eq:ModelIII2loopn3k-1PQ})}
\label{fig:MIIIUV-loop-corrections} 
\end{figure}
This arises from the two-loop beetle diagram in Fig.~\ref{fig:MIIIUV-loop-corrections}.
We have a trivial loop specific factor of $\Xi\sim \left( {16\pi^2}\right)^{-2}$, since we have one light Higgs in each loop with no IR divergence and hence, no logarithmic enhancement factor. Using Eqs. (\ref{eq:generalaxionquality})-(\ref{teta-ma}) with $(\mathfrak{a},\mathfrak{b},\mathfrak{c})=(9,0,25)$, we have a region of high quality axion given in Fig.~\ref{fig:ModelIIIQualityPlot}. 

\subsubsection*{Case with \boldsymbol{$(n,k)=(3,1^+)$}} 
 Using Eq.~(\ref{eq:cons}) we have $q_S=-31/9$. The leading PQ-violating operator with only $(X,S)$ fields is:
\begin{equation}\label{eq:ModelIII2treePQ}
    V_{\cancel{\text{PQ}}}\supset\frac{S^9\left( X\right)^{31}}{M_{\rm Pl}^{36}}+ \text{h.c},
\end{equation}
which is extremely safe for the quality of the axion. The leading PQ-violating operator, induced by two-loop beetle diagram as in Fig.~\ref{fig:MIIIUV-loop-corrections} is given by:
\begin{equation}\label{eq:ModelIII2loopn3k+1PQ}
    V_{\cancel{\text{PQ}}}^{\text{ind}}\supset \Xi~\frac{S^{9}}{M_{\rm Pl}^5}\left(\frac{X}{\Lambda_{\text{FN}}}\right)^{31} + \text{h.c}.
\end{equation}
 Using Eqs. (\ref{eq:generalaxionquality})-(\ref{teta-ma}) again for $(\mathfrak{a},\mathfrak{b},\mathfrak{c})=(9,0,31),$ also produces a high quality axion as DM for some $f_a$ in the DM favorable range in Eq.~(\ref{eq:axionDMcons}). With a single light Higgs doublet in each loop, the loop specific factor is trivial $\Xi \sim(16\pi^2)^{-2}$ without a logarithmic enhancement. 

\subsection{Model III: Summary of results}
\begin{figure}[!htbp]
\centering
\hspace{-1.0cm}\includegraphics[width=1.0\textwidth]{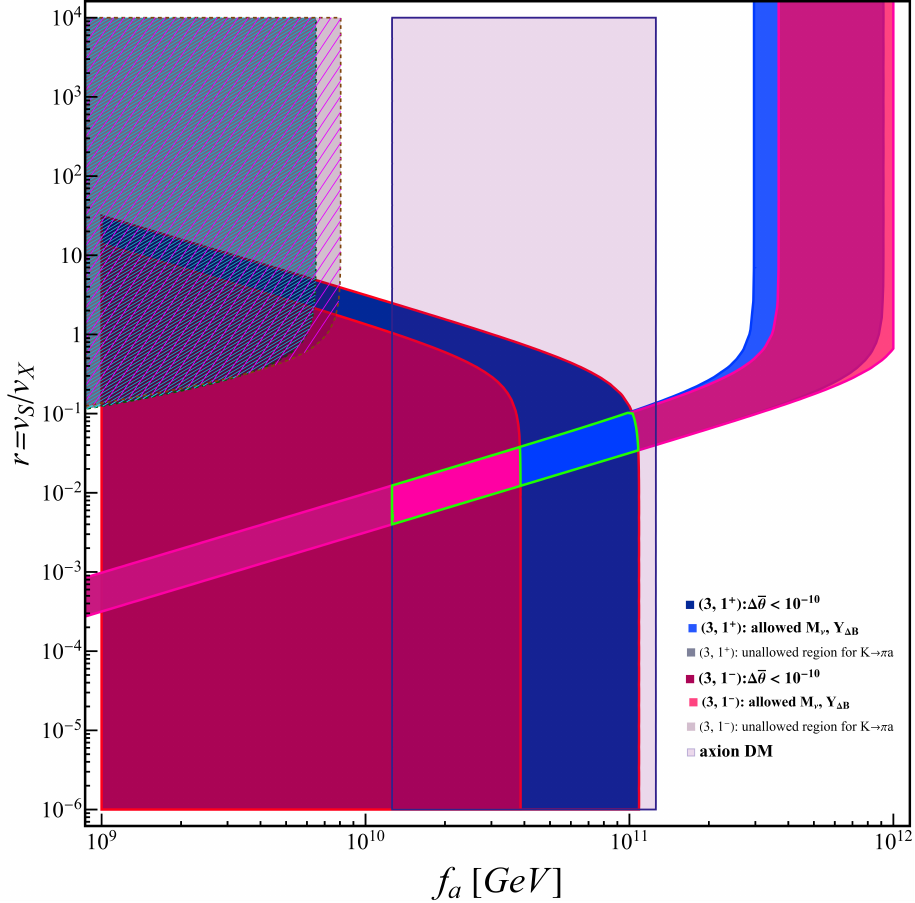}
\caption{Viable range of $(r,f_a)$ and low $\tan \beta\in [1,5]$depicting the region for a high quality axion for the values of $(n,k)=(3,1^{\pm})$ in dark blue and wine red respectively. Their respective brighter bands denote the parameter space for allowed values of $\Lambda_{\text{FN}}$ from Eq.~(\ref{eq:modelIIILambda_FN}) that provide a good fit to neutrino oscillation data and satisfactory baryon asymmetry for each model. The purple region shows the allowed axion DM range from Sec.~\ref{subsec:DMcons} for all models. The hatched regions denote the restricted regions from $K\rightarrow \pi ~a$ bounds from Sec.~\ref{subsec:FVcons}. The green bordered brightest blue and pink regions finally provide the allowed parameter space where all possible features listed above are simultaneously feasible for each case of $(n,k)$ listed in the legends.}
\label{fig:ModelIIIQualityPlot}
\end{figure}

The list of viable $(n,k),r$ and $M_{\text{A}_{\text{H}}}$ values for Model III is given in Table~\ref{tab:ModelIIIqualitycases}.
\begin{table}[!h]
    \hspace{-0.65cm}
    \renewcommand{\arraystretch}{1.5} 
    \begin{minipage}{0.5\textwidth}
    \hspace{0.75cm}\begin{tabular}{|c|c|c|c|c|c|}
        \hline
        $(n,k)$ & $q_S$ &$r$&$M_{\text{A}_\text{H}}\,[\text{GeV}]$ &\text{High Quality} &\text{Axion DM} \\
        \hline
         $(3,1^+)$ & $-31/9$&$[4 \times 10^{\minus 3},0.1]$ & $[5.4 \times 10^4,1.4 \times10^{5}]$&\textcolor{Blue}{\text{\ding{52}}}&\textcolor{Blue}{\text{\ding{52}}} \\
        \cline{1-6}
         $(3,1^-)$ & $-25/9$&$4 \times[ 10^{\minus 3},10^{\minus2}]$ &$[4.4\times 10^3,5\times10^{10}]$ &\textcolor{Blue}{\text{\ding{52}}}&\textcolor{Blue}{\text{\ding{52}}} \\
        \hline
    \end{tabular}
    \end{minipage}
    \hfill
\caption{Allowed values of $(n,k)$, $r$, $q_S$ and the range of the pseudoscalar Higgs mass $M_{\text{A}_\text{H}}$ that solve the flavor puzzle, fit neutrino data and solve strong CP with high quality axion to all orders in Model III within $\tan\beta\in [1,5]$.}\label{tab:ModelIIIqualitycases}
\end{table}
The viable range of $\Lambda_{\text{FN}}$ based on the simultaneous neutrino data and leptogenesis constraints from Eq.~(\ref{eq:ModelIIIBaryonasymm}) in this model is:
\begin{equation}\label{eq:modelIIILambda_FN}
    \Lambda_{\text{FN}}\in [2.0\times10^{13},1.1\times10^{14}] \,\text{GeV}. 
\end{equation}
The region of a high quality axion for both these cases is depicted in Fig.~\ref{fig:ModelIIIQualityPlot}. The wine red ($(n,k)=(3,1^-)$) region is overlaid on the dark blue ($(n,k)=(3,1^+)$) region for axion quality, DM and neutrino fit/ leptogenesis and hence it should be noted that the allowed parameter space for the former is contained in the latter. Additional contributions to the pseudoscalar Higgs mass and their respective new possible cases are presented in Appendix \ref{subsec:pseudoaddI} for Model III. This contribution to $M_{\text{A}_\text{H}}$ from Eq.~(\ref{eq:Amassadd}) is reflected in the case of $(3,1^-)$ in Table~\ref{tab:ModelIIIqualitycases} as well.

\section{Experimental probes of the high quality flavored axion: Model I}\label{sec:experimentalprobes}

Our class of high quality flavored axion models, inevitably acquire flavor dependent couplings to the SM fields. It was seen in Sec.~\ref{subsec:axionSMFVcouplings} that it is not trivial that all couplings are flavor dependent and that there is a constraint of gauge anomaly cancellation that spares the axion-photon coupling  from flavor dependency. In essence, we have a high quality flavored axion model that interpolates between a DFSZ I axion for $r\ll1$ where the axion lives predominantly in the $S$ field and a flaxion/ axiFlavon~\cite{Ema:2016ops, Calibbi:2016hwq} for $r\gg1$ where the axion lives predominantly in the Flavon field $X$. In general, we dub such a high quality flavored accidental axion as the ``\textbf{\emph{Flaccion}}". In this section, we consider a single model, Model I of Sec.~\ref{sec:ModelI}, with flavor charges from Table~\ref{tab:ModelIcharges} and $(n,k)=(3,1^+), q_S=-13/9$ which is valid for large $\tan \beta \in [50,140]$ to demonstrate the phenomenological possibilities of the flaccion.\footnote{This same analysis can be performed for any of the other cases given in Secs.~\ref{sec:ModelI}, \ref{sec:ModelII}, \ref{sec:ModelIII} and \ref{sec:pseudoadd} which exhibit similar properties.} We consider this case in particular due to the possibility of ``astrophobic" matter couplings discussed in Sec.~\ref{sec:axionmattercouplings}.

\subsection{Flavor violation constraints}\label{subsec:FVcons}

The most stringent bounds on $f_a,m_a$ come from the flavor-violating decays: $K^{+}\rightarrow \pi^{+} a$~\cite{E949:2007xyy} and $B^{+}\rightarrow (K,\pi)^{+}a$ and $\mu^+\rightarrow e^+ a \gamma$~\cite{Bolton:1988af} where the latter two are contained in the former bounds and are weaker. We consider three values of $r$ for which we can derive the lower bound on $f_a$ and upper bound on $m_a$ from Eq.~(\ref{eq:FVLowerbound}) with $\kappa_{ds}(\epsilon)= \epsilon$ (assuming reasonably, the $\mathcal{O}(1)$ Yukawa coupling dependence to be exactly $1$) as:
\begin{itemize}
\centering
    \item $r=0.5$: $f_a\geq 4.83 \times 10^9\, \rm GeV$, $m_a\leq 1.17 \times 10^{-3}\,\rm eV$,
    \item $r=0.81$: $f_a\geq 8.15 \times 10^9 \,\rm GeV$, $m_a\leq 7.0 \times 10^{-4}\,\rm eV$,
    \item $r=10$: $f_a\geq 1.4 \times 10^{10}\, \rm GeV$, $m_a\leq 4.06 \times 10^{-4}\,\rm eV$.
\end{itemize}
Such flaccion models can be tested at the  NA62~\cite{NA62:2025upx} experiment at CERN with up to date bounds shown as the benchmark exclusion bounds in Figs.~\ref{fig:ModelIQualityPlot},~\ref{fig:ModelIIQualityPlot},~\ref{fig:ModelIIIQualityPlot} and ~\ref{fig:gaegay} as the hatched-grayed regions. Near future experiments like KOTO at J-PARC~\cite{KOTO:2024zbl} and HIKE~\cite{HIKE:2023ext} at CERN, the latter of which is expected to probe even lower, up-to $\text{Br}(K^+ \rightarrow \pi^+ a) \lesssim 1.0 \times 10^{-12}$. The bounds on $f_a$ from $Br(\mu^+ \rightarrow e^+ a)$ are expected to be improved by the MEG II~\cite{Renga:2018fpd} and $\mu\rightarrow3e$~\cite{Blondel:2013ia} experiments whose details can be found in Ref.~\cite{Calibbi:2020jvd}. These could compete with the astrophysical bounds that we discuss in the next sections and can act as a test of our class of models from flavor violating decay searches.

\subsection{Enhancement or suppression of axion-matter couplings}\label{sec:axionmattercouplings}
\begin{figure}[!htbp]
\centering
\includegraphics[width=0.95\textwidth,trim=1mm 1mm 1mm 1mm, clip]{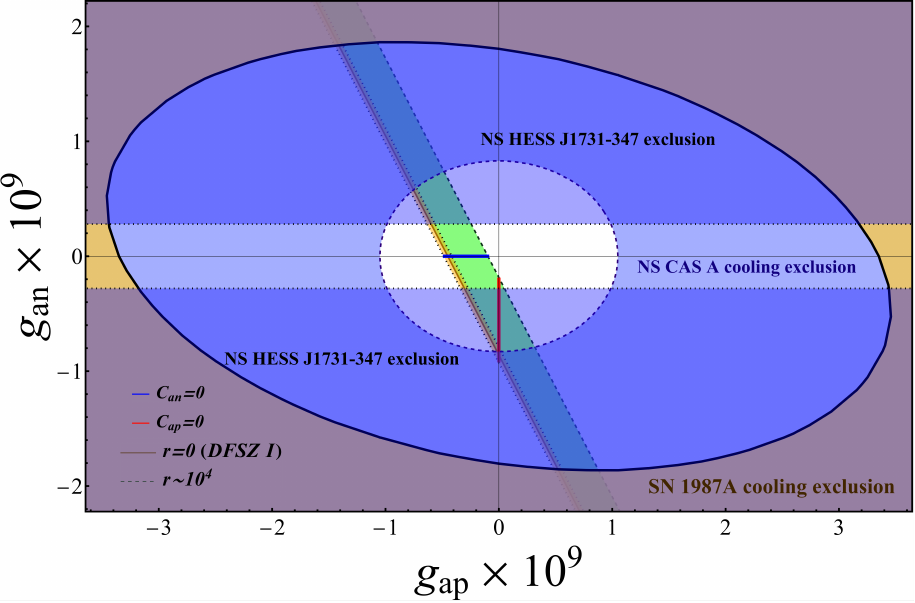}
\caption{Axion-proton $g_{ap}$ vs axion-neutron $g_{an}$ for Model I with $(n,k)=(3,1^+)$. The unshaded region is allowed. The thin orange band enveloped by dotted lines is the allowed  region for the DFSZ I axion and the green region is the allowed region for the Flaccion with the $r=0,10$ boundaries. The various exclusion limits are explained in the text below. The blue and red lines represent the neutrophobic and protophobic axion values respectively.\label{fig:gapgan}}
\end{figure}

The axion-nucleon coupling gets modified to having a dependence on the new parameter $r$ in this class of flavor dependent axion models:
\begin{equation}\label{eq:axion-nucleon-couplings}
    \begin{aligned}[b]
        C_{ap}(r,\tan \beta) &= -0.47(3) + 0.88(3) C_{au}(r,\tan\beta) - 0.39(2) C_{ad}(r,\tan\beta) - C_{a,\text{sea}}(r,\tan\beta) \;, \\
        C_{an}(r,\tan\beta) &= -0.02(3) + 0.88(3) C_{ad}(r,\tan\beta) - 0.39(2) C_{au}(r,\tan\beta) - C_{a,\text{sea}}(r,\tan\beta) \; ,\\
         C_{a,\text{sea}}(r,\tan\beta) &=0.038(5)C_{as}(r,\tan \beta)+0.012(5)C_{ac}(r,\tan \beta)\\
    &+0.009(2)C_{ab}(r,\tan \beta)+0.0035(4)C_{at}(r,\tan \beta),
    \end{aligned}
\end{equation}
where the $C_{au},C_{ad}$ are the leading contributions from an EFT valid at energies lower than $\Lambda_{\text{QCD}}$, the $C_{a,\text{sea}}$ contribution comes from the other heavy quarks~\cite{GrillidiCortona:2015jxo}. Only this time, the bare axion-quark couplings are dependent on $r$ and the flavor charges as seen from Eqs. (\ref{eq:caq})-(\ref{eq:cea}). It is trivial to realize that in the limit of $r\rightarrow0$, we reproduce the DFSZ-I axion in Eq.~(\ref{eq:DFSZIlimitmattercouplings}).\footnote{Although we can theoretically see this convergence to the DFSZ-I axion, $r>10^{-8}$ for $f_a\in[10^8,10^{12}]$~GeV is a lower limit on the scale of $\langle X\rangle$ which cannot be above $M_{\rm Pl}$.} A given model has fixed values of flavor charges but we still have interesting possibilities, as our axion matter couplings depend on this new parameter $r$. For instance, the axion electron coupling $C_{ae}$ also gets modified to depend on $r$ and the flavor charges in Table~\ref{tab:ModelIcharges} for Model I in Sec.~\ref{sec:ModelI} leading to:
\begin{equation}\label{CaeModelI
}
    C_{ae}(r,\tan\beta)=\frac{27 \tan^4\beta-13 r^2 \left(3 \tan^4\beta+8 \tan^2\beta+4\right)}{\left(169 r^2+81\right) \left(\tan^2\beta+1\right)^2}.
\end{equation}
Therefore, there are points in the parameter space for specific values of $r$ and ranges of $\tan \beta$ that produce zero coupling of the axion to protons (protophobia), neutrons (neutrophobia) or electrons (electrophobia). These points represent interesting implications as they can avoid certain astrophysical constraints, hence becoming astrophobic:
\begin{itemize}
    \centering
    \item  \textbf{Protophobia}: $C_{ap}= 0, r\simeq 1.4099,\,\,\tan \beta\in[50,140]$, 
    \item \textbf{Neutrophobia}: $C_{an}= 0, r\simeq 0.5088,\,\,\tan \beta\in[50,140]$,
    \item \textbf{Electrophobia}: $C_{ae}= 0, r\simeq 0.8316,\,\,\tan \beta\in[50,140]$. 
\end{itemize}
As we can see, these points are very model dependent and it is not guaranteed they exist and happen all at the same value of $r$. We define the dimensionless axion-matter and photon couplings respectively as:
\begin{equation}\label{eq:gap,gan,gae definition}
    \begin{aligned}[b]
        & g_{ap}=\frac{C_{ap} m_p}{f_a},\,g_{an}=\frac{C_{an} m_n}{f_a},\,g_{ae}=\frac{C_{ae} m_e}{f_a},\,g_{a\gamma}=\frac{\alpha_{\rm EM}} {2\pi }\frac{C_{a\gamma}}{f_a},
    \end{aligned}
    \end{equation}
and plot the couplings across various planes after rescaling them to appropriate values. In Fig.~\ref{fig:gapgan}, we plot the $g_{ap}$ and $g_{an}$ with the following common astrophysical bounds arising from SuperNova (SN) and Neutron Star (NS) cooling in Supernova Remnants (SNR)~\cite{Caputo:2024oqc,ParticleDataGroup:2024cfk}. 
\begin{itemize}
    \item The most stringent bound on $g_{an}$ comes from the axion bremsstrahlung cooling of a young NS in the HESS J1731-347 SNR ~\cite{Beznogov:2018fda}: $|g_{an}|\leq 2.8 \times 10^{-10} (90\% \, \rm C.L)$ denoted by the grayed out region above and below the horizontal dotted lines in Fig. \ref{fig:gapgan}. 
    \item The observation of the rapid cooling of a Neutron star in the Cassiopeia A (CAS A) Supernova Remnant (SNR) puts the most stringent bound on both the $g_{ap},g_{an}$ couplings~\cite{Leinson:2014cja,Hamaguchi:2018oqw}: $1.6 g_{an}^2+g_{ap}^2\leq 1.1 \times 10^{-18}$
    \item The most relaxed bound comes from the neutrino data from SN 1987A excluding the cooling of the SN through axion emission~\cite{Carenza:2019pxu}:
$g_{an}^2+0.29g_{ap}^2+0.27 g_{an}g_{ap} \lesssim3.25 \times 10^{-18}$.
\end{itemize}
The green band represents the total parameter space of axion nucleon couplings for the flaccion, the DFSZ-I region in Fig.~\ref{fig:gapgan} is plotted for $\tan \beta \in [50,150]$, which is the range of validity of this model. For larger $r$, there is no significant increase in the size of the couplings as there is a sharp cutoff for $r>10$. The various values of $r$ are depicted with the red and blue lines to illustrate the span of this case of Model I and to showcase the asymptotic approach for large $r$. This is also in the sweet spot for the range of $r$ depicted in Fig.~\ref{fig:ModelIQualityPlot}. 

\begin{figure}[!htbp]
\centering
\includegraphics[width=0.95\textwidth]{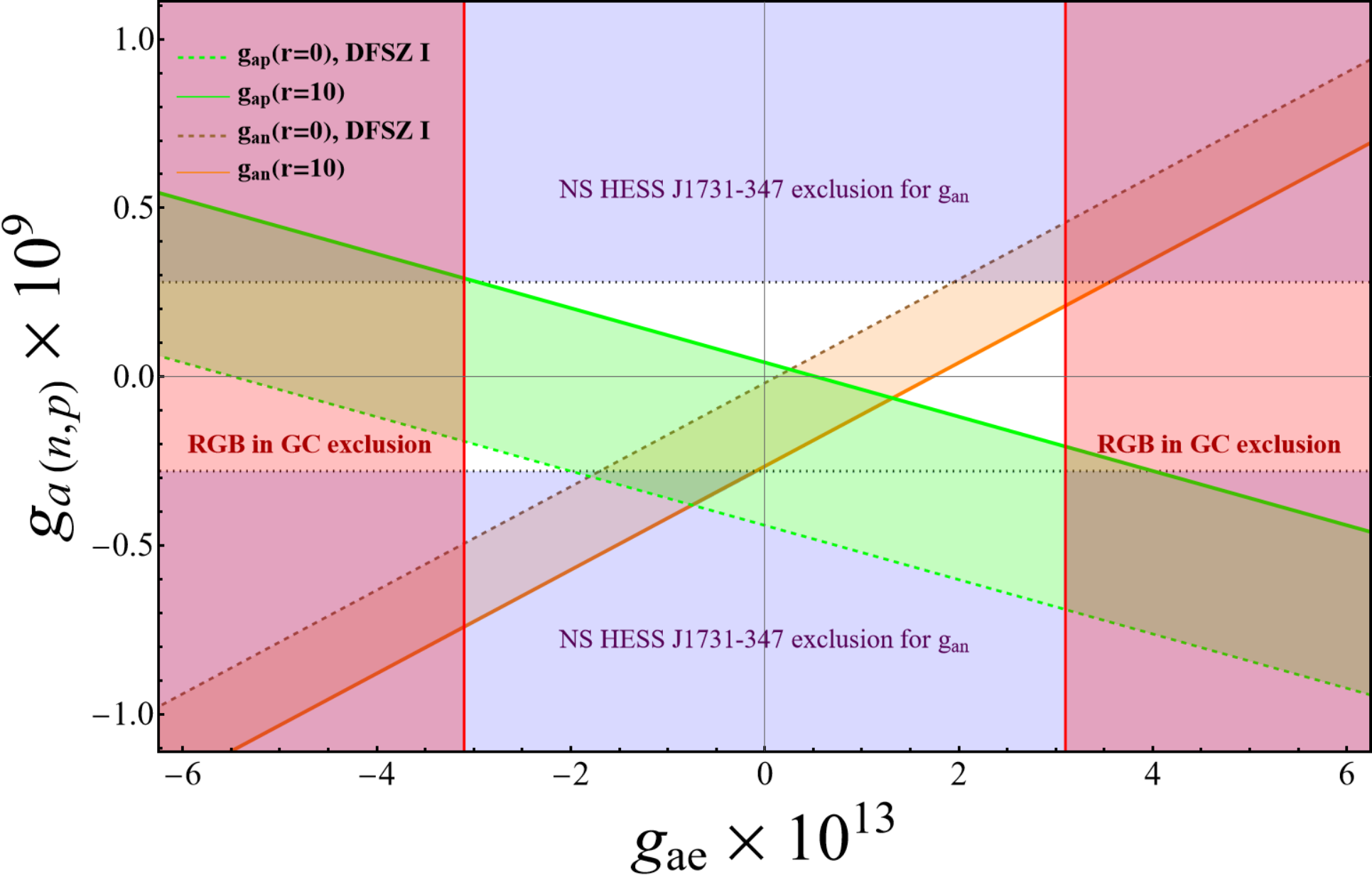}
\caption{Axion-nucleon couplings $g_{ap}$ (green band), $g_{an}$ (orange band) vs axion electron coupling $g_{ae}$ . The dashed lines are DFSZ-I and the flaccion region is extended for increasing values of $r$ into the green and orange bands for $g_{ap}$ and $g_{an}$ respectively. The astrophysical exclusions from the NS in HESS J1731-347 for the $g_{an}$ coupling is shown in light purple. The red region is the exclusion from the cooling of RGB Stars in GC.\label{fig:gaegaNu}}
\end{figure}

\begin{figure}[!htbp]
\centering
\includegraphics[width=1.0\textwidth,
    trim=1mm 0mm 0mm 1mm, clip]{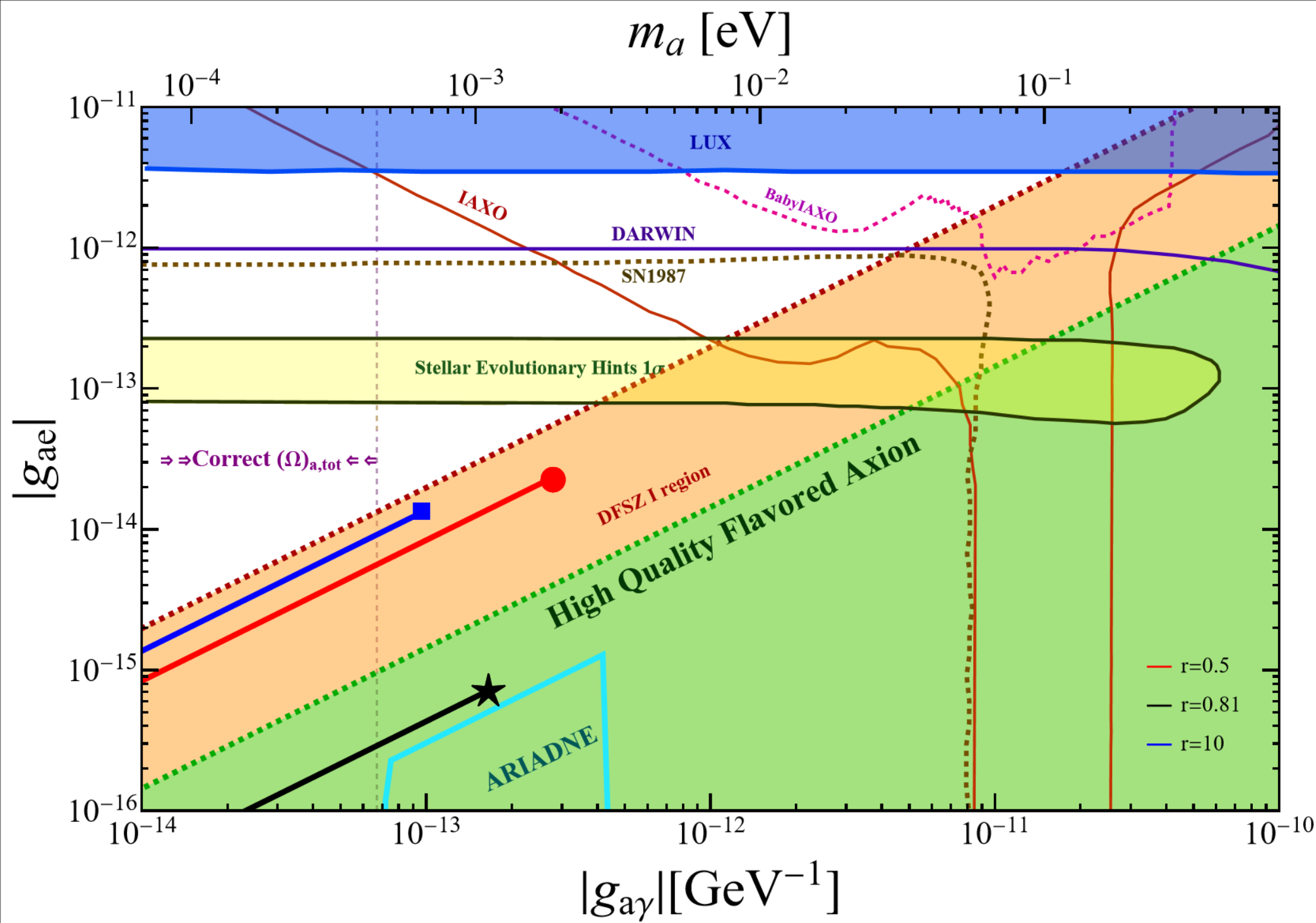}
\caption{Axion-Electron $g_{ae}$ vs Axion-photon coupling $g_{a\gamma}$ for Model I with $(n,k)=(3,1^+)$ for varying $m_a$ at $\tan \beta = 50$. The Red ($r=0.5$), Blue ($r=10$) and Black ($r=0.81$) lines are the flaccions cut off at their flavor bounds on $m_a$ represented by the circle, square and star respectively. The orange region is the DFSZ-I axion range for the whole range of allowed values of $\tan \beta$. The green region is our new flavored axion extension. The various experimental sensitivities are explained in the text.} \label{fig:gaegay}
\end{figure}

The axion-nucleon couplings are plotted with respect to axion-electron couplings in Fig.~\ref{fig:gaegaNu} for various values of $r$ represented by the green and orange band for the departure from the DFSZ-I axion case for $g_{ap}$ and $g_{an}$ couplings respectively. The NS HESS J1731-347 SNR bound does not apply for the $g_{ap}$ coupling and the strongest constraint for the $g_{ae}$ coupling comes from the cooling of a Red Giant Branch (RGB) star in the $M3$~\cite{Straniero:2018fbv} and $M5$~\cite{Viaux:2013lha} Globular Clusters (GC): $ |g_{ae}|\leq 3.1 \times 10^{-13} \, (95\% \rm\, C.L) $. 
Finally, the axion-photon coupling is identical to the DFSZ-I case but we plot the new ``electrophobic" axion-electron coupling with respect to the axion photon coupling to distinguish it from the conventional DFSZ-I axion in Fig.~\ref{fig:gaegay}. The green region is the expanded flaccion parameter space from the DFSZ-I orange region. For the $r=0.5$ red line in the DFSZ-I region we can observe the axion-electron coupling start to decrease. Increasing $r\rightarrow0.8316$ to the electrophobic point allows us to reach lower into the green region denoted by a near electrophobic $r=0.81$ black line and finally after crossing the electrophobic point (where $g_{ae}=0$), we venture back upwards towards the DFSZ-I region depicted by the $r=10$ blue line. We project various experimental sensitivities from Helioscope experiments including IAXO~\cite{IAXO:2019mpb} and BabyIAXO~\cite{Abeln:2020eft} for both the $g_{a\gamma}$ and $g_{ae}$ couplings~\cite{ciaran_o_hare_2020_3932430} to contrast where such flaccions might be discovered. The stellar evolutionary hints region includes the combined White Dwarf Luminosity Function (WDLF), WD Pulsation and RGB stars at $1\sigma$~\cite{DiVecchia:2019ejf}. Large detectors for WIMPS like LUX~\cite{LUX:2017glr} and liquid XENON detectors like DARWIN~\cite{DARWIN:2016hyl} have the capacity to probe $g_{ae}$ at these relevant sensitivities through the axio-electric effect~\cite{Derevianko:2010kz} whose projections are also included. We also present the projected $g_{ae}$ sensitivity from a ``fifth force" experiment, ARIADNE~\cite{Arvanitaki:2014dfa} based on their allowed $f_a,m_a$ region and its proclivity to probe electrophobic axions. Finally, the SN 1987A experimental bounds on $g_{a\gamma}$ are a collection of conservative bounds with more uncertainty compared to other stellar hints~\cite{Turner:1987by}. These are derived from the exclusion of SN cooling through axions fitted on the DFSZ-I parameter space of the $g_{a\gamma}$ coupling.
Each of the lines with different values of $r$ terminate at their respective upper bound of $m_a$ (circle, square and diamond) owing to flavor violating decay constraints mentioned in Sec.~\ref{subsec:FVcons}. Finally, the correct total axion DM relic abundance limits the viable parameter space given in Sec.~\ref{subsec:DMcons} of such high quality flavored QCD axions to fall within the narrow range in between the two vertical dashed purple lines in Fig.~\ref{fig:gaegay}.
\section{Solution to the cosmological domain wall problem}\label{sec:Domainwallproblem}
An accidental axion as a result of a gauged symmetry, identified from a linear combination of Nambu Goldstone (NG) bosons has a domain wall number equal to the minimum positive integer that corresponds to the residual discrete transformation that can undo a PQ transformation. For this reason, $N_{\text{DW}}\in \mathbb{Z}_{+}$ can be expressed as:
\begin{equation}\label{eq:NDWcommon}
    N_{\text{DW}}= \text{min}\left[\frac{1}{f_a}\sum_{\alpha=u,d,X,S}n_{\alpha} K_\alpha v_{\alpha}\right], n_{\alpha}\in \mathbb{Z}
\end{equation}
where $K_{\alpha}$ is given in Eqs. (\ref{eq:axion})-(\ref{eq:comp4}). In other words, the domain wall number is the dimension of the quotient group $G/ZH$ where $G$ is the unbroken gauge group, $H$ is the broken symmetry of the degenerate vacuum and $Z$ is the center of the unbroken gauge group $G$~\cite{Ernst:2018bib}. The expression in Eq.~(\ref{eq:NDWcommon}) when simplified fully results in a numerator and denominator respectively given as:
\begin{equation}\label{Num_Denom}
\begin{aligned}[b]
\text{Num}(N_{\text{DW}})&=n N_g \left(\pm k\,v_u^2v_d^2v_S^2(q_S(n_u+n_d+n \,n_S)\pm k \,n_Sq_X)+v^2v_S^2v_X^2 (n_S\, q_X-n_X\,q_S)\right. \\
&\left.-v_u^2v_d^2v_X^2(n^2q_Sn_X+n\,q_X (n_u+n_d)\pm k\, n\,q_X\,n_X) \right)\\
\text{Den}(N_{\text{DW}})&=q_X(k^2v_u^2v_d^2v_S^2+v^2v_S^2v_X^2+n^2v_u^2v_d^2v_X^2)
\end{aligned}
\end{equation}
where, $N_g$ is the number of generations of colored fermions in the model.\footnote{$N_g$ was given to be $3$ in Secs.~\ref{subsec:axionfermioncouplings} and \ref{subsec:axiongluonphotoncouplings}. Here, we have extended that to an arbitrary number of colored fermions in the PQ color anomaly to emphasize the anomaly cancellation because of the gauged flavor symmetry.} For $N_{\text{DW}}$ to remain an integer, the 3 terms in the numerator should be the same integer multiple $\mathfrak{x}\in \mathbb{N}$ of the corresponding terms in the denominator, which gives us the following constraints:
\begin{equation}\label{eq:NDWcons}
 \begin{aligned}[b]
    \pm N_g\,n\,k\,(q_S(n_u+n_d+n \,n_S)\pm k \,n_Sq_X)& =\mathfrak{x}\,k^2\,q_X\\
     N_g\,n\,(n_S\,q_X-n_X\,q_S)=\mathfrak{x} \,q_X \\
     -N_g\,n\,(n^2q_Sn_X+n\,q_X (n_u+n_d)\pm k\, n\,q_X\,n_X) &=\mathfrak{x} \,n^2\,q_X
 \end{aligned}   
\end{equation}
The minimum positive value of $\mathfrak{x}$ gives the domain wall number of these class of accidental flavored axion models. Thus our equations reduce to 
\begin{equation}
    \begin{aligned}[b]
         &\mathfrak{x}= N_g\,n\,\left(\frac{q_S}{q_X}n_X-n_S\right),\\
         &\pm k\, n_X+n_u+n_d+n\,n_S=0
    \end{aligned}
\end{equation}
The second equation arising from the pseudoscalar mass term in Eqs. (\ref{eq:pot})-(\ref{eq:cons}) can always be satisfied with an arbitrary choice of $n_u,n_d,n_X,n_S\in\mathbb{Z}$. Thus the domain wall number of the accidental axion is given by: 
\begin{equation} \label{Eq:NDW}
    N_{\text{DW}}=\text{min} \Bigg[N_g n \left( \frac{q_S}{q_X}n_X -n_S \right)\Bigg], N_{\text{DW}},N_g,n\in \mathbb{N};k,n_X,n_S\in \mathbb{Z},
\end{equation}
The requirement for a high-quality axion grants us a rational $q_S$, such that the numerator and denominator are co-primes. Say, $q_S=\pm\frac{l}{m},~\text{gcd}(l,m)=1;~ l,m\neq1$ \footnote{If $\text{gcd}(l,m)=k$, then $l=\widetilde{l}/k,m =\widetilde{m}/k\implies N_{\text{DW}}=1$ automatically and the $k-1$ false vacua again degenerate into 1 indistinguishable vacuum. The $k$-fold degeneracy is lifted by the Planck-suppressed operators $S^{\widetilde{l}}X^{\widetilde{m}}$ along the same lines as the Barr-Seckel model~\cite{Barr:1992qq}.}(using the same notation from Sec.~\ref{subsec:axionquality} and gcd being ``Greatest Common Divisor"). This implies the added condition of $\text{gcd}(p,q)=1$ as a necessity for high quality axion from Eqs. (\ref{eq:cons})-(\ref{eq:lm}). So, in the phenomenologically compatible normalization of $q_X=\pm1$:
\begin{equation}\label{eq:NDW1}
     N_{\text{DW}}=\text{min}\left[\frac{N_g n}{m} \left(n_X\,l -n_S\,m \right)\right].
\end{equation}
Using ``Bezout's" identity from modular arithmetic, the second factor of the above equation can always be set to 1 since for $(l,m)$ co-primes, $\exists ~~ n_X,n_S\in \mathbb{Z}\, \ni\, (n_X\,l-n_S\,m)=1$. With the expressions for $l,m$ established in Eq.~(\ref{eq:lm}), without loss of generality, we end up with: 
\begin{equation}
    N_{\text{DW}}=\text{min}\left[\frac{N_g}{\mathfrak{q}}\right].
\end{equation}
In all our models from Secs.~\ref{sec:ModelI}, \ref{sec:ModelII} and \ref{sec:ModelIII} constrained by flavor phenomenology, from a gauged flavor symmetry, $\mathfrak{q}=N_g=3$ is the inevitable solution owing to the anomaly cancellation of 3 generations of quarks and the fractional hypercharge of the SM quarks.\footnote{In the case when $N_g\neq \mathfrak{q}$, we can prove without loss of generality that $\text{min}\left[\frac{N_g n}{m} \left(n_X\,l -n_S\,m \right)\right]=\text{min}\left[N_g  \left((\pm k\mathfrak{q}+\mathfrak{p}) -n~n_S\, \right)\right]=N_g$ since $\text{gcd}(\pm k\mathfrak{q}+p,n)=\text{gcd}(\pm k\mathfrak{q}+p,q)=1$. This corresponds to the typical case of a DFSZ-I axion model where the DWs are connected again by a gauge symmetry and their degeneracies are lifted by the argument in the previous footnote. See also Ref.~\cite{Barr:1982uj}.} Therefore, we always end up with $N_{\text{DW}}=1$ for all such models with a flaccion with three generations of SM fermions. For a general calculation of $N_{\rm DW}$ with the heavy Higgs UV doublets as well, we refer to the discussions in Appendix.~\ref{sec:NDWgen}, where we show that $N_{\rm DW}$ is unchanged compared to the case of effective theory description.

If PQ symmetry breaks after inflation we end up with a network of axionic strings and domain walls~\cite{Sikivie:1982qv,Vilenkin:2000jqa}. We should note that having $N_{\text{DW}}=1$ does not automatically solve the cosmological domain wall problem if the PQ symmetry breaks after inflation~\cite{Barr:1992qq}. We have an anomaly-free gauge symmetry $U(1)_X$ and a global symmetry $U(1)_S$ with a color anomaly, and the axion is a linear combination of the phases of these two symmetries effectively. So, we have two types of strings that form through the Kibble-Zurek mechanism~\cite{1976JPhA....9.1387K}, ``global" and ``local" strings which we denote by their winding numbers $(w_X,0),(0,w_S)$ and $(w_X,w_S)$ respectively. These produce windings that are invariant under rotations for the $(X,S)$ fields with string tensions $E_X,E_S$ for the global strings $(n_X~l,0),(0,n_S~m)$ respectively:
\begin{equation}
\begin{aligned}[b]    
\label{eq:DWPsolutioncond}
& \Delta\theta_X =2\pi l n_X, E_X\propto (n_Xlv_X)^2 \\
& \Delta\theta_S =2\pi m n_S, E_S\propto (n_Smv_S)^2 \\
& \Delta \overline{\theta}=2\pi (n_Xl-n_Sm).
\end{aligned}
\end{equation}
Without loss of generality, the $U(1)$ symmetry broken by the greater of the two VEVs, gets gauged first and $(n_X~l,0)$ global strings form at a temperature $T\sim v_X$. Now, when the second scalar develops a VEV at a temperature $T\sim v_S<v_X$, the other global strings of the type $(0,n_S~ m)$ are formed whose minimal energy is obtained for $n_S=1$. The total winding energy of these string configurations is $E_{\text{wind}} \propto (n_Xl-n_Sm)^2$ and such strings will choose a minimal tension corresponding to $n_X=1$. So, the number of walls bounded by $(n_X~l,n_S~m)$ strings\footnote{True gauge strings~\cite{Niu:2023khv} (when $|n_X\,l-n_S\,m|_{min}=0$) solutions may also exist but their contribution to axion DM relic density abundances appears to be sub-leading to the global string decay modes and are hence not realistically considered here.}
\begin{equation}
\begin{aligned}[b]
 N_{W}[(n_X\,l,n_S\,m)]&=\text{min}\left(|n_X\, l-n_S\,m|\right)\\
 N_{W}[(0,m)]&=|m|\\
 N_{W}[(l,0)]&=|l|.   
\end{aligned}    
\end{equation}
So, if we only had global symmetries, the conditions $|m|=1 \text{ or }|l|=1 $ would have been sufficient to solve the domain wall problem. But we have gauge symmetries and thus an additional condition of$\text{min}\left(|n_X\, l-n_S\,m|\right)=1$ is required, which means that $l,m$ being relatively prime is a necessary condition to solve the domain wall problem. In our class of models, this requirement is met automatically as a consequence of the gauged flavor symmetry's requirement to protect the axion from Planck suppressed corrections. For the complete solution, domain walls that form after the axion mass is turned on by QCD effects at $T\sim\Lambda_{\text{QCD}}$, should terminate in such strings of minimal configuration~\cite{Barr:1992qq}.

\section{Summary and outlook}
\label{sec:Conclusions}
A bottom up framework that addresses several key puzzles of the SM is presented in this work. We employ a gauged abelian flavor symmetry to explain the mass and mixing hierarchies of all the fermions in the SM; including the neutrinos, whose masses are generated by a type-I seesaw at the same flavor scale $\Lambda_{\rm FN}$ while simultaneously explaining the baryon asymmetry of the universe through leptogenesis. The strong CP problem is solved without a domain wall problem by an axion, emerging from an accidental PQ symmetry arising from gauging the flavor symmetry, thereby protecting the axion potential against perturbative quantum gravitational corrections. 

This axion, as a consequence of emerging accidentally from a flavor symmetry has flavor dependent couplings to matter, prompting discovery in future flavor violating decay searches. We dub such a high quality axion, the ``\emph{\textbf{Flaccion}}"; a flavored accidental axion. The flaccion also serves as a viable DM candidate. Furthermore, we study the effects of the UV completion of flavor models on the quality of the axion and conclude that the flaccion is safe from all order of quantum gravitational corrections in perturbation theory. This is especially non-trivial where we have identified loops of higher order (flower, insect and angel diagrams) that can be leading contributions. We thus manage to present three such models exhibiting more interesting features on top of the above mentioned:

\begin{itemize}
    \item \textbf{Model I:} The flavor sector is UV completed using Higgs doublets that generate the mass and mixing hierarchies.  The neutrino mass texture obtained here has a $(2,2)$ cofactor zero prediction with both normal and inverted ordering fits for the mass and mixing spectra. The flavor charges of this model allow some unique features for the axion coupling to matter whereby, an electrophobic axion emerges for particular values of the VEVs of the complex singlet scalars $S,X$ present in the model. Other nucleophobic couplings can also occur at various points in the parameter space. This model prefers the scale of the axion $f_a\sim \Lambda_{\rm FN}\sim10^{10}-10^{11}~\text{GeV}$.
    \item \textbf{Model II:}  The flavor sector is UV completed using VLFs, which surprisingly boosts the range of the parameter space for the quality of the flaccion. This facilitates the quality to be high even when the flavor scale  $\Lambda_{\rm FN}$ is near $M_{\rm Pl}$ and safely prefers a high flavor scale at $\Lambda_{\rm FN}\sim 10^{16}-10^{17}$  GeV safely evading any Landau poles for the hypercharge gauge coupling before $M_{\rm Pl}$.
    \item \textbf{Model III:} This has a flavor charge structure that is compatible with an $SU(5)$ GUT embedding resulting in TeV scale RHNs and low scale resonant leptogenesis. The heavy flavor sector is again UV completed using Higgs doublets and prefers an intermediate flavor scale at $\Lambda_{\rm FN}\sim 10^{13}-10^{14}$ GeV with similar features to Model I. 
\end{itemize}
All the relevant model parameters and the features of the SM they can explain that have been presented in this work are tabulated in Table~\ref{tab:summary}. Additional cases in each model are listed in Appendices (\ref{subsec:pseudoaddI}, \ref{subsec:pseudoaddII}).
\begin{table}[!htbp]
\centering
\renewcommand{\arraystretch}{1.9}
\resizebox{1.0\textwidth}{!}{
\begin{tabular}{|>{\centering\arraybackslash}m{3cm}||c|c||c|c|c||c|c|}
\hline
 & \multicolumn{2}{c|}{Model I} & \multicolumn{3}{c|}{Model II} & \multicolumn{2}{c|}{Model III} \\
\hline
 $(n,k)$& $(3,1^+)$ & $(3,1^-)$ &  $(2,2^-)$ & $(3,1^+)$ & $(3,1^-)$ & $(3,1^+)$ & $(3,1^-)$ \\
\hline
High quality axion & \textcolor{Blue}{\text{\ding{52}}}  & \textcolor{Blue}{\text{\ding{52}}} & \textcolor{Blue}{\text{\ding{52}}} & \textcolor{Blue}{\text{\ding{52}}} & \textcolor{Blue}{\text{\ding{52}}} & \textcolor{Blue}{\text{\ding{52}}} & \textcolor{Blue}{\text{\ding{52}}}  \\
\hline
Charged fermion masses/mixing  & \textcolor{Blue}{\text{\ding{52}}} &\textcolor{Blue}{\text{\ding{52}}} &  \textcolor{Blue}{\text{\ding{52}}} & \textcolor{Blue}{\text{\ding{52}}} & \textcolor{Blue}{\text{\ding{52}}} & \textcolor{Blue}{\text{\ding{52}}} & \textcolor{Blue}{\text{\ding{52}}}\\
\hline
Neutrino masses/mixing      & \textcolor{Blue}{\text{\ding{52}}} &\textcolor{Blue}{\text{\ding{52}}}& \textcolor{Blue}{\text{\ding{52}}} & \textcolor{Blue}{\text{\ding{52}}} & \textcolor{Blue}{\text{\ding{52}}} & \textcolor{Blue}{\text{\ding{52}}} & \textcolor{Blue}{\text{\ding{52}}} \\
\hline
Axion DM           & \textcolor{Blue}{\text{\ding{52}}} &\textcolor{Red}{\text{\ding{56}}} &\textcolor{Blue}{\text{\ding{52}}}& \textcolor{Blue}{\text{\ding{52}}}  & \textcolor{Blue}{\text{\ding{52}}} & \textcolor{Blue}{\text{\ding{52}}} & \textcolor{Blue}{\text{\ding{52}}} \\
\hline
Leptogenesis       & \textcolor{Blue}{\text{\ding{52}}} &\textcolor{Blue}{\text{\ding{52}}}   & \textcolor{Blue}{\text{\ding{52}}} & \textcolor{Blue}{\text{\ding{52}}} & \textcolor{Blue}{\text{\ding{52}}} & \textcolor{Blue}{\text{\ding{52}}} & \textcolor{Blue}{\text{\ding{52}}}  \\
\hline
\end{tabular}
}
\caption{Viability of flaccion models to address current peculiarities in the SM.}
\label{tab:summary}
\end{table}
A similar approach needs to be pursued for the KSVZ-type~\cite{Kim:1979if,Shifman:1979if} invisible axion where the axion is protected by the use of a similar gauged $U(1)_{F}$ flavor symmetry in a future work. The topology of such flaccion models offers us complimentary flavor violating probes, conjoined with gravitational wave signatures from first order phase transitions and topological defects. These would produce unique complimentary signatures that could be tested at future experimental collaborations. We also plan to study the origin of inflation in such a class of flaccion models making effective use of the presence of the two additional characteristic scalar fields.

\acknowledgments The work of KSB and SCC are supported in part by US Department of Energy grant number DE-SC0016013. 
KSB and ZT wish to acknowledge the Center for Theoretical Underground Physics and Related Areas (CETUP*) and the Institute
for Underground Science at Sanford Underground Research Facility (SURF) for hospitality and
for providing a stimulating environment during the 2025 Summer Workshop on Neutrino Physics.
SCC is grateful to Carlos Tamarit for fruitful discussions on axionic domain walls and thanks the Mainz Institute of Theoretical Physics (MITP) of the Cluster of Excellence PRISMA+ (Project ID 390831469) for their warm hospitality and support. We are also grateful to Robert Ziegler for helpful suggestions on flavor bounds for the axion.

\appendix
\section{Appendix: UV completion of the flavor models}\label{sec:AppendixA}
\subsection{Model II: Vector-like fermionic flavor UV completion} \label{subsec:ModelIIUVcompletion}
\begin{figure}[!htbp]
    \centering
    \begin{minipage}{0.39\textwidth}
        \centering
        \includegraphics[width=\linewidth]{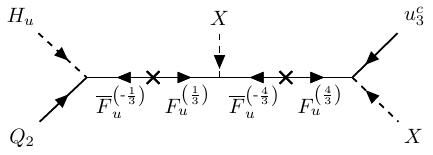}
    \end{minipage}
    \hfill
    \begin{minipage}{0.39\textwidth}
        \centering
        \includegraphics[width=\linewidth]{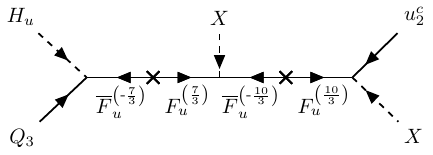}
    \end{minipage}
    \includegraphics[width=0.6\textwidth]{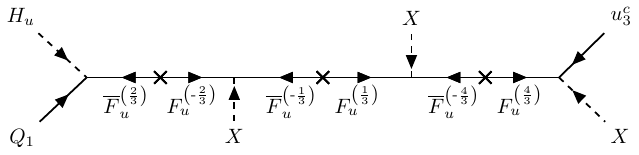}
    \includegraphics[width=0.8\textwidth]{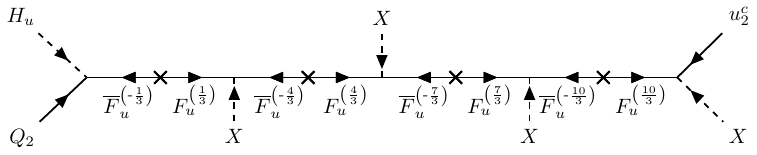}
    \includegraphics[width=0.8\textwidth]{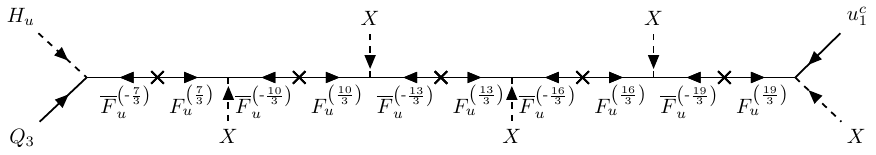}
    \includegraphics[width=0.8\textwidth]{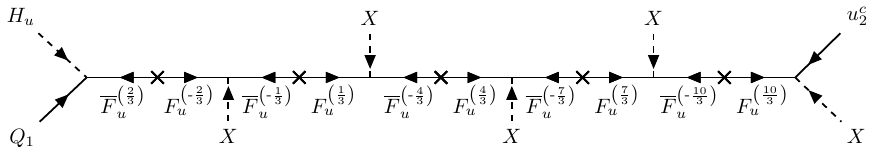}
    \includegraphics[width=1.0\textwidth]{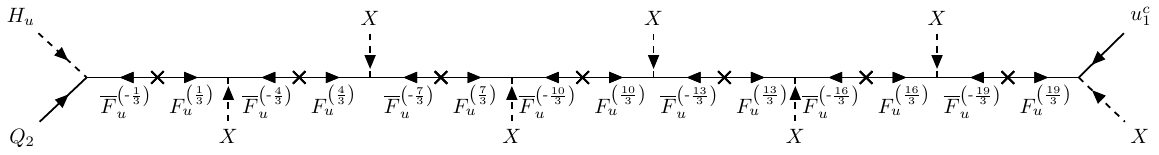}
    \includegraphics[width=1.0\textwidth]{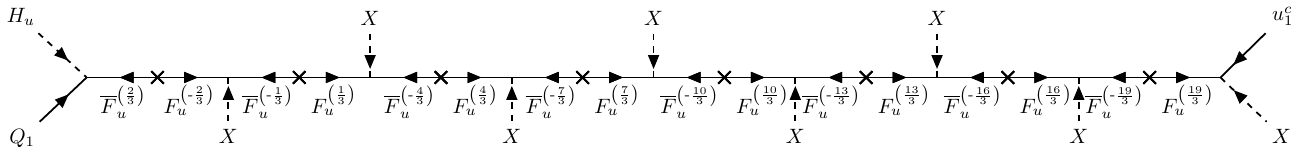}
    \caption{Diagrams generating the Up quark Yukawa sector of Model II given in Eq.~(\ref{eq:modelIIUV}).}
    \label{fig:modelIIuUV} 
\end{figure}

\begin{figure}[!htbp]
\centering
    \centering
    \begin{minipage}{0.49\textwidth}
        \centering
        \includegraphics[width=\linewidth]{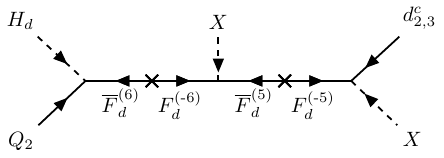}
    \end{minipage}
    \hfill
    \begin{minipage}{0.49\textwidth}
        \centering
        \includegraphics[width=\linewidth]{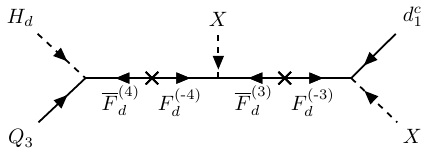}
    \end{minipage}
    \includegraphics[width=0.7\textwidth]{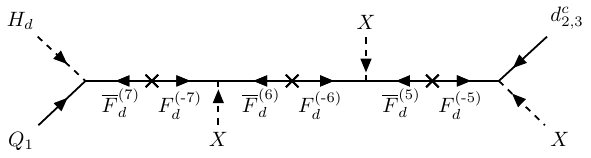}
    \includegraphics[width=0.8\textwidth]{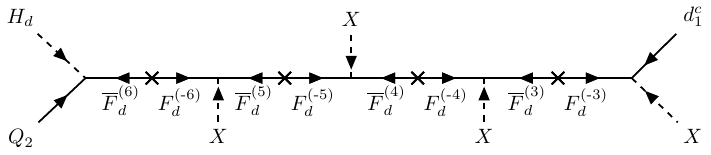}
    \includegraphics[width=1.0\textwidth]{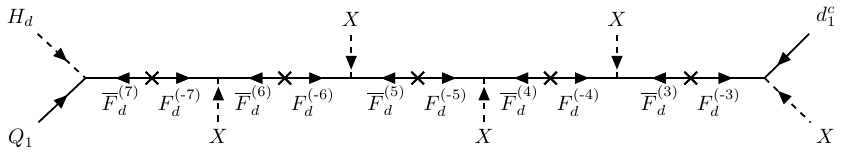}
\caption{Diagrams generating the down quark Yukawa sector of Model II given in Eq.~(\ref{eq:modelIIUV}).}
\label{fig:modelIIdUV} 
\end{figure}

\begin{figure}[!htbp]
\centering
 \begin{minipage}{0.35\textwidth}
        \centering
        \includegraphics[width=\linewidth]{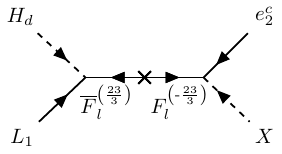}
        \end{minipage}
        \hfill
        \begin{minipage}{0.35\textwidth}
        \centering
        \includegraphics[width=\linewidth]{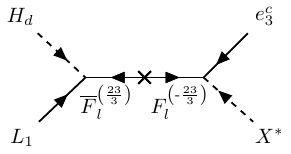}
    \end{minipage}
    \hfill
    \begin{minipage}{0.55\textwidth}
        \centering
        \includegraphics[width=\linewidth]{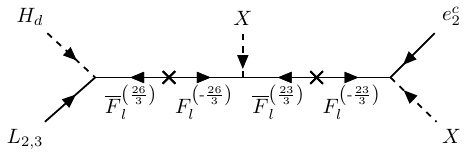}
    \end{minipage}
    \includegraphics[width=0.9\textwidth]{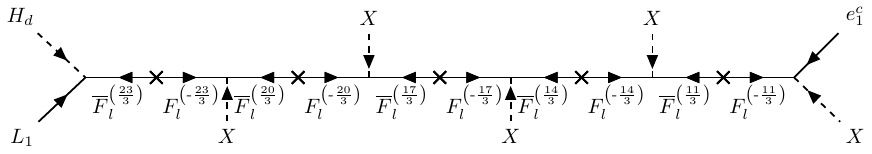}
    \includegraphics[width=1.0\textwidth]{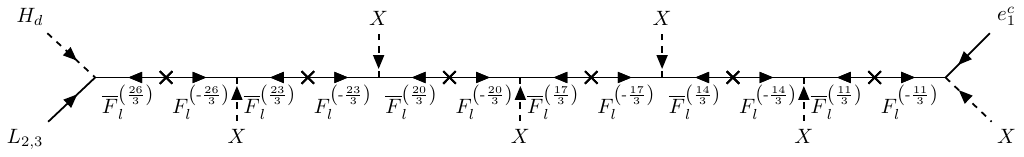}
    
\caption{Diagrams generating the lepton Yukawa sector of Model II given in Eq.~(\ref{eq:modelIIUV}).}
\label{fig:modelIIlUV} 
\end{figure}

\begin{figure}[!htbp]
\centering
\begin{minipage}{0.35\textwidth}
        \centering
        \includegraphics[width=\linewidth]{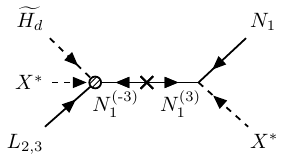}
        \centering
        \includegraphics[width=\linewidth]{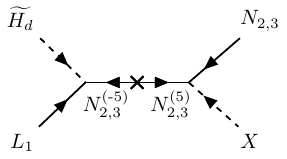}
    \end{minipage}
    \hfill
\begin{minipage}{0.55\textwidth}
        \centering
        \includegraphics[width=\linewidth]{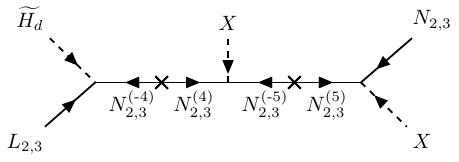}
        \includegraphics[width=\linewidth]{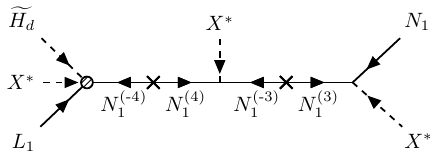}
    \end{minipage}
\caption{Diagrams generating the Dirac neutrino Yukawa sector of Model II given in Eq.~(\ref{eq:modelIIUV}) where the shaded blob vertices are Planck suppressed.}
\label{fig:modelIInuDUV} 
\end{figure}

\begin{figure}[!htbp]
    \centering
    \includegraphics[width=0.3\textwidth]{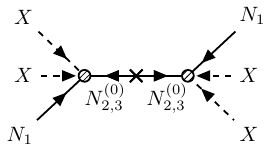}
    \includegraphics[width=0.65\textwidth]{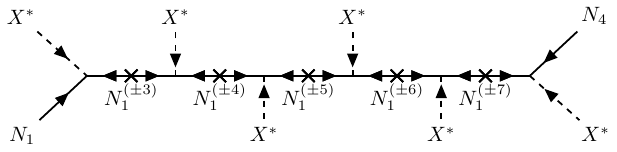}
    \includegraphics[width=0.75\textwidth]{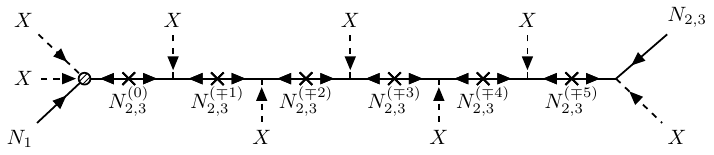}
    \includegraphics[width=1.0\textwidth]{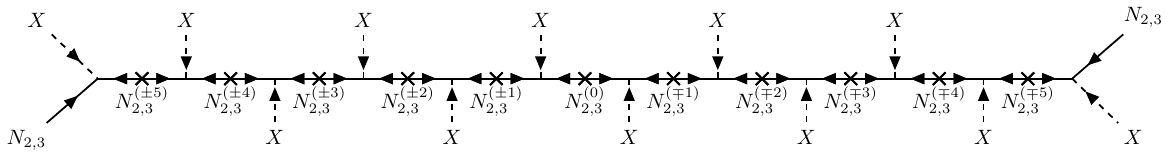}
\caption{Diagrams generating the Majorana neutrino Yukawa sector of Model II given in Eq.~(\ref{eq:modelIIUV}) where the shaded blob vertices are Planck suppressed.}
\label{fig:modelIInunuUV} 
\end{figure}

In this appendix, we provide the UV completion for the texture in Eq.~(\ref{eq:modelIIUV}).\footnote{For early attempts of UV-completing FN models with fermions, see Refs.~\cite{Calibbi:2012yj,Calibbi:2012at,Rathsman:2023cic}.} The Up quark sector can be UV completed with eight up-type VLF pairs 
\begin{equation}\label{eq:ModelIIuUVfields}
\begin{aligned}[b]
& \hspace{0.25cm}F_u^{\left(\minus\frac{2}{3}\right)}, F_u^{\left(\frac{1}{3}\right)},F_u^{\left(\frac{4}{3}\right)},F_u^{\left(\frac{7}{3}\right)},F_u^{\left(\frac{10}{3}\right)},F_u^{\left(\frac{13}{3}\right)},F_u^{\left(\frac{16}{3}\right)},F_u^{\left(\frac{19}{3}\right)},\\
&\overline{F}_u^{\left(\frac{2}{3}\right)}, \overline{F}_u^{\left(\minus\frac{1}{3}\right)},\overline{F}_u^{\left(\minus\frac{4}{3}\right)},\overline{F}_u^{\left(\minus\frac{7}{3}\right)},\overline{F}_u^{\left(\minus\frac{10}{3}\right)},\overline{F}_u^{\left(\minus\frac{13}{3}\right)},\overline{F}_u^{\left(\minus\frac{16}{3}\right)},\overline{F}_u^{\left(\minus\frac{19}{3}\right)},
\end{aligned}  
\end{equation}
and their conjugate counterparts $\overline{F}_u$ with opposite signed flavor charges seen in Fig.~\ref{fig:modelIIuUV}. As a general notation: the SM Quantum numbers of these VLFs are $F^{(q)}_{u,d,l}:(\{3,3,1\},1,Y_{u,d,l})$, $\overline{F}^{(q)}_{u,d,l}:(\{\bar{3},\bar{3},1\},1,Y_{u,d,l})$ respectively
Similarly we need five down-type VLFs to generate the down quark masses shown in Fig.~\ref{fig:modelIIdUV}: 
\begin{equation}\label{eq:ModelIIdUVfields}
\begin{aligned}[b]
& 
F_d^{\left(-3\right)},F_d^{\left(-4\right)},F_d^{\left(-5\right)},F_d^{\left(-6\right)},F_d^{\left(-7\right)},\\
&\hspace{0.5cm}\overline{F}_d^{\left(3\right)},\overline{F}_d^{\left(4\right)},\overline{F}_d^{\left(5\right)},\overline{F}_d^{\left(6\right)},\overline{F}_d^{\left(7\right)}.
\end{aligned}  
\end{equation}
For the charged lepton sector we need six additional electron type VLFs shown in Fig.~\ref{fig:modelIIlUV}:
\begin{equation}\label{eq:ModelIIlUVfields}
\begin{aligned}[b]
& F_l^{\left(\minus\frac{11}{3}\right)},F_l^{\left(\minus\frac{14}{3}\right)},F_l^{\left(\minus\frac{17}{3}\right)},F_l^{\left(\minus\frac{20}{3}\right)},F_l^{\left(\minus\frac{23}{3}\right)},F_l^{\left(\minus\frac{26}{3}\right)},\\
& \hspace{0.5cm} \overline{F}_l^{\left(\frac{11}{3}\right)},\overline{F}_l^{\left(\frac{14}{3}\right)},\overline{F}_l^{\left(\frac{17}{3}\right)},\overline{F}_l^{\left(\frac{20}{3}\right)},\overline{F}_l^{\left(\frac{23}{3}\right)},\overline{F}_l^{\left(\frac{26}{3}\right).}
\end{aligned}  
\end{equation}
The Dirac Neutrino sector does not require additional down type VLFs for their mass generation and is shown in Fig.~\ref{fig:modelIInuDUV}.
In order to generate the RHN Majorana sector that provides exactly two massive light neutrinos, we need to use the following field configuration to produce an invertible Majorana effective mass matrix:
\begin{equation}\label{eq:ModelIINUVfields}
\begin{aligned}[b]
&N_{2,3}^{\left(\pm5\right)},N_{2,3}^{\left(\pm4\right)},N_{2,3}^{\left(\pm3\right)},N_{2,3}^{\left(\pm2\right)},N_{2,3}^{\left(\pm1\right)},N_{2,3}^{\left(0\right)}\\
&\hspace{0.5cm}N_1^{(\pm3)},N_1^{(\pm4)},N_1^{(\pm5)},N_1^{(\pm6)},N_1^{(\pm7)}
\end{aligned}  
\end{equation}
and is shown in Fig.~\ref{fig:modelIInunuUV} where the ``$\mp$" in the third and fourth diagrams means that the VLF chain goes from $N_i^{(-n)}$ on the left to $N_i^{(n)}$  on the right of the mass insertion and ``$\pm$" means the VLF chain proceeds from $N_i^{(n)}$ on the left to $N_i^{(-n)}$  on the right of the mass insertion. We need copies of the VLFs depicted by the subscripts $i=1,2,3$ for avoiding low ranked effective mass matrices for the RHNs which participate in the light neutrino mass generation.
This can be seen if we explicitly integrate out the FN fields and perform the seesaw as illustrated in Eq.~(\ref{eq:FNRHNfunYuk}). 
The same goes for the $N_1,N_4$ entries which can be still generated at the renormalizable level and provide a significant contribution to the RHN mass spectrum, which require the additional fields in Eq.~(\ref{eq:ModelIINUVfields}) labeled with the 1 subscript. This ensures that both the Dirac and Majorana mass matrices are consistently generated by integrating out the FN fields. This model also remains perturbative up to the Planck scale.  This can be seen from the beta function coefficients, which are given by
\begin{equation}
b_3 = \frac{5}{3},~~~b_2 = -3,~~~b_1 = \frac{283}{9},
\end{equation}
which leads to a value of $\alpha_1(M_{\rm Pl}) = 0.03$ with $\Lambda_{\rm FN} = 3.9 \times 10^{16}$ GeV, which is in its allowed range, see Eq.~(\ref{eq:ModelIIparams}).

\subsection{Model III: Higgs doublet flavor UV completion}\label{subsec:ModelIIIUVcompletion}
\begin{figure}[!htbp]
\centering
  \begin{minipage}{0.4\textwidth}
        \centering
        \includegraphics[width=\linewidth]{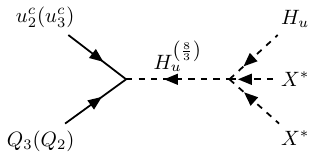}
    \end{minipage}
    \hfill
    \begin{minipage}{0.55\textwidth}
        \centering
        \includegraphics[width=\linewidth]{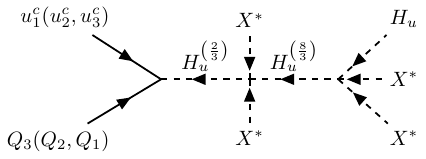}
\end{minipage}
\caption{Diagrams generating  up-type quark Yukawa operators of Model III given in Eq.~(\ref{eq:modelIIIUV}).}
\label{fig:modelIIIuUV}
\end{figure}

\begin{figure}[!htbp]
\centering
\begin{minipage}{0.45\textwidth}
        \centering
        \includegraphics[width=\linewidth]{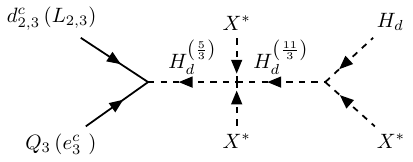}
    \end{minipage}
    \hfill
    \begin{minipage}{0.5\textwidth}v
        \centering
        \includegraphics[width=\linewidth]{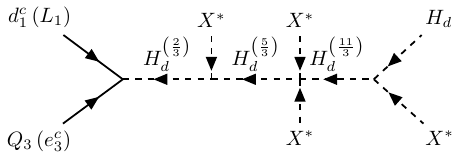}
\end{minipage}
\hfill
    \begin{minipage}{0.6\textwidth}
        \centering
        \includegraphics[width=\linewidth]{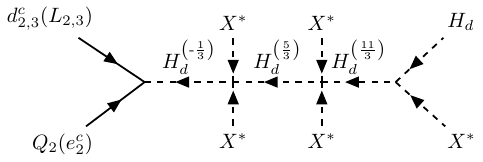}
\end{minipage}
\hfill
    \begin{minipage}{0.65\textwidth}
        \centering
        \includegraphics[width=\linewidth]{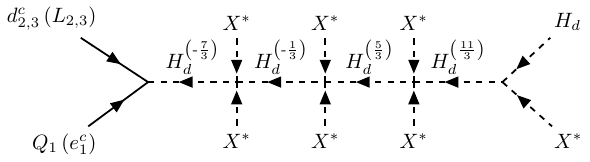}
\end{minipage}
\caption{Diagrams generating down type quark and charged lepton Yukawa operators for Model III given in Eq.~(\ref{eq:modelIIIUV}).}
\label{fig:ModelIIIdUV}
\end{figure}

\begin{figure}[!htbp]
\centering
\begin{minipage}{0.55\textwidth}
        \centering
        \includegraphics[width=\linewidth]{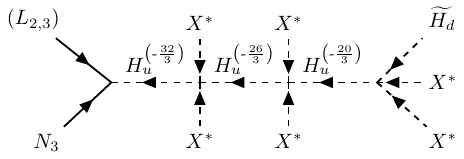}
    \end{minipage}
    \hfill
    \begin{minipage}{0.55\textwidth}
        \centering
        \includegraphics[width=\linewidth]{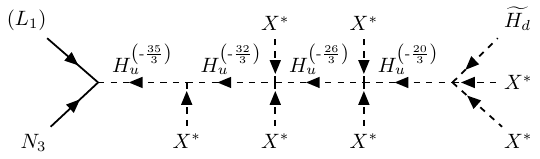}
\end{minipage}
\hfill
    \begin{minipage}{0.9\textwidth}
        \centering
        \includegraphics[width=\linewidth]{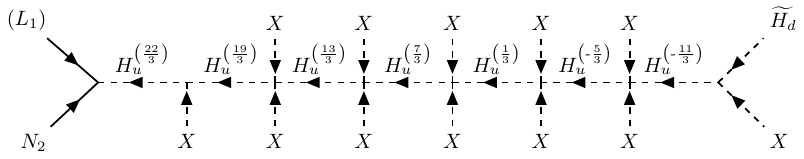}
\end{minipage}
\hfill
    \begin{minipage}{0.95\textwidth}
        \centering
        \includegraphics[width=\linewidth]{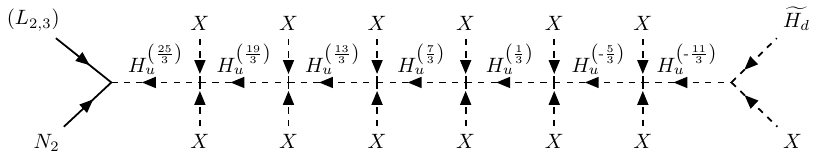}
\end{minipage}
\caption{Diagrams generating down type quark and charged lepton Yukawa operators for Model III given in Eq.~(\ref{eq:modelIIIUV}).}
\label{fig:ModelIIInuDUV}
\end{figure}

\begin{figure}[!htbp]
\centering
        \centering
        \includegraphics[width=0.3\textwidth]{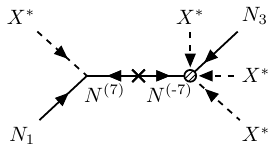}
        \includegraphics[width=1.0\linewidth]{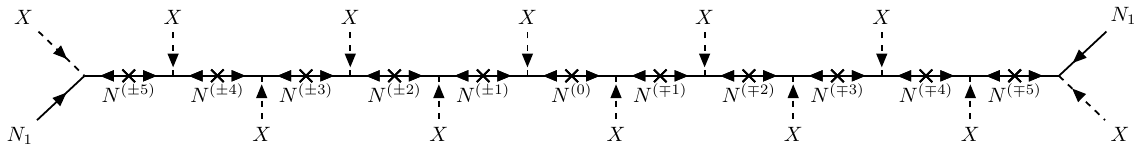}
\caption{Diagrams generating the Majorana neutrino Yukawa sector of Model III given in Eq.~(\ref{eq:modelIIIUV}) where the shaded blob vertex is Planck suppressed.}
\label{fig:modelIIInunuUV} 
\end{figure}

The UV completion for the texture in Eq.~(\ref{eq:modelIIIUV}) is constructed with additional heavy Higgs doublets at $\Lambda_{\rm FN}$. We require two additional up-type Higgs doublets to generate the up quark masses shown in Fig.~\ref{fig:modelIIIuUV}:
\begin{equation}\label{eq:ModelIIIUVufields}
    \{H_u^{\left(\frac{2}{3}\right)},H_u^{\left(\frac{8}{3}\right)}\}\supset\{\overline{\varepsilon}^4,\overline{\varepsilon}^2\}H_u
\end{equation}
We require five additional down-type Higgs doublets to generate the down quark and charged lepton masses shown in Fig.~\ref{fig:ModelIIIdUV}:
\begin{equation}\label{eq:ModelIIIUVdfields}
    \{H_d^{\left( \minus\frac{7}{3}\right)},H_d^{\left(\minus\frac{1}{3}\right)},H_d^{\left( \frac{2}{3}\right)},H_d^{\left( \frac{5}{3}\right)},H_d^{\left( \frac{11}{3}\right)}\}\supset\{\varepsilon^7,\varepsilon^5,\overline{\varepsilon}^4,\overline{\varepsilon}^3,\overline{\varepsilon}\}H_d
\end{equation}
We require eleven additional up-type Higgs doublets (or $\widetilde{H_d}$ type) to generate the Dirac Neutrino Yukawa shown in Fig.~\ref{fig:ModelIIInuDUV}:
\begin{equation}\label{eq:ModelIIIUVuDfields}
\begin{aligned}[b]    
   & \{ H_u^{\left( \minus\frac{32}{3}\right)}, H_u^{\left( \minus\frac{26}{3}\right)},H_u^{\left( \minus\frac{20}{3}\right)}, H_u^{\left( \minus\frac{11}{3}\right)}, H_u^{\left( \minus\frac{5}{3}\right)}, H_u^{\left( \frac{1}{3}\right)}, H_u^{\left( \frac{7}{3}\right)}, H_u^{\left( \frac{13}{3}\right)}, H_u^{\left( \frac{19}{3}\right)}, H_u^{\left( \frac{22}{3}\right)}, H_u^{\left( \frac{25}{3}\right)}\}\\
    & \hspace{3.cm}\equiv \{\overline{\varepsilon}^6,\overline{\varepsilon}^4,\overline{\varepsilon}^2,\varepsilon ,\varepsilon^3,\varepsilon^5,\varepsilon^7,\varepsilon^9,\varepsilon^{11},\varepsilon^{12},\varepsilon^{13}\}\widetilde{H_d}
\end{aligned}
\end{equation}
where the $\varepsilon$ factors basically mean the chain of $H_{u,d}^{(q)}$ that ends in $H_{u,d}$ with incremental couplings to the flavon $X$. Here, we can use the fact that four of the $H_u$ doublets can be identified to be
\begin{equation}
    \{H_u^{\left( \minus\frac{11}{3}\right)}, H_u^{\left( \minus\frac{5}{3}\right)}, H_u^{\left( \frac{1}{3}\right)}, H_u^{\left( \frac{7}{3}\right)}\}\equiv \{\widetilde{H_d}^{\left( \frac{11}{3}\right)}, \widetilde{H_d}^{\left( \frac{5}{3}\right)}, \widetilde{H_d}^{\left( \minus\frac{1}{3}\right)}, \widetilde{H_d}^{\left( \minus\frac{7}{3}\right)}\}
\end{equation} 
which are already listed $H_d$ doublets in Eq.~(\ref{eq:ModelIIIUVdfields}). Therefore we only require seven additional $H_u$ type doublets required for the Dirac Yukawa sector.
For the Majorana neutrino sector, we need two terms to be UV completed. The first is to generate the $y^N_{13},y^N_{31}$ elements using one Planck suppressed vertex and one other vertex with $N_1N^{(\minus7)}X^*$ illustrated in Fig.~\ref{fig:modelIIInunuUV}.
Finally, similar to Model II, we need a single long chain diagram to generate the Majorana neutrino mass elements connecting $N_{1},N_1$. This can be facilitated by adding 12 additional VLFs:
\begin{equation}
\begin{aligned}[b]
&N^{\left(\pm7\right)},N^{\left(\pm5\right)},N^{\left(\pm4\right)},N^{\left(\pm3\right)},N^{\left(\pm2\right)},N^{\left(\pm1\right)},N^{\left(0\right)}\\
\end{aligned}  
\end{equation}
Here the $(\mp)$ in the second diagram in Fig.~\ref{fig:modelIIInunuUV} means the VLF chain goes from $N^{(-n)}$ on the left to $N^{(n)}$  on the right of the mass insertion like the earlier case of Model II.
In this UV completion of Model III, the beta function coefficients for the evolution of the three gauge couplings take values given by
\begin{equation}
b_3 = -7,~~~b_2 = -\frac{2}{3},~~~b_1 = \frac{28}{3}~.
\end{equation}
This leads to a value $\alpha_1(M_{\rm Pl}) = 0.02$, where we have used $\Lambda_{\rm FN} = 5 \times 10^{13}$ GeV, consistent with Eq.~(\ref{eq:modelIIILambda_FN}). 

\section{Additional cases of high quality axion models}\label{sec:pseudoadd}
\subsection{Additional sources of pseudoscalar Higgs mass from the UV theory: New cases for Models I and III}\label{subsec:pseudoaddI}
Upon UV completion using Higgs doublets in Models I and III, additional contributions to the Higgs pseudoscalar mass apart from the operator in Eq.~(\ref{eq:pot}) arise at the flavor scale involving the FN Higgs doublets. This is readily apparent for the case of $(n, k)=(3,1^-)$ in Models I and III from the operators in Eq.~(\ref{Hud-n3k-1}) and the first term of Eq.~(\ref{eq:ModelIIIPQV}) respectively. These operators provide the dominant contribution to the Higgs pseudoscalar mass but they exhibit lesser powers of inverse $M_{\rm Pl}$ than the corresponding operator in Eq.~(\ref{eq:pot}). We can categorize such operators with lower Planck suppression which is dependent on $n$ as follows:
\begin{enumerate}
    \item $\boldsymbol{n=4}$: $d=6$ operators of the form $S^4H_u^{(q_1)}H_d^{(q_2)}/M_{\rm Pl}^2$.
    \item $\boldsymbol{n=3}$: $d=5$ operators of the form $S^3H_u^{(q_1)}H_d^{(q_2)}/M_{\rm Pl}$.
    \item $\boldsymbol{n=2}$: $d=5$ operators of the form $S^2X^{(*)}H_u^{(q_1)}H_d^{(q_2)}/M_{\rm Pl}$.
    \item $\boldsymbol{n=2}$: $d=4$ renormalizable operators of the form $S^2H_u^{(q_1)}H_d^{(q_2)}$.
    \item $\boldsymbol{n=1}$: $d=4$ renormalizable operators of the form $S~X^{(*)}H_u^{(q_1)}H_d^{(q_2)}$.
    \item $\boldsymbol{n=1}$: $d=3$ cubic operators of the form $S~H_u^{(q_1)}H_d^{(q_2)}.$
\end{enumerate}
 When Eq.~(\ref{eq:pot}) was the leading term, $M_{\text{A}_{\text{H}}}<100$ GeV for $|k|>1$. The above types of operators when present, for specific values of $(n,k)$ and the specific UV completion fields listed in Secs.~\ref{subsec:ModelIaxionquality} and \ref{subsec:ModelIIIUVcompletion} act as the leading source of the Higgs pseudoscalar mass and allow more values of $k$ by evading the lower bound of the pseudoscalar Higgs mass constraint. This in turn, opens up possible cases with higher $k$ values. Utilizing the wisdom gained in analyzing various cases in detail in Models I and III, we can eliminate all the cases except for the first two category of operators i.e. all categories with $n<3$ do not have a high quality axion because of the general loop contribution to the axion potential in Eq.~(\ref{PQvil-Hud2}) for Model I and Eq.~(\ref{eq:ModelIIIPQvil-Hud}) for Model III. Therefore, we list the additional possible cases in Model I for the first two categories of $n=4, 3$, considering the UV-completion heavy Higgs fields $\left( H_u^{(q_1)},~H_d^{(q_2)}\right)$ from Eqs. (\ref{eq:compHu})-(\ref{eq:compHd}) below
\begin{table}[H]
    \centering
    \renewcommand{\arraystretch}{1.5}
    \begin{tabular}{|c||c|>{\columncolor{green!10}}c|c|c|>{\columncolor{green!10}}c|>{\columncolor{green!10}}c|>{\columncolor{green!10}}c|>{\columncolor{green!10}}c|>{\columncolor{green!10}}c|c|}
        \hline
        $(n,k)$  &$(3,7^{-})$& $(4,7^{-})$ & $(3,6^{-})$   & $(3,2^{-})$  &  $(3,2)$ & $(3,4)$ &$(4,5^{-})$& $(4,3^{-})$ & $(4,1^{-})$ \\
        \hline
        $q_S$ & $11/9$ & $11/12$ & $8/9$  & $-4/9$  & $-16/9$ & $-22/9$ & $5/12$&$-1/12$ &$-7/12$ \\
        \hline
        $q_1$ & $-1$ & $-1$ & $0$ & $0$  & $2$ & $4$ & $-1$&$-1$ & $-1$\\
        \hline
        $q_2$ & $-8/3$ & $-8/3$ & $-8/3$ & $4/3$  & $10/3$ & $10/3$ & $-2/3$&$4/3$ & $10/3$ \\
        \hline
    \end{tabular}
    \caption{Additional high quality cases that satisfy neutrino fit in Model I for $n=3,4$, the flavor charge of $S$, $q_S$ for each case and the flavor charges of the up-type $(q_1)$ and the down-type $(q_2)$ Higgs doublets that induce the leading pseudoscalar mass. The columns in green allow axion as DM.}
    \label{tab:ModelIadd}
\end{table}
The properties for these additional cases in Table~\ref{tab:ModelIadd} can be straightforwardly deduced following the analyses in the illustrative cases in Sec.~\ref{subsec:ModelIaxionquality}. Similarly, we take into account the Higgs UV completion heavy fields in Eqs. (\ref{eq:ModelIIIUVufields})-(\ref{eq:ModelIIIUVdfields})-(\ref{eq:ModelIIIUVuDfields}) for Model III and list the additional possible cases for $n=3$ in Table~\ref{tab:ModelIIIadd1}
\begin{table}[H]
    \centering
    \renewcommand{\arraystretch}{1.3}
    \begin{tabular}{|c||c|c|c|c|c|c|c|>{\columncolor{green!10}}c|>{\columncolor{green!10}}c|}
        \hline
        $k$  & $11^{-}$ & $9^{-}$ & $8^{-}$ & $7^{-}$ & $6^{-}$ & $5^{-}$ &  $4^{-}$ & $3^{-}$ & $2^{-}$  \\
        \hline
        $q_S$ & $5/9$ & $-1/9$ & $-4/9$ & $-7/9$ & $-10/9$ & $-13/9$  & $-16/9$ & $-19/9$ & $-22/9$ \\
        \hline
        $q_1$ & $2/3$ & $2/3$ & $2/3$ & $2/3$ & $8/3$ & $8/3$  & $2/3$ & $8/3$ & $8/3$  \\
        \hline
        $q_2$ & $-7/3$ & $-1/3$ & $2/3$ & $5/3$ & $2/3$ & $5/3$  & $14/3$ & $11/3$& $14/3$ \\
        \hline
    \end{tabular}
    \caption{Additional high quality cases that satisfy neutrino fit in Model III for $n=3$, the flavor charge of $S$, $q_S$ for each case and the flavor charges of the up-type $(q_1)$ and the down-type $(q_2)$ Higgs doublets that induce the leading pseudoscalar mass. The columns in green allow axion as DM.}
    \label{tab:ModelIIIadd1}
\end{table}
Additionally for $n=4, d=6$ operators, we have the following cases for Model III in Table~\ref{tab:ModelIIIadd2}
\begin{table}[H]
    \centering
    \renewcommand{\arraystretch}{1.3}
    \begin{tabular}{|c||>{\columncolor{green!10}}c|>{\columncolor{green!10}}c|>{\columncolor{green!10}}c|>{\columncolor{green!10}}c|>{\columncolor{green!10}}c|>{\columncolor{green!10}}c|}
        \hline
        $k$  & $11^{-}$ & $9^{-}$ & $7^{-}$  & $5^{-}$ & $3^{-}$  & $1^{-}$ \\
        \hline
        $q_S$ & $5/12$ & $-1/12$  & $-7/12$  & $-13/12$   & $-19/12$  & $-25/12$\\
        \hline
        $q_1$ & $2/3$ & $2/3$  & $2/3$  & $8/3$   & $8/3$ & $14/3$ \\
        \hline
        $q_2$ & $-7/3$ & $-1/3$  & $5/3$  & $5/3$  & $11/3$& $11/3$\\
        \hline
    \end{tabular}
    \caption{Additional high quality cases that satisfy neutrino fit in Model III for $n=4$, the flavor charge of $S$, $q_S$ for each case and the flavor charges of the up-type $(q_1)$ and the down-type $(q_2)$ Higgs doublets that induce the leading pseudoscalar mass. The columns in green allow axion as DM.}
    \label{tab:ModelIIIadd2}
\end{table}
Again, we can deduce from the operator in Eq.~(\ref{eq:ModelIII2loopn3k-1PQ}) and its equivalents in each of the cases in Tables~\ref{tab:ModelIIIadd1}~-~\ref{tab:ModelIIIadd2}, that the columns in green also have axion as DM in addition to having a high quality axion producing the correct neutrino fit and baryon asymmetry. The mass of the Higgs pseudoscalar arising for all the cases possible in Tables~\ref{tab:ModelIadd} and \ref{tab:ModelIIIadd1},~\ref{tab:ModelIIIadd2} are independent of the $k$ values and can be written as
\begin{equation}\label{eq:Amassadd}
    M_{\text{A}_\text{H}}^2\simeq \lambda^{'}_{\text{HS}}\frac{1+\tan^2\beta}{\tan \beta}\left| \frac{-3\,n\sqrt{q_X^2 + q_S^2r^2} }{q_X\sqrt{2}}\right|^{3}\frac{|f_a|^{3,4}}{M_{\text{Pl}}^{(1,2)}}\epsilon^{(h_u-q_1)}\epsilon^{(h_d-q_2)},
\end{equation}
which are the contributions from the first two $n=3,d=5$ and $n=4,d=6$ operator categories respectively. The flavor phenomenology however might be interesting due to the unique values of $q_S$ charges in each case and can consequentially lead to exact cancellation in couplings as was observed for the astrophobic cases in Model I in Sec.~\ref{sec:experimentalprobes}.

\subsection{Higher flavor scale: New cases for Model II}\label{subsec:pseudoaddII}
The higher value of $\Lambda_{\rm FN}$ in Model II (see Eq.~(\ref{eq:modelIILambda_FN})), allows these additional following cases: $n=1,k=\pm4,\pm5,\pm6,\pm7;~n=2,k=\pm3,\pm4,\pm5,\pm6;~n=3,k=\pm2,\pm3;~n=4,k=\pm 1$. These arise solely from more values of $(n,k)$ satisfying the $M_{\text{A}_{\text{H}}}<100~\text{GeV}$ bound from Eqs. (\ref{eq:pot})-(\ref{Amass}). Thus, in Table~\ref{tab:ModelIIadd} we list all $(n,k)$ selections giving acceptable values of $M_{A_H}$, high quality axion, axion DM, the correct neutrino fit and baryon asymmetry.
\begin{table}[H]
    \centering
    \renewcommand{\arraystretch}{1.2}
    \small 
    \resizebox{\textwidth}{!}{
    \begin{tabular}{|c||>{\columncolor{green!10}}c|>{\columncolor{green!10}}c|>{\columncolor{green!10}}c|>{\columncolor{green!10}}c|>{\columncolor{green!10}}c|>{\columncolor{green!10}}c|>{\columncolor{green!10}}c|>{\columncolor{green!10}}c|>{\columncolor{green!10}}c|}
        \hline
        $(n,k)$ & $(1,7^{-})$ & $(1,6^{-})$ & $(1,5^{-})$ & $(2,6^{-})$ & $(2,4^{-})$  & $(3,3^{-})$ & $(3,2^{-})$ & $(3,2^{+})$ & $(3,3^{+})$ \\
        \hline
        $q_S$ & $40/3$ & $37/3$ & $34/3$  & $37/6$ & $31/6$  & $28/9$ & $25/9$ & $13/9$ & $10/9$ \\
        \hline
    \end{tabular}
    }
    \caption{Additional high quality cases in Model II, with $(n,k)$ values, $q_S$ flavor charges. All the columns in green solve all the puzzles in the SM listed in Table~\ref{tab:summary} simultaneously.}
    \label{tab:ModelIIadd}
\end{table}
The list of additional cases in Tables~\ref{tab:ModelIadd}, \ref{tab:ModelIIIadd1}, \ref{tab:ModelIIIadd2},~\ref{tab:ModelIIadd} are a summary of the final results of the analyses of Secs.~\ref{subsec:ModelIaxionquality}, \ref{subsubsec:ModelIIAxionquality} and \ref{subsec:ModelIIIaxionquality} respectively performed case by case. For conciseness, specific details of these analyses have been omitted as they are very similar to the ones in the illustrations. 

\section{Domain Wall number for Multi-Doublet UV completion}\label{sec:NDWgen}
We generalize the computation of $N_{\rm DW}$ in Sec.~\ref{sec:Domainwallproblem} for the case of an arbitrary number scalars in the cases of Higgs flavor UV completion. For each scalar we still use the  parameterization $\varphi_i=(v_i/\sqrt{2})\exp\left(i\eta_i/v_i\right)$. We consider the Higgs UV completion of Model I, see Eqs.~(\ref{eq:compHu})-(\ref{eq:compHd}), which consists of additional six Higgs doublets with masses of $\mathcal{O}(\Lambda_{\rm FN})$. The following analysis is valid equivalently for the UV completion of Model III specified  in Eqs.~(\ref{eq:ModelIIIUVufields})-(\ref{eq:ModelIIIUVdfields}) as well. The full scalar content of Model I is
\begin{equation}
\varphi_i=\{H_u, H_d, H_u^{(2)}, H_u^{(0)},H_u^{(-1)}, H_d^{\left(\frac{-8}{3}\right)}, H_d^{\left(\frac{4}{3}\right)}, H_d^{(\frac{10}{3})}, X, S\}.
\label{eq:HFN-scalaers}
\end{equation}
For their VEVs\footnote{Although the heavy Higgses have masses of $\mathcal{O}(\Lambda_{\rm FN})$, it can still be established generically that each of them has a slightly different mass owing to their independent quartic couplings. Thus the analysis when simplified to all the heavy Higgses of same masses, results in Eq.~(\ref{eq:fa0}) identically.} and phases we have respectively
$$
v_i=\{v_u, v_d, v_{u_1} , v_{u_2} , v_{u_3}, v_{d_1} , v_{d_2} , v_{d_3}, v_X, v_S  \},
$$
\begin{equation}
\eta_i=\{\eta_u, \eta_d,\eta_u^{(2)}, \eta_u^{(0)},\eta_u^{(-1)}, \eta_d^{\left(\frac{-8}{3}\right)}, \eta_d^{\left(\frac{4}{3}\right)}, \eta_d^{\left(\frac{10}{3}\right)}, \eta_X, \eta_S\},
\label{eq:HFN-param1}
\end{equation}
where for the VEVs we have defined
\begin{equation}
v_{u_1}=\langle H_u^{(2)}\rangle ,~~v_{u_2}=\langle H_u^{(0)}\rangle ,~~v_{u_3}=\langle H_u^{(-1)}\rangle
,~~v_{d_1}=\langle H_d^{\left(\frac{-8}{3}\right)}\rangle ,~~v_{d_2}=\langle H_d^{\left(\frac{4}{3}\right)}\rangle ,~~v_{d_3}=\langle H_d^{\left(\frac{10}{3}\right)}\rangle .
\label{eq:FN-vevs}
\end{equation}
Here the phase of each FN doublet has the superscript of the corresponding  $U(1)_{F}$ charge, indicating  which doublet they are coming
from. From the couplings of Eq.~(\ref{eq:ModelIfundyukud}) the following $\alpha_i,  i \in \{1, \dots, 6\}$ phase combinations become massive:
$$
\alpha_1=\frac{\eta_u}{v_u}-\frac{\eta_u^{(2)}}{v_{u_1}}-2\frac{\eta_X}{v_X},~~
\alpha_2=\frac{\eta_u^{(2)}}{v_{u_1}}-\frac{\eta_u^{(0)}}{v_{u_2}}-2\frac{\eta_X}{v_X},~~
\alpha_3=\frac{\eta_u^{(0)}}{v_{u_2}}-\frac{\eta_u^{(-1)}}{v_{u_3}}-\frac{\eta_X}{v_X},
$$
\begin{equation}
\alpha_4=\frac{\eta_d}{v_d}-\frac{\eta_d^{\left(\frac{-8}{3}\right)}}{v_{d_1}}-2\frac{\eta_X}{v_X},~~
\alpha_5=\frac{\eta_d^{\left(\frac{4}{3}\right)}}{v_{d_2}}-\frac{\eta_d^{\left(\frac{10}{3}\right)}}{v_{d_3}}+2\frac{\eta_X}{v_X},~~
\alpha_6=\frac{\eta_d}{v_d}-\frac{\eta_d^{\left(\frac{4}{3}\right)}}{v_{d_2}}+2\frac{\eta_X}{v_X}.
\label{eq:FN-phases}
\end{equation}
From (\ref{eq:FN-phases}) we have:
\begin{equation}
\begin{aligned}[b]    
&\frac{\eta_u^{(2)}}{v_{u_1}} = \frac{\eta_u}{v_u} - 2\frac{\eta_X}{v_X} - \alpha_1 ,~~
\frac{\eta_u^{(0)}}{v_{u_2}} = \frac{\eta_u}{v_u} - 4\frac{\eta_X}{v_X} - \alpha_1 - \alpha_2 ,~~
\frac{\eta_u^{(-1)}}{v_{u_3}} = \frac{\eta_u}{v_u} - 5\frac{\eta_X}{v_X} - \alpha_1 - \alpha_2 - \alpha_3 ,\\
&\frac{\eta_d^{\left(\frac{-8}{3}\right)}}{v_{d_1}} = \frac{\eta_d}{v_d} - 2\frac{\eta_X}{v_X} - \alpha_4  ,~~
\frac{\eta_d^{\left(\frac{4}{3}\right)}}{v_{d_2}} = \frac{\eta_d}{v_d} + 2\frac{\eta_X}{v_X} - \alpha_6  ,~~
\frac{\eta_d^{\left(\frac{10}{3}\right)}}{v_{d_3}} = \frac{\eta_d}{v_d} + 4\frac{\eta_X}{v_X} - \alpha_5 - \alpha_6   .
\label{eta-ud-phases}
\end{aligned}
\end{equation}
The quark Yukawa couplings with the Higgs doublets are of the form:
\begin{equation}
\left( Q_1, Q_2, Q_3\right) 
\left(
  \begin{array}{ccc}
    0 & H_u^{(\minus1)} & 0 \\
    0 & H_u^{(0)} & H_u^{(2)} \\
    H_u^{(\minus1)} & H_u^{(2)} & H_u \\
  \end{array}
\right)
\left(
  \begin{array}{c}
    u^c_1 \\
    u^c_2 \\
   u^c_3\\
  \end{array}
\right)\!,~
\left( Q_1, Q_2, Q_3\right) 
\left(
  \begin{array}{ccc}
    H_d^{\left(\frac{10}{3}\right)}  &0& 0 \\
    0 & H_d^{\left(\frac{\minus8}{3}\right)}  & H_d^{\left(\frac{\minus8}{3}\right)}  \\
    0 &H_d & H_d \\
  \end{array}
\right)
\left(
  \begin{array}{c}
    d^c_1 \\
    d^c_2 \\
   d^c_3\\
  \end{array}
\right)
\label{FN-q-yuk}
\end{equation}

With the quark field redefinitions:
$$
Q_1\to e^{i\left( \frac{\eta_u^{(2)}}{v_{u_1}}-\frac{\eta_u^{(-1)}}{v_{u_3}}\right)}Q_1,~~
Q_2\to e^{i\left( \frac{\eta_u}{v_u}-\frac{\eta_u^{(2)}}{v_{u_1}}\right)}Q_2,~~
Q_3\to Q_3,
$$
$$
u^c_1\to e^{-i \frac{\eta_u^{(-1)}}{v_{u_3}}}u^c_1,~~
u^c_2\to e^{-i \frac{\eta_u^{(2)}}{v_{u_1}}}u^c_2,~~
u^c_3\to e^{-i \frac{\eta_u}{v_u}}u^c_3,~~
$$
\begin{equation}
d^c_1\to e^{i\left( \frac{\eta_u^{(-1)}}{v_{u_3}}-\frac{\eta_u^{(2)}}{v_{u_1}}-\frac{\eta_d^{\left(\frac{10}{3}\right)}}{v_{d_3}}\right)}d^c_1,~~
d^c_{2,3}\to e^{-i \frac{\eta_d}{v_d}}d^c_{2,3},
\label{FN-rephase}
\end{equation} 
and taking into account (\ref{eta-ud-phases}) one can verify that
all the phases of (\ref{eq:HFN-param1}) are rotated away from the quark Yukawa couplings (\ref{FN-q-yuk}). 

With these phase shifts the loop induced $\eta_iG\widetilde G$ coupling will be proportional
to the $SU(3)_c^2\times (phase)$ anomaly. The latter is proportional to the sum of the corresponding quark phases:
$$
2\frac{\eta_u^{(2)}}{v_{u_1}}-2\frac{\eta_u^{(-1)}}{v_{u_3}}+2\frac{\eta_u}{v_u}
-2\frac{\eta_u^{(2)}}{v_{u_1}}-\frac{\eta_u^{(-1)}}{v_{u_3}}
-\frac{\eta_u^{(2)}}{v_{u_1}}-\frac{\eta_u}{v_u}+\frac{\eta_u^{(-1)}}{v_{u_3}}-\frac{\eta_u^{(2)}}{v_{u_1}}-\frac{\eta_d^{\left(\frac{10}{3}\right)}}{v_{d_3}}-
2\frac{\eta_d}{v_d}=
$$
$$
=-2\frac{\eta_u^{(2)}}{v_{u_1}}-2\frac{\eta_u^{(-1)}}{v_{u_3}}+\frac{\eta_u}{v_u}-\frac{\eta_d^{\left(\frac{10}{3}\right)}}{v_{d_3}}-
2\frac{\eta_d}{v_d}\to -3\left( \frac{\eta_u}{v_u}+\frac{\eta_d}{v_d}-\frac{10}{3}\frac{\eta_X}{v_X}\right) \to
$$
\begin{equation}
\to -3\left( \frac{K_u}{v_u}+\frac{K_d}{v_d}-\frac{h_u+h_d}{q_X}\frac{K_X}{v_X}\right) a . 
\label{33-phase}
\end{equation}
In (\ref{33-phase}) we have taken into account (\ref{eta-ud-phases}) and at the last stage extracted the axion field from $\eta_u, \eta_d$
and $\eta_X$ and set $\frac{10}{3}=\frac{h_u+h_d}{q_X}$. We see that the $aG\widetilde G$ coupling has the same structure as the last term of Eq.~(\ref{eq:fa0}).

Therefore,  the domain wall number $N_{\rm DW}$ can be   determined from the equation
\begin{equation}
\frac{n_a}{3}\!\left( \!\frac{K_u}{v_u}\!+\!\frac{K_d}{v_d}\!-
\!\frac{h_u\!+h_d}{q_X}\frac{K_X}{v_X}\!\right)^{\!-1}
\!\!\!=\!\sum_i n_iv_iK_i,
\label{HFN-ns}
\end{equation}
where $n_a$ is axion's winding number and $N_{\rm DW}={\rm min}|n_a|\neq 0$ while the axion field is the following superposition from Eq.~(\ref{eq:axion}):
\begin{equation}
a=\sum_i K_i\eta_i,~i\in\{u,d,u_1,u_2,u_3,d_1,d_2,d_3,X,S\} .
\label{HFN-ax}
\end{equation}
Subsequently, Eqs. (\ref{HFN-ns}) and (\ref{HFN-ax}) are valid for all models we have considered irrespective of the number of heavy Higgs UV completion fields as is required by construction. In the basis of Eq.~(\ref{eq:HFN-param1}), the corresponding  $n_i$  and $K_i$ appearing in  Eqs.~(\ref{HFN-ns}) and (\ref{HFN-ax}) are:

$$
K_i=\{K_u, K_d, K_{u_1} , K_{u_2} , K_{u_3}, K_{d_1} , K_{d_2} , K_{d_3}, K_X, K_S  \},
$$
\begin{equation}
n_i=\{n_u, n_d, n_{u_1} , n_{u_2} , n_{u_3}, n_{d_1} , n_{d_2} , n_{d_3}, n_X, n_S  \}.
\label{K-n}
\end{equation}
The pseudoscalar $A_H$, which gains its mass via the term in Eq.~(\ref{eq:pot}), is the same as earlier in Eq.~(\ref{eq:A}). 
However, the NG states are now the following superpositions:
\begin{equation}
G_Z=N_Z\left( v_u\eta_u-v_d\eta_d+\sum_{i=1}^3v_{ui}\eta_u^{(q_{ui})}-\sum_{i=1}^3v_{di}\eta_d^{(q_{di})}\right) ,~~~
G_F=N_F\sum q_iv_i\eta_i .
\label{FN-golds}
\end{equation}
Orthogonality conditions of the axion state Eq.~(\ref{HFN-ax}) with the states given in Eqs.~(\ref{eq:A})-(\ref{eq:FN-phases}) and (\ref{FN-golds}) read
\begin{equation}
M_K\vec \eta_K=\vec w_K K_S,
\label{ort-FN}
\end{equation}
where 
$$
M_K=\left(
      \begin{array}{ccccccccc}
        v_{u_1}v_X & 0 & -v_{u}v_X & 0 & 0 & 0 & 0 & 0 & -2v_uv_{u_1} \\
        0 & 0 & v_{u_2}v_X & -v_{u_1}v_X & 0 & 0 & 0 & 0 & -2v_{u_1}v_{u_2} \\
        0 & 0 & 0 & v_{u_3}v_X  & -v_{u_2}v_X  & 0 & 0 & 0 & -v_{u_2}v_{u_3} \\
        0 & v_{d_1}v_X & 0 & 0 & 0 & -v_{d}v_X & 0 & 0 & -2v_dv_{d_1}  \\
        0 & 0 & 0 &0 & 0 & 0 & v_{d_3}v_X & -v_{d_2}v_X & 2v_{d_2}v_{d_3} \\
        0 & v_{d_2}v_X &0 & 0 & 0 & 0& -v_dv_X & 0 & 2v_dv_{d_2} \\
        v_dv_Xv_S & v_uv_Xv_S & 0 & 0 & 0 & 0 & 0 & 0 & \pm kv_uv_dv_S \\
        v_u & -v_d & v_{u_1} & v_{u_2} & v_{u_3} & -v_{d_1}  & -v_{d_2} & -v_{d_3} & 0 \\
        4v_u & -\frac{2}{3}v_d  &2v_{u_1}  & 0 & -v_{u_3}  &-\frac{8}{3}v_{d_1}& \frac{4}{3}v_{d_2} & \frac{10}{3}v_{d_3} & v_X 
      \end{array}
    \right),
$$
$$
\vec \eta_K^{~T}=\left( K_u, K_d, K_{u_1} , K_{u_2} , K_{u_3}, K_{d_1} , K_{d_2} , K_{d_3}, K_X \right) ,
$$
\begin{equation}
\vec w_K^{~T} =\left( 0,0,0,0,0,0,-nv_uv_dv_X,0,-q_Sv_S \right) .
\label{MK-etaK--wK}
\end{equation}
These relations allow us to express the 9 components of $\vec \eta_K$ in terms of $K_S$ as:
\begin{equation}
\vec \eta_K =M_K^{-1}\vec w_K K_S .
\label{Ks-FN}
\end{equation}
In addition, we have the normalization condition
\begin{equation}
\sum_iK_i^2=1,
\label{norm-K}
\end{equation}
which allows to determine all $K_i$. Plugging them into Eq.~(\ref{HFN-ns}), 
one can verify that this can be satisfied for arbitrary values of the VEVs, if the following conditions are satisfied:
$$
n_a\!=\!3(n_u +n_d)\!-\!10 n_X,~~~n_S=\frac{1}{10n}\left( kn_a-(10+3k)(n_u+n_d)\right),
$$
$$
n_{u_1}\!=\! \frac{1}{5}(n_a+ 2 n_u-3n_d)  ,~~~
n_{u_2}\!=\! \frac{1}{5}(2n_a-n_u-6n_d)   ,~~~ 
n_{u_3}\!=\!   \frac{1}{2}(n_a-n_u-3n_d)   ,
$$
\begin{equation}
n_{d_1}\!=\! \frac{1}{5}(n_a-3 n_u+2n_d)  ,~~~
n_{d_2}\!=\! \frac{1}{5}(3n_u+8n_d-n_a)  ,~~~
n_{d_3}\!=\! \frac{1}{5}(6n_u+11n_d-2n_a).
\label{eqs-FNn}
\end{equation}
For a given $(n,k)$, Eq.~(\ref{eqs-FNn}) can have solutions for integer $n_i$'s. For instance:
$$
{\rm For} ~(n,k)=(3, 1^{+}):\hspace{0.5cm} n_a\!=\! 1,~n_d\!=\! 7-n_u,~n_X\!=\! 2,~ n_S\!=\! -3,~n_{u_1}\!=\! n_u-4,
$$
\begin{equation}
n_{u_2}\!=\! n_u-8,~n_{u_3}\!=\! n_u-10,~
n_{d_1}\!=\! 3-n_u,~n_{d_2}\!=\! 11-n_u,~n_{d_3}\!=\! 15-n_u.
\label{NDW-n3k1}
\end{equation}

$$
{\rm For} ~(n,k)=(3, 1^{-}):\hspace{0.5cm} n_a\!=\! 1,~n_d\!=\! -n_u- 13,~n_X\!=\! -4,~ n_S\!=\! 3,~n_{u_1}\!=\! n_u+8,
$$
\begin{equation}
n_{u_2}\!=\! n_u+16,~n_{u_3}\!=\! n_u+20,~
n_{d_1}\!=\! -5-n_u,~n_{d_2}\!=\! -21-n_u,~n_{d_3}\!=\! -29-n_u.
\label{NDW-n3k-1}
\end{equation}
We have given the solutions with minimal positive integer $n_a=1$, which means that $N_{\rm DW}=1$ and is unaltered from Eq.~(\ref{Eq:NDW}).

Note that the result (\ref{eqs-FNn}) is valid also for cases when pseudoscalar Higgs gets mass via the UV sector. For instance, for the $(n,k)=(3,1^{-})$ case, the
relevant coupling is $S^3H_u^{(-1)}H_d^{\left(\frac{10}{3}\right)}$ from Table~\ref{tab:ModelIadd}. For this case, the following combination of the phases gains the mass
\begin{equation}
\alpha_H=3\frac{\eta_S}{v_S}+\frac{\eta_u^{(-1)}}{v_{u_3}}+\frac{\eta_d^{(10)}}{v_{d_3}}.
\label{alH-FN}
\end{equation}
Using Eq.~(\ref{alH-FN}) in Eq.~(\ref{eta-ud-phases}), instead of Eq.~(\ref{eq:A}) we obtain
\begin{equation}
\alpha_H=\frac{\eta_u}{v_u}+\frac{\eta_d}{v_d}-\frac{\eta_X}{v_X}+3\frac{\eta_S}{v_S}-\alpha_1-\alpha_2 -\alpha_3-\alpha_5-\alpha_6.
\label{alH-FN1}
\end{equation}
Using orthogonality of the axion $a$ with this state (and taking into account that $\alpha_i$'s orthogonality with $a$ is already imposed) we will get the  condition equivalent to the orthogonality of $A_H$ and $a$. Therefore, the results will be the same. Similar arguments apply to the cases
given in Appendix~\ref{subsec:pseudoaddI} (see Tables~\ref{tab:ModelIadd},~\ref{tab:ModelIIIadd1} and \ref{tab:ModelIIIadd2}).

\bibliographystyle{JHEP}
\bibliography{biblio}

\end{document}